\documentclass[fleqn,usenatbib]{mnras}

\usepackage[T1]{fontenc}
\usepackage{ae,aecompl}
\usepackage{graphicx}
\usepackage{amsmath,amssymb}
\usepackage{booktabs}
\usepackage{threeparttable}
\usepackage{siunitx}
\usepackage{xcolor}
\usepackage{hyperref}
\usepackage{enumitem}
\usepackage{multirow}

\hypersetup{
  colorlinks=true,
  linkcolor=blue,
  citecolor=blue,
  urlcolor=blue
}

\usepackage{comment}
\usepackage{subcaption}
\newcommand{\dustpy}{\textsc{dustpy}}
\newcommand{\radmc}{\textsc{radmc-3d}}

\newcommand{\alphavisc}{\alpha}
\newcommand{\vfrag}{v_{\mathrm{frag}}}

\newcommand{\twelveco}{\mbox{$^{12}$CO}}
\newcommand{\thirteenco}{\mbox{$^{13}$CO}}
\newcommand{\ceighteeno}{C$^{18}$O}
\newcommand{\ntwohp}{N$_2$H$^+$}

\usepackage[normalem]{ulem}

\usepackage{tikz}
\usetikzlibrary{positioning, arrows.meta, calc}
\title[Outer pressure bumps beyond the dust disc]{Missing Outer Rings in the exoALMA Sample: Physical Mechanisms Suppressing Dust Ring Formation Despite Strong Pressure Bumps}

\author[Lilian Luo et al.]{
Lilian Luo,$^{1}$\thanks{E-mail: lilian.luo.22@ucl.ac.uk},
Paola Pinilla$^{1}$,
Jochen Stadler$^{2}$
\\
$^{1}$Mullard Space Science Laboratory, University College London, Holmbury St Mary, Dorking, Surrey RH5 6NT, UK\\
$^{2}$ European Southern Observatory, Karl-Schwarzschild-Str. 2, 85748 Garching bei München, Germany\\
}

\date{Accepted XXX. Received YYY; in original form ZZZ}
\pubyear{2026}

\begin{document}
\label{firstpage}
\pagerange{\pageref{firstpage}--\pageref{lastpage}}
\maketitle

\begin{abstract}
The exoALMA survey has revealed candidate gas pressure bumps at radii beyond the observed millimetre continuum discs in several protoplanetary systems, raising the question of whether these regions could host undetected dust structures. We investigate this for nine of the exoALMA discs by reconstructing gas surface density profiles from midplane pressure gradient measurements derived from $^{12}$CO and $^{13}$CO ALMA Band 7 observations. We evolve the dust using \dustpy\ across a grid of turbulence and fragmentation parameters ($\alphavisc \in \{{10^{-4},10^{-3}}\}$; $\vfrag \in \{{1,10}\} \, \mathrm{m\,s^{-1}}$), followed by radiative transfer calculations with \radmc\ to give predictions for the continuum emission. We find that the dominant outcome across the explored parameter space is efficient outer dust trapping, leading to prominent outer continuum rings that should be detectable at ALMA Band 7, in tension with the observations. The $^{13}$CO-based reconstructions almost universally produce detectable outer rings, whereas the $^{12}$CO-based reconstructions show more variation. We further test delayed formation of the outer pressure bump and enhanced outer-disc turbulence as possible suppression mechanisms of ring formation. Enhanced turbulence weakens the predicted outer continuum emission most effectively, although the efficiency of the mechanism depends on the system explored and on where the turbulence transition is placed. Delayed bump formation generally does not prevent rapid outer dust accumulation.
\end{abstract}


\section{Introduction}
\label{sec:intro}
High-resolution millimetre continuum observations with ALMA have shown that protoplanetary discs commonly have substructures in the form of rings, gaps, spirals and azimuthal asymmetries \citep[e.g.][]{Andrews_2018, ALMA_part_2015, Huang_2018, Long_2018,Cieza_2021, vanderMarel_2019, Jennings_2022, Guerra_alvarado2025, Bosschaart_2026}. 

Early surveys of the brightest and most spatially extended discs, such as the DSHARP survey  \citep{Andrews_2018, Huang_2018} demonstrated that these features are widespread. Similar structures were also identified across discs in Taurus and Ophiuchus star forming regions \citep{Long_2018, Cieza_2021}. Subsequent work has moved towards showing the high prevalence of substructures throughout the broader disc population: \citet{vanderMarel_2019} found that ring/gap systems span a wide range of stellar ages and luminosities, while parametric and non-parametric visibility fitting techniques demonstrated that substructures could be recovered at sub-beam resolution, including in systems with fainter and more compact emission \citep[e.g.][]{Long_2019, Jennings_2022}. More recently, \citet{Bosschaart_2026} found that rings and gaps are a near-universal trait of extended discs ($R_{dust}> 30$\,au) in nearby star-forming regions. However, this picture may not extend to the entire disc population; \citet{Guerra_alvarado2025} found that two-thirds of the Lupus discs are compact ($R_{dust} < 30$\,au) and look smooth at $\sim$4\,au resolution, suggesting that substructures could be more prominent primarily in the larger and more massive systems.
The AGE-PRO large programme \citep{Zhang_2025}, which targeted a representative sample spanning the disc lifetime in the Ophiuchus ($\lesssim1\,$Myr), Lupus ($\sim1-3\,$Myr) and Upper Scorpius region ($\sim2-6\,$Myr), provided some further insights. Visibility fitting of the continuum emission of the resolved  discs revealed substructures in $\sim$80 \% in the Ophiuchus discs, providing support for early formation of substructures \citep{Vioque_2025}, in agreement with the results from \cite{Hsieh_2025} showing that dust substructures are already common in  Class I discs. Substructures were similarly detected in the majority of well-resolved discs across the regions in the AGE-PRO sample; however, many of the compact discs (particularly in Lupus) are only marginally resolved, and some showed tentative visibility variations suggestive of unresolved structure \citep{Vioque_2025}. Consistent with this, comparisons with dust evolution modelling suggest that the evolution of spectral index and continuum flux of the AGE-PRO sample are better reproduced by models with dust traps \citep{Kurtovic_2025, Luo_2026}. This could imply that substructures are common even in compact discs but are hidden below current resolution limits. The nature and prevalence of substructures in the faint and compact discs are therefore still open questions. 

These observations have motivated extensive efforts to link observed dust structures with underlying disc physics. Disc substructures are commonly interpreted as signatures of dust trapping at local gas pressure maxima, where the radial inward drift of solids is halted \citep[e.g.][]{Weidenschilling_1977, Whipple_1972}. These pressure maxima can arise through a diverse set of physical mechanisms, including gravitational interactions with planets embedded in the disc \citep{Pinilla_2012, Dong_2015, Dipierro_2015}, self-gravitating spiral arms \citep{Dipierro_2015b}, and magnetohydrodynamic effects such as zonal flows and dead-zone edges \citep[e.g.][]{Johansen_2009, Flock_2015} (see \citet{Bae_2023} for a comprehensive review of proposed mechanisms). In such regions, particles accumulate and grow, producing the ring-like features seen in millimetre continuum emission. Theoretical and numerical studies have shown that these pressure perturbations can efficiently concentrate dust and create long-lived substructures \citep[e.g.][]{Pinilla_2012,Dullemond_2018,Stammler_2019}. Observational support for dust trapping has emerged from ALMA observations, where gas kinematic perturbations and deviations from Keplerian rotation are found to be spatially associated with continuum rings and gaps, consistent with underlying pressure structure \citep[e.g.][]{Pinte_2018,Teague_2018,Rosotti_2020, Izquierdo_2023}.

Line-intensity maps of CO isotopologues are only an indirect probe of the midplane pressure structure, since the mapping from the gas column density to pressure is complicated by optical depth and thermal effects. Comparisons of the CO line emission and continuum substructures find that CO gaps tend to coincide with dust gaps but are often shallower and some deep dust gaps lack any clear CO counterpart \citep{Zhang_2021}. However the statistical significance of any continuum and line emission substructure correlation has been questioned \citep{Jiang_2022}.

A more direct tracer is therefore the gas kinematics: deviations of the rotation curve from Keplerian rotation reflect the radial pressure gradient in the disc midplane, and can reveal pressure perturbations even where the corresponding change in gas column density is weak or ambiguous. This motivates the use of rotation curves to locate candidate pressure substructures.

\begin{figure*}
    \centering
    \includegraphics[width=0.96\textwidth]{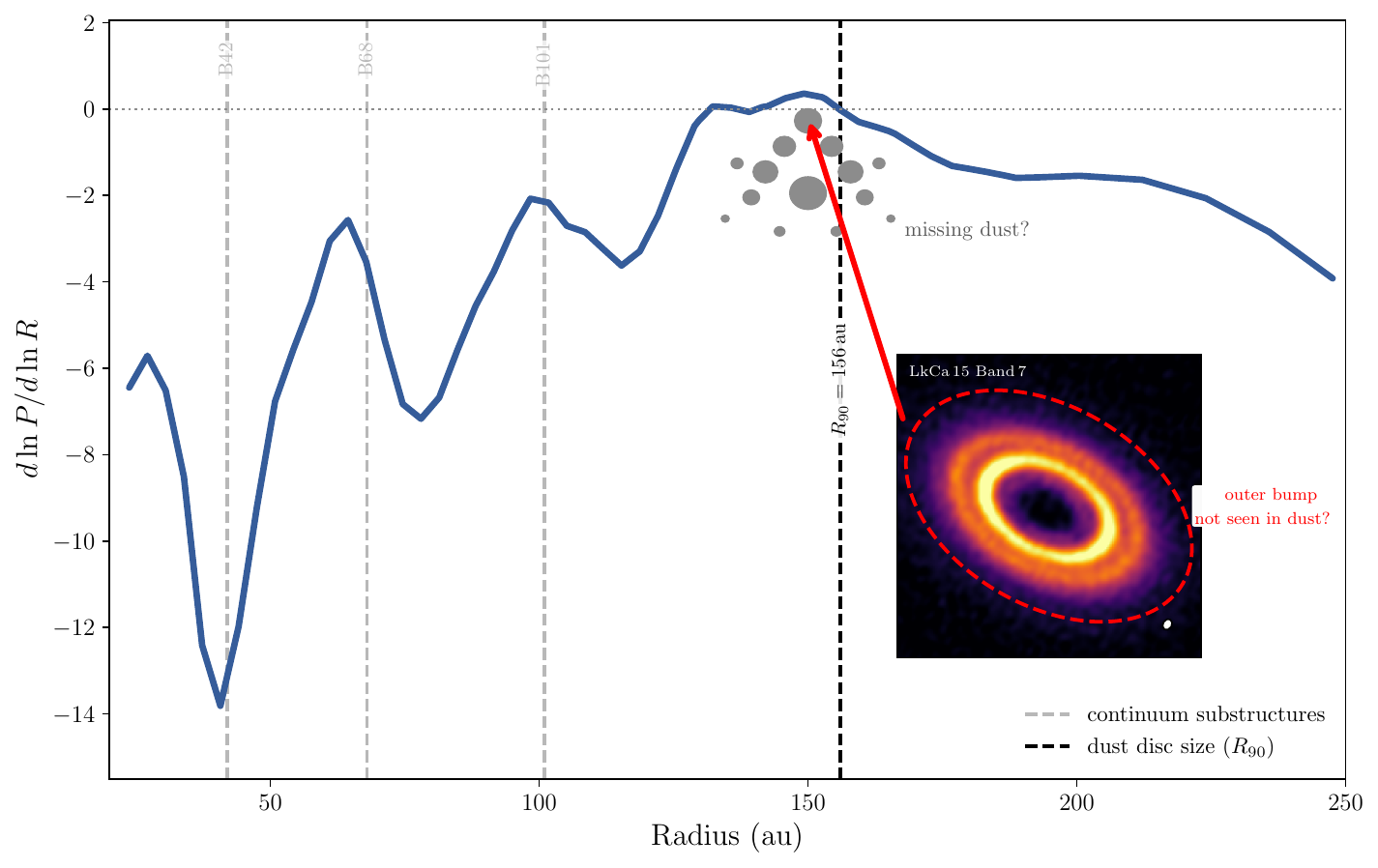}
    \caption{Illustration of the missing dust rings in pressure bumps located outside the dust disc, shown here for LkCa\,15 as a representative example. \textit{inset:} ALMA Band~7 ($\sim$870\,$\mu$m) continuum image. \textit{Plot:} Mid-plane pressure-gradient profile, $d\ln P / d\ln R$, taken from exoALMA \twelveco~ observations and combining small- and large-scale measurements. Local variations in the pressure gradient trace candidate pressure perturbations that may act as dust traps. Within $R_{90, obs}$ (black dashed vertical line) the pressure structures coincide with observed continuum rings (grey dashed vertical lines represent the locations of the observed bright rings), while beyond the dust disc, there is no millimetre emission at the location of the gas pressure bump. }
    \label{fig:Lkca15_bump_gas_vs_dust}
\end{figure*}

The exoALMA survey \citep{2025_Teague} has enabled uniform, high-resolution analyses of gas emission and inferred midplane properties for a sample of nearby protoplanetary discs. By combining high spectral and spatial resolution observations of CO isotopologues, exoALMA gives detailed constraints on the gas kinematics and thermodynamic structure of several bright discs, offering a powerful means of probing embedded perturbations such as those induced by forming planets. Analysis of exoALMA rotation curves has revealed substructures in the radial derivative of the midplane gas pressure in several systems \citep{Stadler_2025}. Many of these features were shown to coincide with known dust continuum rings, supporting the interpretation that local pressure maxima act as efficient dust traps. However, a notable subset of systems exhibits additional pressure-gradient substructures at radii beyond the observed millimetre continuum disc edge. As illustrated in LkCa 15 in Fig.~\ref{fig:Lkca15_bump_gas_vs_dust}, the pressure structures within $R_{90, obs}$ are well matched by the observed continuum rings, while those at large radii lack any corresponding millimetre emission.

These outer gas substructures should, in principle, act as efficient dust traps, yet they do not appear to produce any detectable continuum rings. Understanding why is therefore essential for interpreting the relationship between gas dynamics and dust evolution in protoplanetary discs. This raises a key question: are solids genuinely absent at these radii,
or could dust be present but difficult to detect due to grain growth,
fragmentation, radial drift, or opacity effects? Some observational
evidence for the latter has recently been identified: \citet{Das_2026}
found extended low surface brightness, ``halo", emission in the outer
regions of three systems, including DM\,Tau from the exoALMA sample.
This component accounts for 20--30\% of the total submillimetre flux
and a significant fraction of the total dust mass, yet is cold and
faint enough to be missed entirely in shallower surveys. Whether
efficient trapping at a resolved outer gas substructure produces a
localised continuum ring, more diffuse halo-like emission, or no
detectable signature at all depends on the dust microphysics governing
how solids respond to the bump. This is the question we address through
forward modelling of dust evolution and synthetic continuum emission.

In this work, we consider two questions. First, if the outer pressure substructures inferred from exoALMA kinematics can act as dust traps, do standard dust evolution models predict outer continuum rings that exoALMA should have detected? Second, where such rings are predicted but not observed, which physical conditions render the trap sufficiently permeable, or the trapped dust sufficiently faint (the ``hidden dust'' scenario) to escape detection? To answer these questions, we reconstruct $\Sigma_{\rm g}(r)$ for nine exoALMA discs from the \twelveco\ and \thirteenco\ pressure-derivative profiles of \citet{Stadler_2025}, evolve the dust with \dustpy\ across a grid of turbulence and fragmentation parameters ($\alphavisc \in  {\{10^{-4},10^{-3}}\}$ and $\vfrag \in {\{1,10}\}\,\mathrm{m\,s^{-1}}$), and predict Band~7 continuum emission with \radmc.

This paper is organised as follows. Section~\ref{sec:methods} summarises the modelling pipeline (Fig.~\ref{fig:flowchart}). Section~\ref{sec:results} presents the result of the entire sample, two contrasting discs, and two suppression tests. Section~\ref{sec:discussion} discusses the implications and limitations, and Section~\ref{sec:conclusions} summarises our conclusions.

\section{Methods}
\label{sec:methods}
\subsection{Overview}
\label{sec:methods_overview}

Figure~\ref{fig:flowchart} summarises our modelling pipeline. For each of the nine discs, we reconstruct two independent gas surface density profiles from the \twelveco\ and \thirteenco\ pressure-derivative measurements of \citet{Stadler_2025} (Section~\ref{sec:methods_gas}), evolve the dust in each for four $(\alphavisc, \vfrag)$ combinations (Section~\ref{sec:methods_dustpy}), and post-process every model into a synthetic Band~7 image whose emission beyond the observed continuum edge is classified as \emph{Ring}, \emph{Halo} or \emph{No outer mm} (Section~\ref{sec:methods_rt}). This gives 72 reference models (9 discs $\times$ 2 tracers $\times$ 4 configurations). Independently, we identify candidate outer pressure substructures from the rotation curves (Section~\ref{sec:methods_candidate_regions}); these are used only for comparison with the predicted emission and do not enter the models. Finally, we test two mechanisms that could suppress outer ring formation; a late-forming outer bump and enhanced outer-disc turbulence, on a subset of discs (Section~\ref{sec:methods_tests}). The stellar and disc properties of the sample are listed in Table~\ref{tab:sample}.

\begin{figure*}
\centering
\begin{tikzpicture}[
  font=\footnotesize,
  every node/.append style={execute at begin node={\hyphenpenalty=10000\exhyphenpenalty=10000}},
  stage/.style={draw=blue!55!black, line width=0.8pt, rounded corners=3pt, fill=blue!6,
               align=center, text width=3.0cm, minimum height=3.0cm, inner sep=5pt},
  input/.style={draw=black!45, rounded corners=3pt, fill=gray!12, align=center, inner sep=4pt, minimum height=1.5cm},
  test/.style={draw=orange!70!black, line width=0.8pt, rounded corners=3pt, fill=orange!10,
               align=left, inner sep=5pt},
  aux/.style={draw=black!60, dashed, rounded corners=3pt, fill=white, align=center, inner sep=5pt},
  arr/.style={-{Stealth[length=2.4mm]}, line width=0.9pt},
  lab/.style={font=\scriptsize, align=center, text=black!75},
  ttl/.style={font=\small\bfseries},
  sec/.style={font=\scriptsize, text=black!55},
]
\node[stage] (gas) {{\small\bfseries 1. Gas reconstruction}\\[4pt]
  $\Sigma_{\rm g}(r)$ from the pressure gradient, normalised to $M_{\rm gas}$\\[4pt]
  {\scriptsize\color{black!55}Sect.~\ref{sec:methods_gas}}};

\node[stage, right=1.3cm of gas] (dust) {{\small\bfseries 2. Dust evolution}\\[4pt]
  \dustpy, 1\,Myr, fixed $\Sigma_{\rm g}$\\[2pt]
  $\alphavisc \in \{10^{-4}, 10^{-3}\}$\\
  $\vfrag \in \{1, 10\}$\,m\,s$^{-1}$\\[4pt]
  {\scriptsize\color{black!55}Sect.~\ref{sec:methods_dustpy}}};

\node[stage, right=1.3cm of dust] (rt) {{\small\bfseries 3. Synthetic imaging}\\[4pt]
  \radmc\ Band~7 images, convolved with the observed beam\\[4pt]
  {\scriptsize\color{black!55}Sect.~\ref{sec:methods_rt}}};

\node[stage, right=1.3cm of rt] (cls) {{\small\bfseries 4. Detectability}\\[4pt]
  Emission above $3\sigma$ beyond $R_{90,\rm obs}$?\\[3pt]
  \textbf{Ring} / \textbf{Halo} / \textbf{No outer mm}\\[4pt]
  {\scriptsize\color{black!55}Sect.~\ref{sec:methods_rt}}};

\draw[arr] (gas) -- node[lab, below=2pt] {$\times$2\\tracers} (dust);
\draw[arr] (dust) -- node[lab, below=2pt] {$\times$4\\configs} (rt);
\draw[arr] (rt) -- node[lab, below=2pt] {= 72\\models} (cls);

\node[input, text width=3.0cm, above=0.8cm of gas] (inP) {\textbf{9 exoALMA discs}\\[2pt]
  $d\ln P_{\rm g}/d\ln r$ (\twelveco, \thirteenco)\\
  $T_{\rm mid}(r)$, $M_{\rm gas}$, $R_{\rm c}$};
\draw[arr] (inP) -- (gas);

\path (rt.north) -- (cls.north) coordinate[pos=0.5] (mid);
\node[input, text width=5.6cm, anchor=south] (inB) at ($(mid)+(0,0.8cm)$)
  {\textbf{exoALMA Band~7 continuum}\\[2pt] $R_{90,\rm obs}$, rms, beam};
\draw[arr] (inB.south -| rt.north) -- (rt.north);
\draw[arr] (inB.south -| cls.north) -- (cls.north);

\node[test, text width=7.0cm, anchor=north] (tests) at ($(dust.south)+(0,-0.8cm)$)
  {{\small\bfseries Suppression tests} (\twelveco\ only)\\[3pt]
   \textit{Late bump:} bump added after 1\,Myr, then +0.5\,Myr\\
   \hspace*{1em}(LkCa\,15, J1615)\\[2pt]
   \textit{Enhanced turbulence:} $\alpha_{\rm out}=10^{-2}$ beyond $R_{\rm dz}$\\
   \hspace*{1em}(LkCa\,15, J1615, AA\,Tau)};
\draw[arr] (tests) -- (dust);

\node[aux, text width=3.0cm, anchor=north] (cand) at ($(cls.south)+(0,-0.8cm)$)
  {\textbf{Candidate outer bumps}\\[2pt]
   Super-Keplerian zones in the rotation curves\\[2pt]
   \textit{Compared with predictions; NOT a model input}\\[2pt]
   {\scriptsize\color{black!55}Sect.~\ref{sec:methods_candidate_regions}}};
\draw[{Stealth[length=2.2mm]}-{Stealth[length=2.2mm]}, dashed, line width=0.8pt] (cand) -- (cls);

\end{tikzpicture}
\caption{Overview of the modelling pipeline. For each of the nine discs, we reconstruct two gas surface density profiles from the \twelveco\ and \thirteenco\ pressure-derivative profiles of \citet{Stadler_2025}, using the midplane temperature of \citet{Galloway-Sprietsma_2025}, the gas mass of \citet{Trapman_2025} and the characteristic radius of \citet{Longarini_2025}. Each profile is evolved with \dustpy\ for four $(\alphavisc, \vfrag)$ configurations and post-processed with \radmc, giving 72 reference models whose Band~7 emission beyond $R_{90,\rm obs}$ \citep{Curone_2025} is classified as Ring, Halo or No outer mm. Candidate outer-bump regions (dashed) are identified from the rotation curves for comparison only and do not enter the models. The suppression tests (orange) modify the gas history or the turbulence profile for selected discs.}
\label{fig:flowchart}
\end{figure*}

\subsection{Identifying candidate regions for outer bumps}
\label{sec:methods_candidate_regions}
 
To identify candidate outer pressure-substructure regions, we used
radial profiles of azimuthal velocity residuals from
exoALMA data products \citep{Stadler_2025}. The residual
$\delta v_\phi = v_{\rm rot} - v_{\rm Kep}$ measures how much
the observed gas rotation deviates from Keplerian. We express
this as a fractional deviation,
\begin{equation}
    \Delta v_\phi = 100 \times \frac{\delta v_\phi}{v_{\rm Kep}},
\end{equation}
with significance
\begin{equation}
    S = \frac{\Delta v_\phi}{\sigma_{\Delta v_\phi}},
    \qquad
    \sigma_{\Delta v_\phi} = 100 \times 
    \frac{\sigma_{\delta v_\phi}}{v_{\rm Kep}}.
\end{equation}
Here, $\sigma_{\delta v_\phi}$ is the statistical uncertainty on the azimuthal velocity residual provided in the \citet{Stadler_2025} data products, reflecting the azimuthal scatter of the rotation curve within each radial annulus. We adopt the rotation curves measured from the high surface brightness sensitivity image cubes ($0.30''$ beam; \citealt{Loomis_2025}), which offer the higher sensitivity to trace $\delta v_\phi$ in the faint outer disc.
We searched only beyond $R_{90}$, the radius enclosing 90\,\%
of the Band~7 continuum flux \citep{Curone_2025}, since inside $R_{90}$ the pressure maxima broadly coincide with continuum rings \citep[][Fig.~\ref{fig:Lkca15_bump_gas_vs_dust}]{Stadler_2025}, whereas our question concerns pressure substructures without a continuum counterpart. As a spatial
resolution criterion, we used the effective beam size
$\theta_{\rm beam} = \sqrt{\theta_{\rm maj}\theta_{\rm min}}$
(the geometric mean of the beam axes, converted to au), requiring
candidate regions to be wider than one beam.
 
We selected candidate regions as follows. We identified all
connected zones beyond $R_{90}$ where the gas is super-Keplerian
($\Delta v_\phi > 0$). A zone was kept only if (i) it contains at
least one point with $\Delta v_\phi > 1$\,\% and $S > 3$,
ensuring a physically meaningful and statistically significant
velocity excess, and (ii) its total width exceeds one beam,
$\Delta r > \theta_{\rm beam}$, ensuring the feature is
spatially resolved. Our use of super-Keplerian rotation as the selection criterion is physically motivated: the sign of $\Delta v_\phi$ follows that of $\mathrm{d}\ln P_{\rm g}/\mathrm{d}\ln r$, so that super-Keplerian rotation ($\Delta v_\phi > 0$) corresponds to a positive pressure gradient. A genuine pressure bump, where the radial drift of solids reverses, produces a transition from sub- to super-Keplerian and back to
sub-Keplerian rotation with increasing radius. However, we note that sub-Keplerian bumps in the gradient will still perturb the pressure and can locally decelerate the inward drift of solids, producing a ``traffic jam'' pile-up \citep[e.g.][]{vanderMarel2023}, but without reversing the drift.

This classification depends on the adopted stellar mass through $v_{\rm Kep}$, so marginal features could shift across the threshold (Appendix~\ref{app:candidates}). The criterion is used only to identify and report candidate regions; the dust evolution models are driven by the full pressure-gradient profile and are hence unaffected by how a region is labelled.
 
\begin{figure*}
    \centering
    \includegraphics[width=0.96\textwidth]{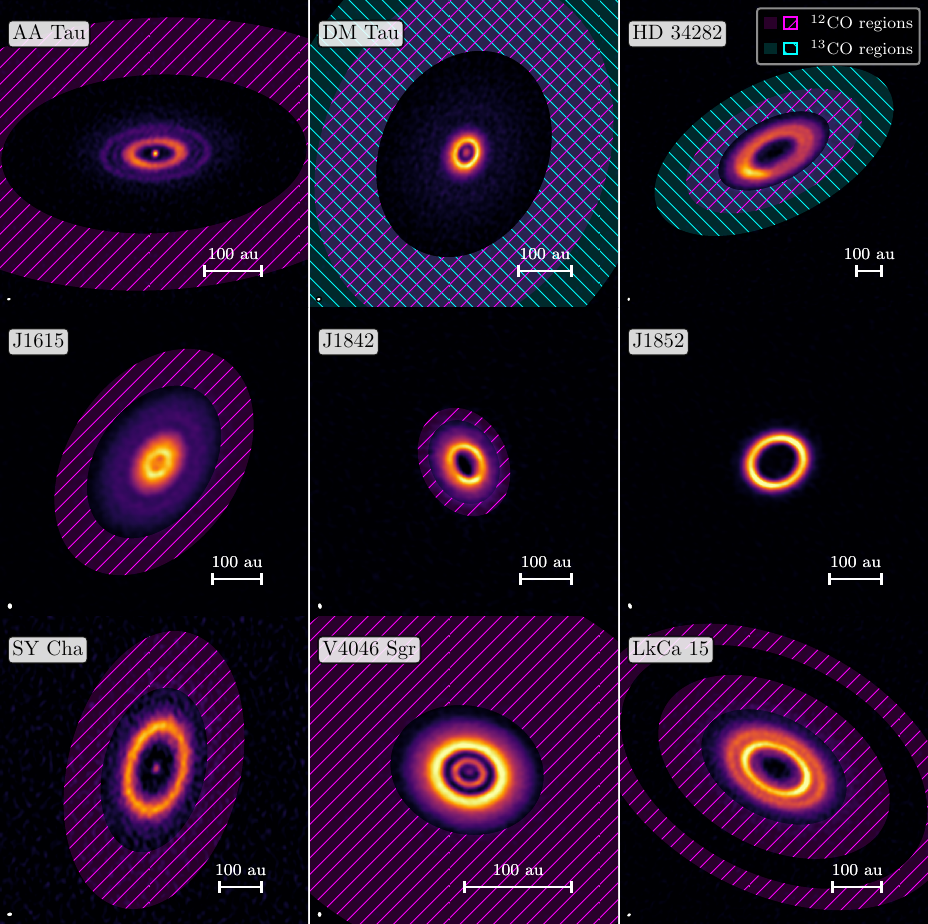}
    \caption{Gallery of exoALMA continuum images (using the robust parameter of $-0.5$) for the nine discs modelled in this work \citep[data from][]{Curone_2025}. Shaded regions
    indicate candidate outer pressure-substructure zones identified
    from super-Keplerian velocity deviations in the \citet{Stadler_2025}
    rotation curve profiles, exterior to the observed dust disc;
    pink and blue regions correspond to \twelveco\ and \thirteenco\
    selections respectively. The synthesised beam is shown in the
    lower-left corner of each panel.}
    \label{fig:exoALMA_gallery}
\end{figure*}

Applying this criterion to the \twelveco\ and \thirteenco\
rotation curves yields markedly different numbers of candidate
regions (Fig.~\ref{fig:exoALMA_gallery}). \twelveco\ identifies
regions in 8 of the 9 systems, while \thirteenco\ identifies
regions in only 2 (DM\,Tau and HD\,34282). In the remaining
systems, except for J\,1852, \thirteenco\ velocity deviations are present beyond $R_{90}$ but fall below the $S > 3$ significance threshold.
This could reflect lower signal-to-noise ratio (SNR) in the
\thirteenco\ profiles rather than a genuine absence of outer
pressure structure. The velocity profiles and selected zones are shown in Appendix~\ref{app:candidates}.

\subsection{Reconstructing 
  \texorpdfstring{$\Sigma_{\rm g}(r)$}{Sigma\_g(r)} 
  from pressure derivatives}
\label{sec:methods_gas}

We present our modelling approach in which the gas surface density profile input to \dustpy\ is derived directly from the observed midplane pressure-derivative profiles of \citet{Stadler_2025}:
\begin{equation}
    \frac{d\ln P_{\rm g}}{d\ln r}(r),
\end{equation}
rather than from a parametric prescription. This allows the dust evolution to be constrained by the actual kinematic structure of each disc.

We reconstruct $\Sigma_{\rm g}(r)$ in four steps, detailed in Appendix~\ref{app:reconstruction}: (i) we blend the small- and large-scale pressure-derivative profiles of \citet{Stadler_2025} and extend them beyond the observed radial range with a \citet{Lynden-Bell_1974} profile; (ii) we integrate to obtain $P_{\rm g}(r)$ up to a multiplicative constant; (iii) we convert pressure to surface density assuming a Gaussian vertical density profile and the disc-specific midplane temperature of \citet{Galloway-Sprietsma_2025}; and (iv) we normalise the profile to the total gas mass. The normalisation is necessary because the pressure gradient constrains only the shape of $P_{\rm g}(r)$, not its amplitude. We adopt the chemically estimated gas masses of \citet{Trapman_2025}, which are independent of the kinematic analysis used to derive the pressure gradients. These are typically a factor of $\sim$2-3 lower than the dynamical masses of \citet{Longarini_2025}; since the predicted ring brightness scales with the gas normalisation, our predictions are conservative in this respect.
 
We produce two independent $\Sigma_{\rm g}(r)$ profiles per 
system using the \thirteenco\ and \twelveco\ pressure-derivative 
measurements from \citet{Stadler_2025} as separate model inputs. 
Since \thirteenco\ is less optically thick than \twelveco, it 
traces deeper into the disc, and should in principle more accurately reflect the 
thermal and density structure of the midplane where dust 
trapping occurs \citep{Galloway-Sprietsma_2025}. 

\subsection{Dust evolution with \dustpy}
\label{sec:methods_dustpy}

We evolve dust in the reconstructed gas backgrounds using
\dustpy\ \citep{Stammler_2022}. \dustpy\ solves the Smoluchowski coagulation-fragmentation equation for the dust surface
density distribution $\Sigma_d(r,a)$, simultaneously
modelling radial drift, turbulent diffusion, and collisional
grain growth and fragmentation \citep{Birnstiel_2010}.
The gas surface density is held fixed throughout: gas evolution is switched off, so the reconstructed pressure bumps are imposed directly through $\Sigma_{\rm g}(r)$ rather than maintained through a radially varying $\alphavisc$ \citep[e.g.][]{Stammler_2023}; feedback from dust onto the gas is neglected.

Dust is initialised with a dust-to-gas mass ratio of 0.01
and an MRN grain size distribution \citep{Mathis_1977}
between $a_{\rm min} = 0.1$\,$\mu$m and $a_{\rm max} =
1.1$\,$\mu$m. The initial distribution is set up with the \dustpy\ option \texttt{allowDriftingParticles} set to \texttt{True}, so the full initial dust reservoir is retained at all radii, including grains that would already be in the drift limit at $t=0$.

At the low gas surface densities of the outer disc, even small grains have relatively large Stokes numbers, $\mathrm{St} \simeq \pi a \rho_{\rm s}/(2\Sigma_{\rm g})$ in the Epstein regime, where $a$ is the grain radius and $\rho_{\rm s}$ the grain material density, and begin to drift from the start of the simulation; this early depletion of the outer dust reservoir is followed self-consistently in our models.

We evolve all models for 1\,Myr. This should be interpreted as a characteristic dust evolution timescale following the formation of the bumps, rather than the actual disc age. The ages in our sample are poorly constrained and span a wide range ($\sim$1--24\,Myr)\citep{Huang_2017}, and the relevant timescale for dust accumulation in the bump is set by local drift and growth timescales rather than the age of the disc. Therefore, our models should not be intended to represent the full evolutionary history of each disc, but instead to assess whether these observed pressure structures are capable of producing detectable outer continuum emission, if given sufficient time for dust to accumulate. 

For each model we record the dust surface density distribution
$\Sigma_d(r,a)$, from which we derive the local
fragmentation and drift barriers ($a_{\rm frag}$,
$a_{\rm drift}$) to characterise the grain size distribution
(see Figs.~\ref{fig:lkca15_initial_dustpy}
and~\ref{fig:j1615_initial_dustpy}). Full model parameters are
summarised in Table~\ref{tab:DustPy_params}. 

We explore four combinations of turbulence parameter and
fragmentation velocity,
\begin{equation}
    \alphavisc \in \{10^{-4},\,10^{-3}\},
    \qquad
    \vfrag \in \{1,\,10\}\,\mathrm{m\,s^{-1}},
\end{equation}
spanning ranges motivated by observational and laboratory
constraints. Dust settling analyses of highly inclined discs
suggest $\alphavisc \lesssim 10^{-3}$
\citep[e.g.][]{Villenave_2022, Villenave_2025}, consistent
with gas evolution modelling of the AGE-PRO sample which
favours $\alphavisc \sim 2$--$4 \times 10^{-4}$
\citep{Tabone_2025}. For fragmentation velocity, silicate
grains fragment near $\vfrag \sim 1\,\mathrm{m\,s^{-1}}$
while icy grains may survive up to $\sim 10\,\mathrm{m\,s^{-1}}$,
though recent experiments suggest even icy aggregates can be
fragile \citep[e.g.][]{Gundlach_2015, Musiolik_2019}.
Models with $\alphavisc = 10^{-4}$ and
$\vfrag = 1\,\mathrm{m\,s^{-1}}$ were found to best
reproduce observed dust distributions in the AGE-PRO sample \citep{Luo_2026};
consistent conclusions have been reported by
\citet{Jiang_2024} and \citet{Tong_2025} for other discs. The broader parameter grid is included to assess the sensitivity of the outer ring predictions to these choices, and in particular
to explore scenarios where efficient grain growth concentrates
mass into large, poorly emitting grains that would be undetectable
at millimetre wavelengths despite being present in significant
quantities.

\begin{table*}
    \centering
    \caption{Parameters for the \dustpy\ dust evolution models.
    Varied parameters are in bold. *For systems (DM\,Tau, HD\,34282) the inner boundary is set to 20\,au, due to computational issues, 
    though this should not affect the outer disc region of interest.}
    \label{tab:DustPy_params}
    \begin{tabular}{lp{0.55\linewidth}}
        \hline
        \textbf{Parameter} & \textbf{Value} \\
        \hline
        Radial grid ($r_{\rm min}$) & 0.5\,au (20\,au for select systems*) \\
        Radial grid ($r_{\rm max}$) & Set by \citet{Stadler_2025} data extent \\
        Radial grid points & Scaled with $r_{\rm max}$ \\
        Initial dust size distribution & MRN power-law \citep{Mathis_1977} \\
        MRN slope & 3.5 \\
        Grain size grid ($a_{\rm min}$, $a_{\rm max}$) & $10^{-5}$\,cm, $10$\,cm \\
        \textbf{Fragmentation velocity} ($\vfrag$) & $1$, $10\,\mathrm{m\,s^{-1}}$ \\
        \textbf{Turbulence parameter} ($\alphavisc$) & $10^{-4}$, $10^{-3}$ \\
        CO tracer & $^{12}$CO, $^{13}$CO (independent models) \\
        Dust evolution time & 1\,Myr \\
        Gas surface density $\Sigma_{\rm g}(r)$ & Reconstructed from \citet{Stadler_2025}
            pressure derivatives; normalised to chemically 
            estimated gas masses from \citet{Trapman_2025} \\
        Gas evolution & Off ($\Sigma_{\rm g}$ fixed) \\
        \texttt{allowDriftingParticles} & \texttt{True} \\
        Midplane temperature $T_{\rm mid}(r)$ & Power-law fit from 
            \citet{Galloway-Sprietsma_2025} \\
        Characteristic radius ($R_c$) & From \citet{Longarini_2025} \\
        Initial dust-to-gas ratio & 0.01 \\
        Dust opacity & \citet{Ricci_2010} \\
        \hline
    \end{tabular}
\end{table*}

\begin{table*}
\centering
\caption{Stellar and disc properties of the nine exoALMA systems modelled in this work. $M_{\rm gas}$ is the chemically estimated gas mass adopted as the normalisation for the reconstructed $\Sigma_{\rm g}(r)$ profiles (Section~\ref{sec:methods_gas}). $M_\star$ and $d$ are from \citet{2025_Teague}; $T_{\rm eff}$, $L_\star$ and $M_{\rm gas}$ from \citet{Trapman_2025}; $R_{\rm c}$ from \citet{Longarini_2025}.}
\label{tab:sample}
\begin{tabular}{lcccccc}
\toprule
Disc & $d$ (pc) & $M_\star$ ($M_\odot$) & $T_{\rm eff}$ (K) & $L_\star$ ($L_\odot$) & $M_{\rm gas}$ ($M_\odot$) & $R_{\rm c}$ (au) \\
\midrule
AA Tau & 135 & 0.79 & 4350 & 1.10 & 0.024 & 156 \\
DM Tau & 144 & 0.45 & 3715 & 0.24 & 0.102 & 240 \\
HD 34282 & 309 & 1.61 & 9520 & 1.00 & 0.148 & 370 \\
J1615 & 156 & 1.14 & 4350 & 0.60 & 0.207 & 167 \\
J1842 & 151 & 1.07 & 4780 & 0.80 & 0.009 & 231 \\
J1852 & 147 & 1.03 & 4780 & 0.60 & 0.057 & 87 \\
SY Cha & 182 & 0.77 & 3792 & 0.47 & 0.017 & 112 \\
V4046 Sgr & 72 & 1.73 & 4370 & 0.50 & 0.035 & 99 \\
LkCa 15 & 157 & 1.14 & 4365 & 1.00 & 0.055 & 150 \\
\bottomrule
\end{tabular}
\end{table*}

\subsection{Synthetic continuum images and detectability}
\label{sec:methods_rt}
We post-process \dustpy\ outputs using \radmc\
\citep{Dullemond_2012} to generate synthetic Band~7
($\sim$870\,$\mu$m) continuum emission maps, using the following approach:

\subsubsection*{3D density structure}

The 1D radial dust surface density distribution
$\Sigma_d(r,a)$ from \dustpy\ is expanded onto a 3D
spherical grid assuming a Gaussian vertical dust profile \citep{Pohl_2016},
\begin{equation}
    \rho_d(R, z, a) = \frac{\Sigma_d(R,a)}{\sqrt{2\pi}\,H_d(R,a)}
    \exp\!\left(-\frac{z^2}{2H_d^2(R,a)}\right),
\end{equation}
where $H_d(a)$ is the dust scale height for each grain
size bin, computed following the standard settling
prescription \citep{Dubrulle_1995},
\begin{equation}
    H_d(a) = H_g\,\sqrt{\frac{\alphavisc}{\alphavisc + \mathrm{St}(a)}},
\end{equation}
with $H_g$ the gas scale height and $\mathrm{St}(a)$ the
Stokes number of grains of size $a$.

\subsubsection*{Dust temperature}

The dust temperature is computed with a \radmc\ thermal Monte Carlo calculation, using the stellar effective temperature and luminosity of each system listed in Table~\ref{tab:sample}. This temperature accounts for irradiation and shadowing by the modelled inner disc, and is computed independently of the midplane temperature $T_{\rm mid}(r)$ used in the dust evolution. For LkCa\,15, we confirmed that adopting the \dustpy\ $T_{\rm mid}(r)$ instead does not change the classification of the outer emission at any of the wavelengths tested.

\subsubsection*{Dust opacities}

We adopt dust opacities from \citet{Ricci_2010},
computed using \texttt{optool} \citep{optool} for a
composition of 21\% graphite, 36\% amorphous carbon,
28\% water ice, and 15\% silicate with 40\% porosity.
Ricci opacities have been shown to reproduce
millimetre fluxes and spectral indices of disc populations
better than the DSHARP prescription
\citep{Stadler_2022, Delussu_2024}, motivating our
choice.

\subsubsection*{Synthetic images}

Inclined synthetic images are generated at the observed
inclination and position angle of each disc and convolved
with a 2D Gaussian matching the observational beam.
We also generate synthetic images at longer wavelengths (1.3--8.6\,mm) for selected systems, to evaluate whether outer dust concentrations might become more detectable at longer wavelengths.

\subsubsection*{Radial profiles and detectability}

Deprojected radial intensity profiles are extracted
from the convolved synthetic images by rotating by the
position angle, dividing the declination axis by
$\cos(i)$, and azimuthally averaging in radial bins of one-fifth of the synthesised beam FWHM, i.e. $\approx$ 5 bins per beam.

Detectability is assessed against the observed sensitivity. Since the synthetic images are computed in absolute units at the distance of each source and convolved with the observed beam, we compare them directly with the rms noise of the observed Band~7 images, $\sigma_{\rm obs}$ \citep{Curone_2025}: an outer emission feature is considered detectable if the synthetic intensity exceeds $3\sigma_{\rm obs}$ beyond $R_{90,\rm obs}$.

The synthetic disc radius $R_{90,\rm syn}$ is then
calculated from the masked model image. The synthetic images are noiseless and are not processed through the interferometric $uv$-sampling and imaging of the observations, so this criterion represents an idealised detection limit: emission just above $3\sigma_{\rm obs}$ would thus correspond to a marginal detection in real data, and very extended, low-surface-brightness emission may be partly filtered out by the interferometer.

For LkCa\,15, the inner 20\,au is masked in the synthetic 
images prior to profile extraction, as the inner disc 
continuum emission is poorly constrained
\citep{Sierra_2025}.

We classify the resulting outer emission into three categories. \emph{Ring} signifies a distinct local intensity maximum at $r > R_{90,\rm obs} + 0.5\,\theta_{\rm beam}$; \emph{halo} denotes emission detectable above the threshold but with no local maximum, the profile declining monotonically from the observed continuum disc edge, together with $R_{90,\rm syn} > R_{90,\rm obs} + \theta_{\rm beam}$; and \emph{no outer mm} indicates no emission extending meaningfully beyond $R_{90,\rm obs}$. Here $\theta_{\rm beam} = \sqrt{\theta_{\rm maj}\,\theta_{\rm min}}$ is the geometric mean beam FWHM evaluated per disc. $R_{90,\rm syn}$ is computed from all emission above $3\sigma$. We adopt the term \emph{halo} by analogy with the extended low-surface-brightness emission reported by \citet{Das_2026}, albeit here it refers to a morphological classification of our synthetic images.

\subsection{Suppression tests: late bump and enhanced turbulence}
\label{sec:methods_tests}

In order to explore potential physical mechanisms that could suppress
the formation of a detectable dust ring despite the presence
of an outer gas pressure bump, we perform two additional
tests on the case-study systems LkCa\,15 and J\,1615, and additionally test enhanced-turbulence on AA\,Tau, for comparison. We focus on these discs as they bracket the range of behaviours in the \twelveco\ sample: LkCa\,15 shows the strongest configuration dependence, with two of the four parameter combinations already falling below the detection threshold in the outer disc, whereas J\,1615 produces detectable emission in every configuration. In both, the outer bump lies well inside the radial extent of the kinematic data, so it can easily be removed and reintroduced without extrapolating the reconstructed profile.
AA\,Tau produces detectable outer rings in every configuration and therefore represents the strongest tension with the observations; we do not adopt it as a primary case study because its reconstructed $\Sigma_{\rm g}$ profile increases with radius, which is atypical of the sample.

\subsubsection*{Late bump introduction}

Whether these outer pressure structures are long-lived, as would be expected for gaps induced by planets, or transient features formed by e.g. recent infall or zonal flows, is unknown. If the bump formed only recently, the outer disc may host little solid material: either insufficient time has passed for the drifting grains to reach and accumulate at the pressure maximum, or most solids have already drifted inward past that location at the stage before the bump formed. Therefore we explore a late-bump scenario to test whether delayed formation of the outer pressure structures can suppress the formation of a detectable continuum ring.

If the bump is induced by a
forming planet, the planet must first grow sufficiently
massive to open a gap and perturb the gas pressure profile,
which occurs on timescales comparable to the local orbital
and viscous timescales \citep{Pinilla_2015}. Alternatively,
late infall of material from the surrounding envelope onto
the outer disc could generate a new pressure feature at
large radii \citep{Das_2026}. \citet{Zhao_2025} demonstrated that an infall induced pressure bump can efficiently trap dust and trigger streaming instability, forming a planetesimal belt, but only while the bump persists before being smoothed by viscous diffusion at the local viscous timescale.

\begin{figure*}
    \centering
    \includegraphics[width=0.48\textwidth]{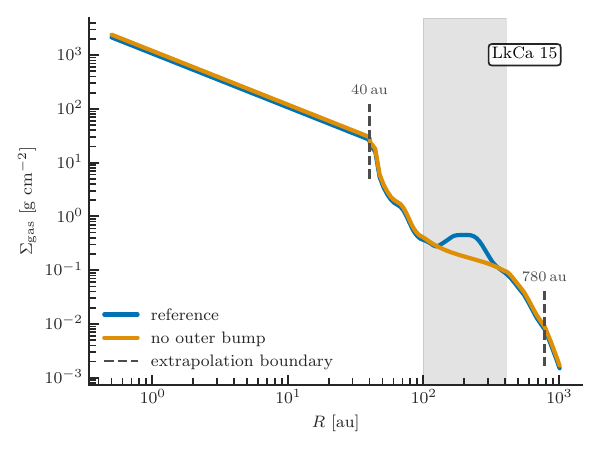}
    \hfill
    \includegraphics[width=0.48\textwidth]{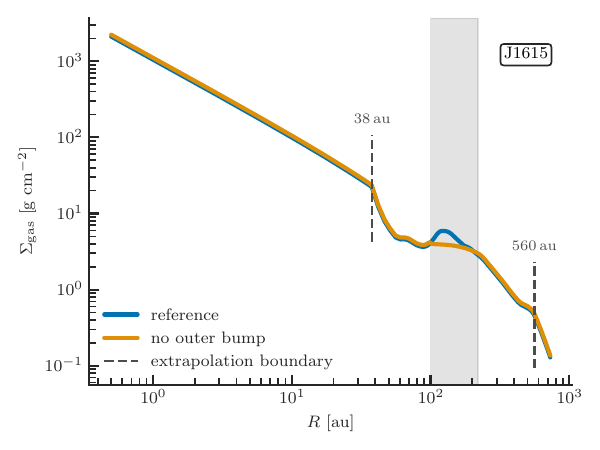}
    \caption{Gas surface density profiles used in the late-bump 
test for LkCa\,15 (left) and J\,1615 (right), derived from 
the exoALMA \twelveco\ pressure-derivative reconstructions 
(Section~\ref{sec:methods_gas}). The \textit{reference} 
profile (solid blue line) is the full reconstructed $\Sigma_{\rm g}(r)$ 
including the outer pressure bump. The \textit{no outer bump} profile (yellow solid line) is constructed using a smooth outer bridge. The shaded region hence indicates the bridge 
interval where the bump is removed. The dashed black lines indicate the edges of the radial range covered by the observed pressure-derivative profiles (labelled in au); inside and beyond these radii the profile is extended analytically (Appendix~\ref{app:reconstruction}).  In the late-bump test, 
\dustpy\ is first run for 1\,Myr using the no outer bump
profile as the gas background; the outer bump is then 
introduced by switching to the reference profile and the 
system is evolved for a further 0.5\,Myr (see Section ~\ref{sec:methods_tests}).}
    \label{fig:Lkca15_j1615_late_bump_gas_surf_density}
\end{figure*}

To test this, we first evolve the disc for $t = 1$\,Myr
with a gas surface density where the outer pressure bump has been removed. The bump region is replaced by a smooth cubic spline in log-log space, connected to the observed profile on either side of the bump and matching the local slope at both boundaries (shaded region in
Fig.~\ref{fig:Lkca15_j1615_late_bump_gas_surf_density}). The bridged interval spans $100$--$410$\,au for LkCa\,15 and $100$--$220$\,au for J\,1615. In both cases the smoothed (no-bump) profile is renormalised to conserve the total disc gas mass.

We then introduce the perturbed outer profile,
reconstructed as described in
Section~\ref{sec:methods_gas}, and evolve for an additional
$0.5$\,Myr. The gas surface density profiles before and
after bump introduction are shown in
Fig.~\ref{fig:Lkca15_j1615_late_bump_gas_surf_density}.
This test follows a similar approach to \citet{Pinilla_2015},
who modelled the response of a dust disc to a pressure
bump introduced after significant prior dust evolution.

\subsubsection*{Enhanced outer-disc turbulence}
Regions of low ionisation in the disc midplane suppress the
magnetorotational instability \citep[MRI;][]{Balbus_1991},
creating so-called dead zones where turbulence is weak and
$\alphavisc$ is low. Beyond the dead zone, where ionisation
from cosmic rays or stellar radiation is sufficient to
sustain MRI, turbulence can be substantially higher.
The location and extent of dead zones depend on the disc
column density, the ionisation sources, and the magnetic
field structure \citep[e.g.][]{Dzyurkevich2013, Flock_2015,  Delage_2022}.

Enhanced turbulence exterior to the main dust disc could
suppress the formation of a detectable outer continuum
ring through two related effects: (i) stronger turbulent
diffusion counteracts the concentration of solids at the
pressure maximum, reducing trapping efficiency; and (ii)
higher turbulent velocities increase grain collision speeds,
promoting fragmentation and limiting grain growth to smaller
sizes with lower millimetre opacity
\citep{Birnstiel_2010, Pinilla_2012}.

To test this, we implement a step-like turbulence profile,
\begin{equation}
    \alphavisc(r) =
    \begin{cases}
        \alphavisc_{\rm in}, & r < R_{\rm dz} \\
        \alphavisc_{\rm out}, & r \ge R_{\rm dz},
    \end{cases}
\end{equation}
placing $R_{\rm dz}$ 5\,au interior to the observed continuum disc edge $R_{90,obs}-5$\,au, i.e. 151, 165 and 134\,au for LkCa\,15, J\,1615 and AA\,Tau respectively. We adopt $\alphavisc_{\rm out} \sim 10^{-2}$,
consistent with estimates for MRI-active outer disc regions
\citep{Flock_2015}, while $\alphavisc_{\rm in}$ takes the
reference values from Table~\ref{tab:DustPy_params}. We do not apply $\alphavisc = 10^{-2}$ throughout the disc, since such high turbulence would  suppress grain growth and equally suppress the observed inner rings.
Although this step-function profile is a simplified
phenomenological prescription and real ionisation transitions
are expected to be smoother, it allows us to test whether the magnitude
of turbulence enhancement required to suppress the outer
ring is physically plausible. 
Physically, such a radial increase in turbulence is expected since the gas surface density drops in the outer disc, allowing cosmic rays to penetrate into the midplane and ionise it sufficiently to activate MRI \citep{Flock_2015, Delage_2022}. We note that because the gas surface density is held fixed in our simulations, no pressure bump forms at the $\alpha$-transition itself as it would in models where gas is evolved \citep[see e.g.][]{Pinilla_2016}.

\section{Results}
\label{sec:results}
We present dust evolution and radiative transfer results for 
nine exoALMA discs with reconstructed outer gas pressure 
substructures, representing the subset of the full 15 target 
exoALMA sample for which published pressure derivative data 
products suitable for forward modelling are available\footnote{Data products available at 
\url{https://dataverse.harvard.edu/dataset.xhtml?persistentId=doi:10.7910/DVN/FIDKQV}; 
see also \citet{Stadler_2025_data}.}.

Detailed figures for all nine systems are presented in the Appendix (Figs.~\ref{fig:dustpy_dmtau}--\ref{fig:radmc_v4046}).

\subsection{Sample-wide outcome}
\label{sec:results_survey}

\begin{figure*}
  \centering
  \includegraphics[width=0.96\textwidth]{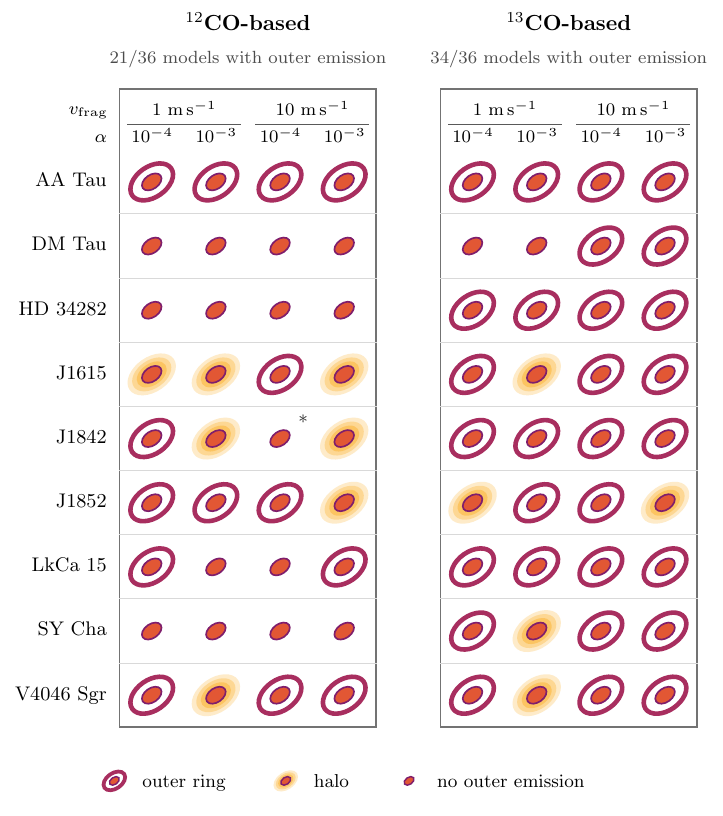}
  \caption{Predicted outer Band~7 emission for the 72 reference models, for the \twelveco-based (left) and \thirteenco-based (right) gas reconstructions. Rows are discs and columns the four $(\alphavisc, \vfrag)$ configurations. Each symbol indicates whether emission above $3\sigma$ beyond $R_{90,\rm obs}$ forms an outer ring, a halo, or is absent (Section~\ref{sec:methods_rt}); the symbols are schematic, not model images. The asterisk marks the J1842 model with no emission above $3\sigma$ at any radius. Numerical values are given in Tables~\ref{tab:12co_summary} and~\ref{tab:13co_summary}.}
  \label{fig:outcomes_reference}
\end{figure*}

Fig.~\ref{fig:outcomes_reference} summarises whether each model produces detectable outer millimetre emission beyond the observed continuum disc edge (numerical values in Appendix~\ref{app:tables}).
Figures~\ref{fig:survey_profiles} and \ref{fig:survey_profiles_13_co}
show the deprojected radial intensity profiles of the synthetic
continuum emission for each disc and dust-evolution configuration,
derived from the \twelveco\ and \thirteenco\ reconstructions
respectively.
 
The dominant outcome across the sample is efficient outer dust trapping: 21 of our 36 \twelveco-based models and 34 of the 36 \thirteenco-based models produce emission detectable above 3$\sigma$ beyond the observed continuum disc edge (9 targets $\times$ 4 parameter combinations per tracer). Across a wide range of $(\alphavisc, \vfrag)$ combinations, the reconstructed gas profiles frequently produce substantial dust pile-ups at large radii, resulting in prominent outer continuum rings in the synthetic Band~7 emission assuming the same angular resolution and sensitivity as the exoALMA observations.
 
A notable result is that the \thirteenco-based reconstructions almost universally produce detectable outer continuum rings across the explored parameter space (Fig.~\ref{fig:outcomes_reference}, right). Even in the few marginal cases, the synthetic continuum disc extends substantially beyond the observed $R_{90,\rm obs}$, indicating efficient outer dust trapping. Most systems and configurations produce an outer peak emission (ring) and all models yield synthetic continuum emission extending well beyond the observed millimetre disc radius.
 
In contrast, the \twelveco-based reconstructions exhibit substantially greater diversity (Fig.~\ref{fig:outcomes_reference}, left). While AA\,Tau, J1615, J1842, J1852, and V4046\,Sgr commonly produce detectable outer rings, several systems and parameter combinations do not generate significant outer continuum emission.

Within the \twelveco\ sample, LkCa\,15 shows the strongest configuration dependence, with two of the four configurations producing no detectable outer emission; we examine it in Section~\ref{sec:results_lkca15_aatau}.

SY\,Cha, DM\,Tau, and HD\,34282 do not produce a detectable outer ring across all explored \twelveco\ configurations. In addition, the model with $\alphavisc = 10^{-4}$, $\vfrag = 10\,\mathrm{m\,s^{-1}}$ for J\,1842 also lacks detectable outer continuum emission; in this model no emission is detected above $3\sigma$ at any radius (the corresponding panel in Fig.~\ref{fig:radmc_j1842} appears bright only because each image is normalised to its own peak), because efficient trapping in the outer bump grows grains to sizes with low Band~7 opacity and prevents dust from leaking through to the inner disc (Fig.~\ref{fig:dustpy_j1842}). In some of these cases, the dust evolution models nevertheless show narrow outer dust rings with efficient growth (see Appendix Figures~\ref{fig:dustpy_j1842} and~\ref{fig:dustpy_sycha}), meaning that the absence of observable Band~7 emission does not necessarily imply the absence of dust trapping in the outer disc.

The radial intensity profiles in Figures~\ref{fig:survey_profiles} and~\ref{fig:survey_profiles_13_co} show that many of the predicted outer continuum rings are spatially associated with the candidate outer pressure-substructure regions inferred from the gas kinematics. In several systems, including AA\,Tau, J1615, and V4046\,Sgr, the synthetic outer continuum rings are located within or near the shaded radial regions corresponding to the velocity-selected outer gas substructures.
 
However, this correspondence is not universal. For example, J1852 produces a prominent outer continuum ring despite lacking a formally significant super-Keplerian velocity region in the observational constraints. This illustrates the point made in Section~\ref{sec:methods_candidate_regions}: the super-Keplerian label is a reporting criterion rather than a prerequisite for trapping, and regions that do not meet this criterion can still host substantial dust concentrations in the models.
 
In several systems, the locations of the \twelveco-based continuum rings also overlap with outer substructures identified in the \thirteenco\ models. Furthermore, both the radial location and prominence of the synthetic outer rings vary between dust evolution configurations, indicating that the detailed dust response depends sensitively on the adopted $(\alphavisc,\vfrag)$ combination.
 
\begin{figure*}
  \centering
  \includegraphics[width=0.96\textwidth]{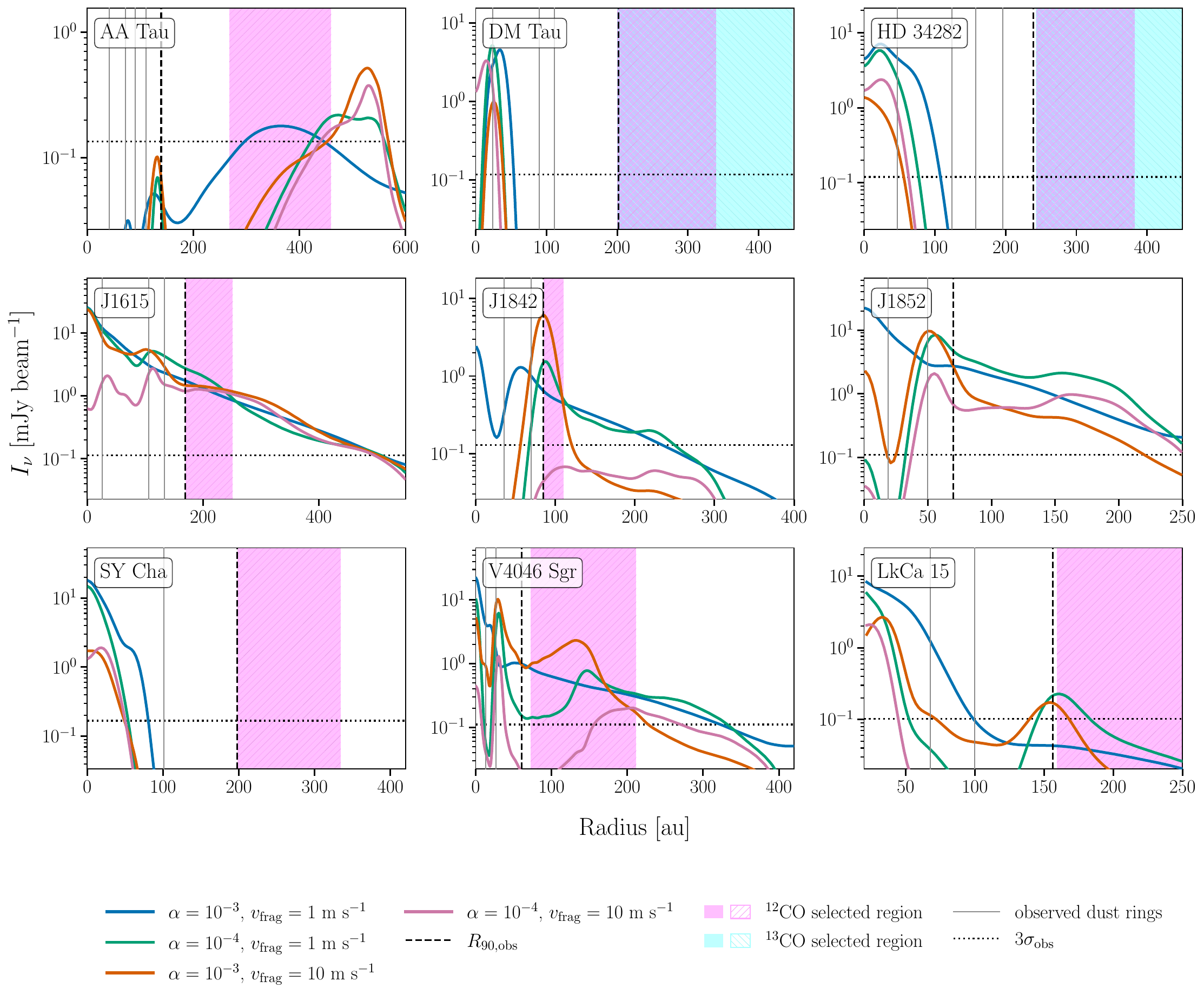}
  \caption{Azimuthally averaged, deprojected radial intensity profiles of the synthetic Band~7 continuum emission for the full disc sample, computed from \radmc\ images convolved to the observed beam of each source. Each panel shows the four dust-evolution model configurations explored in this work, $\alphavisc \in \{10^{-3},10^{-4}\}$ and $\vfrag \in \{1,10\}\,\mathrm{m\,s^{-1}}$, for the \twelveco-based reconstructions. Intensities are shown in absolute units on a logarithmic scale. The horizontal dotted line marks the observed $3\sigma$ sensitivity of each disc, so emission above it would be detectable at the exoALMA sensitivity. The dashed vertical line marks $R_{90,\rm obs}$ and the grey solid lines the observed dust rings \citep{Curone_2025}. Pink (\twelveco) and blue (\thirteenco) shaded regions indicate the candidate outer gas substructures inferred from the kinematic analysis. The inner 20\,au of LkCa\,15 is masked following \citet{Sierra_2025}.}
  \label{fig:survey_profiles}
\end{figure*}

\subsection{Two sample discs: LkCa\,15 and J\,1615}
\label{sec:results_lkca15_aatau}

We now examine the two case-study discs selected in Section~\ref{sec:methods_tests} in more detail.
 
\begin{figure*}
    \centering
    \begin{subfigure}{0.96\textwidth}
        \centering
        \includegraphics[width=\textwidth]{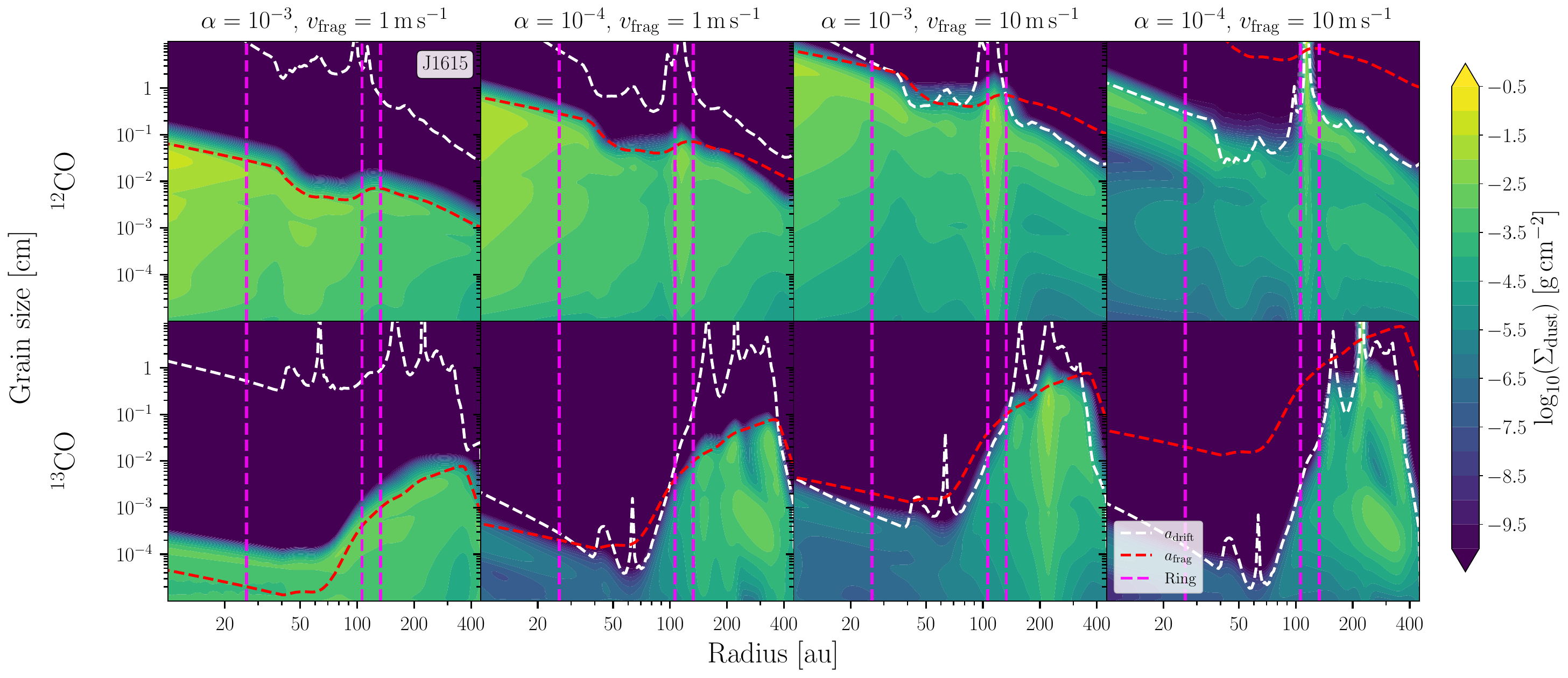}
        \caption{}
        \label{fig:j1615_initial_dustpy}
    \end{subfigure}
 
    \vspace{0.2cm}
 
    \begin{subfigure}{0.96\textwidth}
        \centering
        \includegraphics[width=\textwidth]{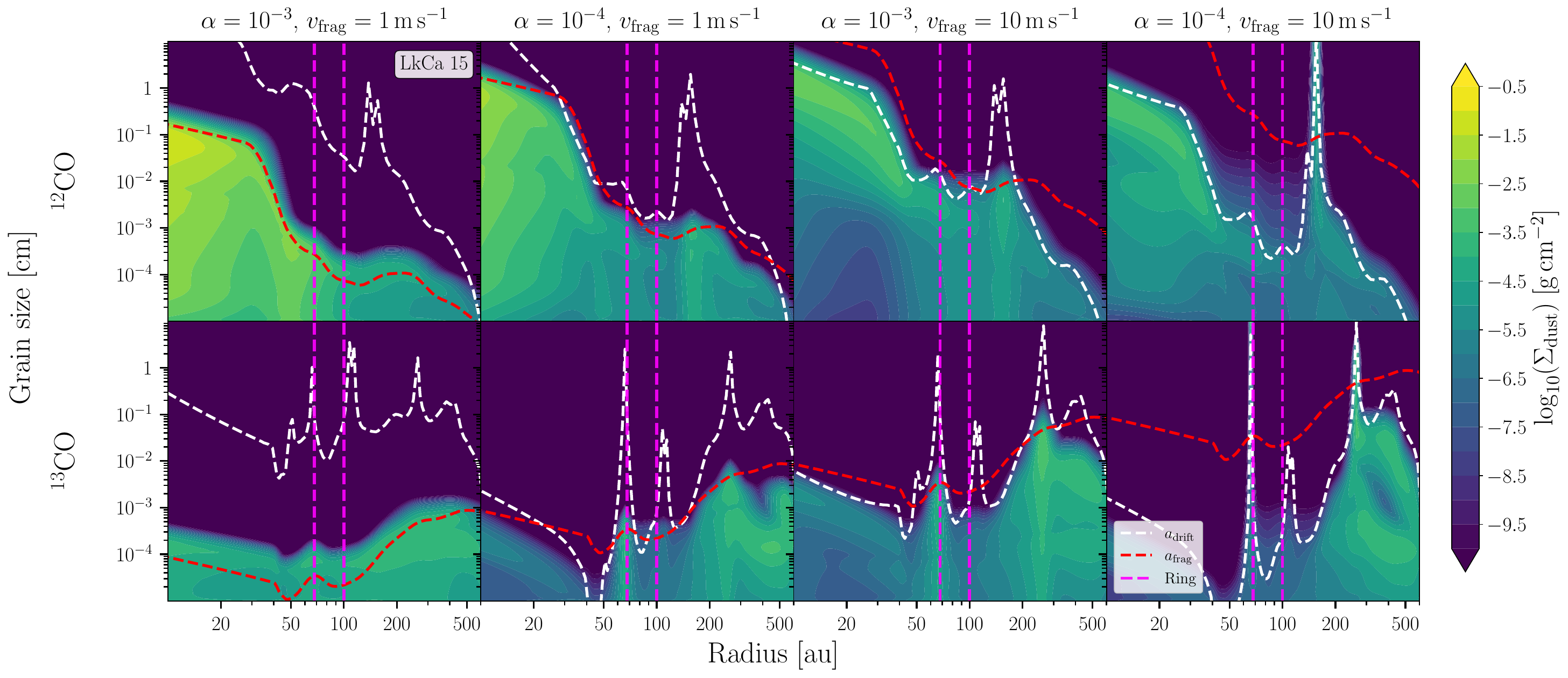}
        \caption{}
        \label{fig:lkca15_initial_dustpy}
    \end{subfigure}
 
    \caption{Dust density distribution after 1\,Myr using the exoALMA gas surface density reconstructions derived from the \twelveco\ and \thirteenco\ observations for J1615 (top panels) and LkCa\,15 (bottom panels). White and red dashed curves indicate the drift and fragmentation barriers, respectively. Vertical dashed lines mark the dust substructures identified in the Band~7 continuum observations \citep{Curone_2025}, with magenta lines corresponding to rings. The four columns represent different dust evolution model configurations spanning $\alphavisc \in \{10^{-4},10^{-3}\}$ and $\vfrag \in \{1,10\}\,\mathrm{m\,s^{-1}}$.}
    \label{fig:initial_dustpy_lkca_j1615}
\end{figure*}

Figure~\ref{fig:initial_dustpy_lkca_j1615} shows the \dustpy\ dust surface density distributions for the four explored $(\alphavisc,\vfrag)$ configurations. Across both systems, the dust evolution behaviour depends strongly on the balance between fragmentation, radial drift, and dust trapping. The low-$\alphavisc$, high-$\vfrag$ configurations generally produce the most efficient grain growth and strongest outer dust concentrations, while higher turbulence and lower fragmentation thresholds yield smaller grain sizes and more diffuse dust distributions.
 
\begin{figure*}
    \centering
    \begin{subfigure}{0.96\textwidth}
        \centering
        \includegraphics[width=\textwidth]{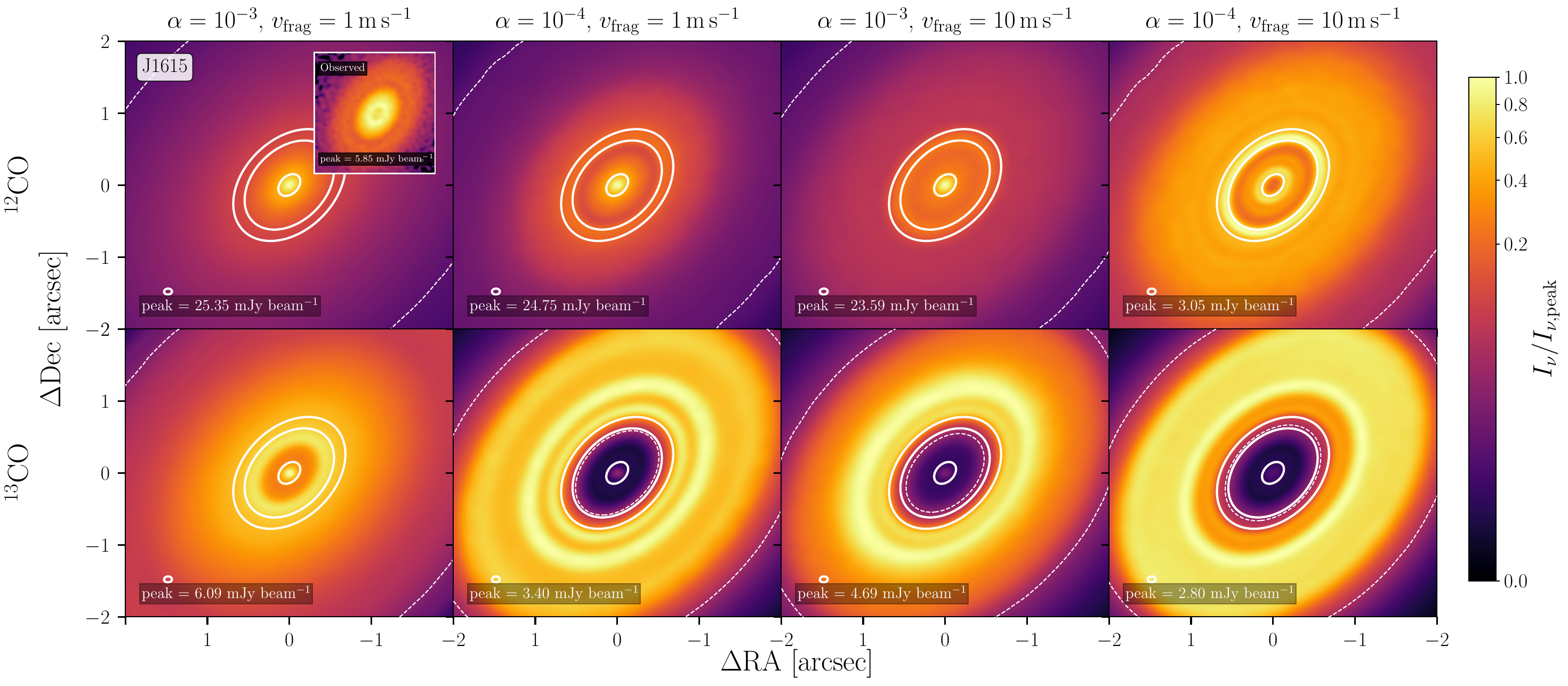}
        \caption{}
        \label{fig:j1615_baseline_cont}
    \end{subfigure}
 
    \vspace{0.5cm}
 
    \begin{subfigure}{0.96\textwidth}
        \centering
        \includegraphics[width=\textwidth]{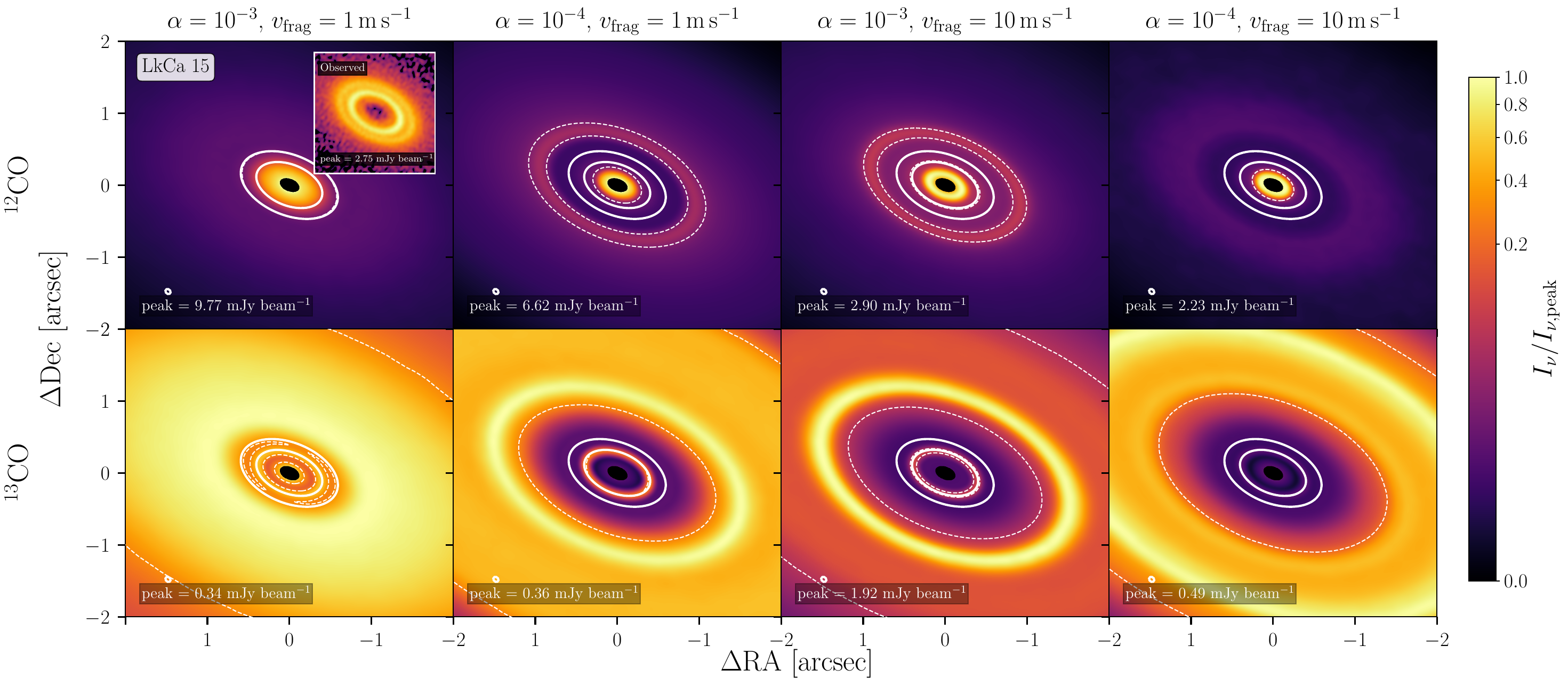}
        \caption{}
        \label{fig:lkca15_baseline_cont}
    \end{subfigure}
 
    \caption{\radmc\ synthetic observations of the \dustpy\ results using the \twelveco\ and \thirteenco\ data of J\,1615 (top panels) and LkCa\,15 (bottom panels). The fluxes are normalised to the peak. The inset shows the observation in Band~7 \citep{Curone_2025}, which is also normalised to peak. The white ellipses indicate dust rings identified in the Band~7 continuum observations \citep{Curone_2025}. The white dashed lines indicate the $3\sigma$ threshold based on the sensitivity of the observations. The four columns represent different dust evolution model configurations spanning $\alphavisc \in \{10^{-4},10^{-3}\}$ and $\vfrag \in \{1,10\}\,\mathrm{m\,s^{-1}}$. The inner 20\,au of LkCa\,15 is masked following \citet{Sierra_2025}.}
    \label{fig:lkca_j1615_baseline_cont}
\end{figure*}

For J\,1615, all four reference configurations for both \twelveco\ and \thirteenco\ models produce substantial outer dust concentrations that remain detectable in the synthetic Band~7 continuum images (see Figs.~\ref{fig:j1615_initial_dustpy} and~\ref{fig:j1615_baseline_cont}). The \twelveco\ models predict an extended halo of continuum emission, whereas the \thirteenco\ models produce bright outer rings. In both cases, the predicted emission lies well beyond the observed millimetre disc radius and remains prominent after beam convolution, making J\,1615 a clear example where the reference models substantially over-predict the observable outer continuum emission.
 
For LkCa\,15 the outcome is configuration-dependent. The model with $\alphavisc = 10^{-4}, \vfrag = 10\,\mathrm{m\,s^{-1}}$ produces a narrow outer dust concentration in the \dustpy\ evolution model (Figure~\ref{fig:lkca15_initial_dustpy}), but the corresponding synthetic Band~7 emission falls below the adopted detection threshold. The model with $\alphavisc = 10^{-3}, \vfrag = 1\,\mathrm{m\,s^{-1}}$ also does not produce an outer ring due to efficient fragmentation and drift of grains. However, the $R_{90,\rm syn}$ for both configurations is substantially more compact than the observed value. In contrast, the model with $\alphavisc = 10^{-4}, \vfrag = 1\,\mathrm{m\,s^{-1}}$ and the model with $\alphavisc = 10^{-3}, \vfrag = 10\,\mathrm{m\,s^{-1}}$ produce detectable outer continuum rings near the observed outer substructure location.
 
This behaviour illustrates that the presence of an outer dust concentration in the dust-evolution models does not necessarily guarantee detectable millimetre continuum emission. In the $\alphavisc=10^{-4},\,\vfrag=10\,\mathrm{m\,s^{-1}}$ case, dust trapping does occur: the pressure bump concentrates solids into a narrow outer ring.
This ring remains undetectable at Band~7 with the exoALMA resolution and sensitivity. The corresponding continuum feature is however well resolved, spanning $\approx 75$\,au or roughly seven beams, so beam dilution is not the limiting factor. Rather, efficient grain growth concentrates most of the mass into grains that are inefficient millimetre emitters, leaving a continuum feature whose peak excess above the local profile baseline is around 20 times below the 3$\sigma$ threshold of the observations. 
 
We tested whether the large grains might instead be detectable at longer wavelengths by computing synthetic images at 1.3\,mm, 1.6\,mm, 3\,mm, 6\,mm and 8.6\,mm (ALMA Bands~6 down to 1). For each band and trial beam size, we measured the peak excess of the outer feature above a locally interpolated radial-profile baseline and calculated the rms required for a $3\sigma$ detection as one-third of this excess. We compared these required sensitivities with equivalent rms values obtained by scaling the Band~7 rms to preserve the same optically thin dust-mass sensitivity, assuming $F_\nu\propto\kappa_\nu B_\nu(T_{\rm dust})$, $\kappa_\nu=2.3\,(\nu/230{\rm \,GHz})^{0.4}\mathrm{\,cm^2g^{-1}}$ and $T_{\rm dust}=20 \, \mathrm{K}$. The required sensitivities remain at least a factor of $\approx20$ more stringent than the corresponding scaled values at every wavelength and beam size tested, so no $3\sigma$ outer emission is recovered at any of these longer wavelengths either. The surface brightness of the outer feature decreases with increasing wavelength, so observing at longer wavelengths provides no net advantage for detecting this feature.

\subsection{Suppression tests}
\label{sec:results_suppression}                 

\begin{figure}
  \centering
  \includegraphics[width=\columnwidth]{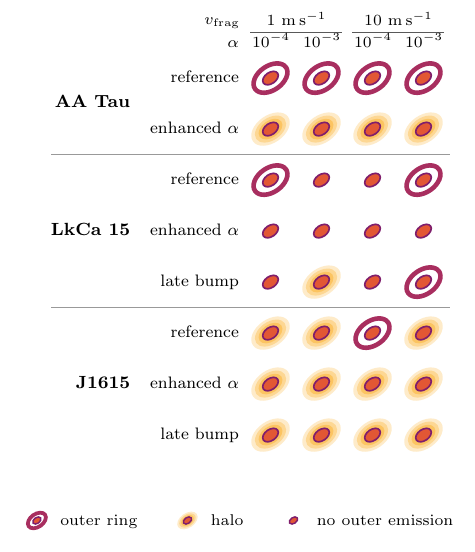}
  \caption{Outcome of the suppression tests for the \twelveco-based models, compared with the reference models. \textit{Late bump}: outer bump introduced after 1\,Myr and evolved for a further 0.5\,Myr; \textit{enhanced $\alpha$}: $\alphavisc_{\rm out}=10^{-2}$ beyond $R_{\rm dz}$ (Section~\ref{sec:methods_tests}). Symbols as in Fig.~\ref{fig:outcomes_reference}. Numerical values are given in Table~\ref{tab:special_models}.}
  \label{fig:outcomes_mechanisms}
\end{figure}
 
Fig.~\ref{fig:outcomes_mechanisms} summarises the outcome of the two suppression tests described in Section~\ref{sec:methods_tests}, applied to the \twelveco-based models.

\subsubsection{Late introduction of the outer bump}  
\label{sec:results_latebump}
Since long-lived pressure bumps efficiently accumulate drifting solids, delayed formation of the outer gas substructure may reduce the dusty material available for trapping and outer ring formation \citep{Pinilla_2015, Stadler_2022}
We first test whether delaying the formation of the outer bump by 1\,Myr (Fig.~\ref{fig:Lkca15_j1615_late_bump_gas_surf_density}) suppresses the outer emission.
 
\begin{figure*}
  \centering
  \includegraphics[width=0.96\textwidth]{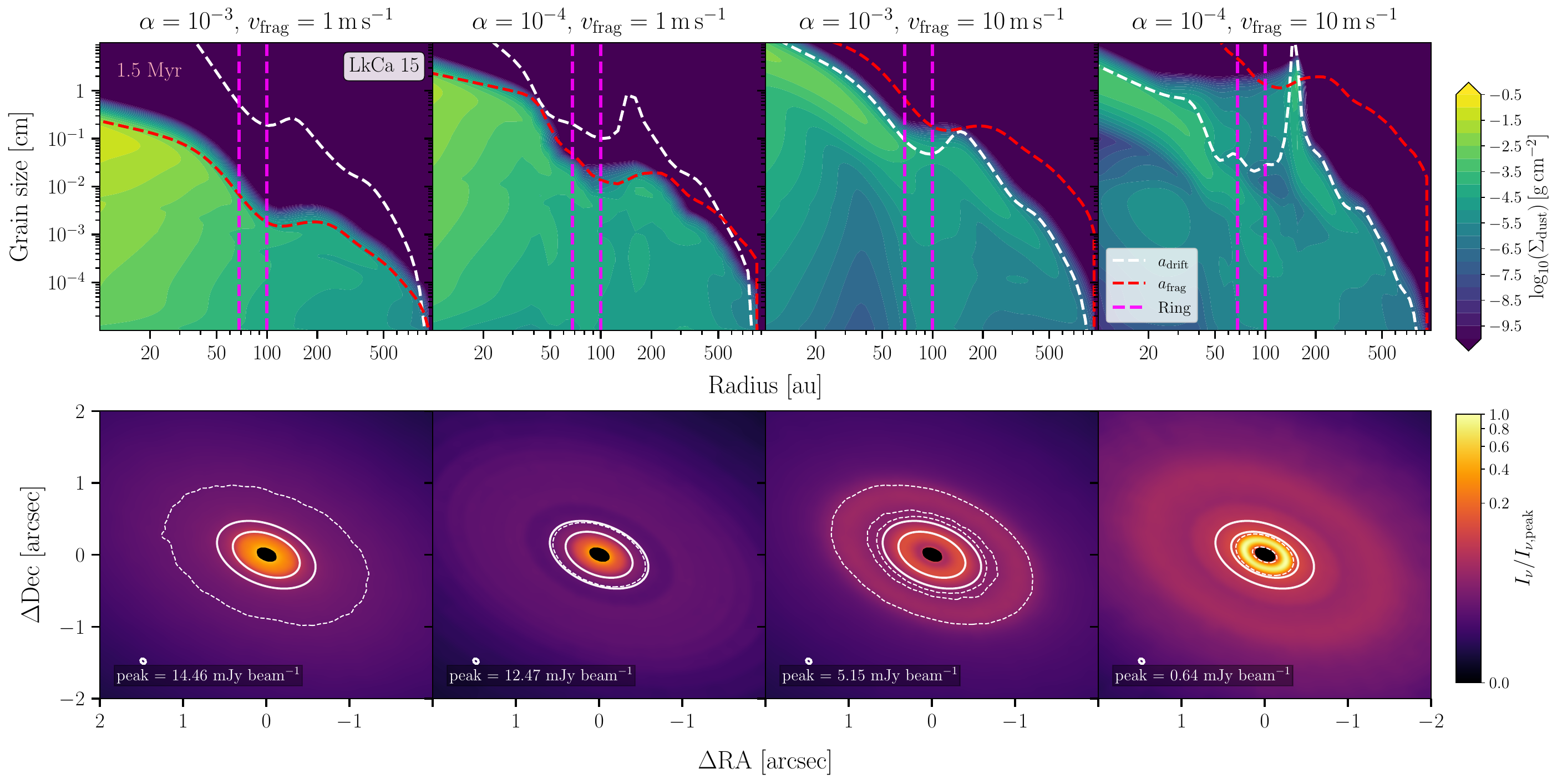}
  \caption{Dust evolution and synthetic Band~7 continuum emission for the late-bump LkCa\,15 models. The outer gas pressure substructure is introduced after 1\,Myr of dust evolution and the system is subsequently evolved for an additional 0.5\,Myr. The panels show the resulting \dustpy\ dust surface-density distributions together with the beam-convolved \radmc\ synthetic continuum images. White dashed contours indicate the observational $3\sigma$ sensitivity threshold.}
  \label{fig:lkca15_latebump}
\end{figure*}

For LkCa\,15, Figs.~\ref{fig:lkca15_latebump} and~\ref{fig:outcomes_mechanisms} show that the late bump prescription does not provide a robust mechanism for suppressing outer continuum emission. For the $\alphavisc=10^{-4},\,\vfrag=1\,\mathrm{m\,s^{-1}}$ configuration, the late-bump introduction suppresses outer ring emission. However, for the other configuration that produces an outer ring in the reference model of LkCa\,15 (i.e. $\alphavisc=10^{-3},\,\vfrag=10\,\mathrm{m\,s^{-1}}$), the ring is not suppressed. We also note that for $\alphavisc=10^{-3},\,\vfrag=1\,\mathrm{m\,s^{-1}}$, a faint halo-like emission becomes detectable, which was not present in the reference model. 
 
Hence we find that detectable outer Band~7 emission can still develop rapidly after the bump is introduced. This is because significant solid material remains in the outer disc during the no-bump phase, either retained at large radii by slow radial drift or already present at the bump location. This material is rapidly swept up into the pressure maximum once the bump is introduced, forming a detectable ring within the additional 0.5\,Myr of evolution.
 
\begin{figure*}
  \centering
  \includegraphics[width=0.96\textwidth]{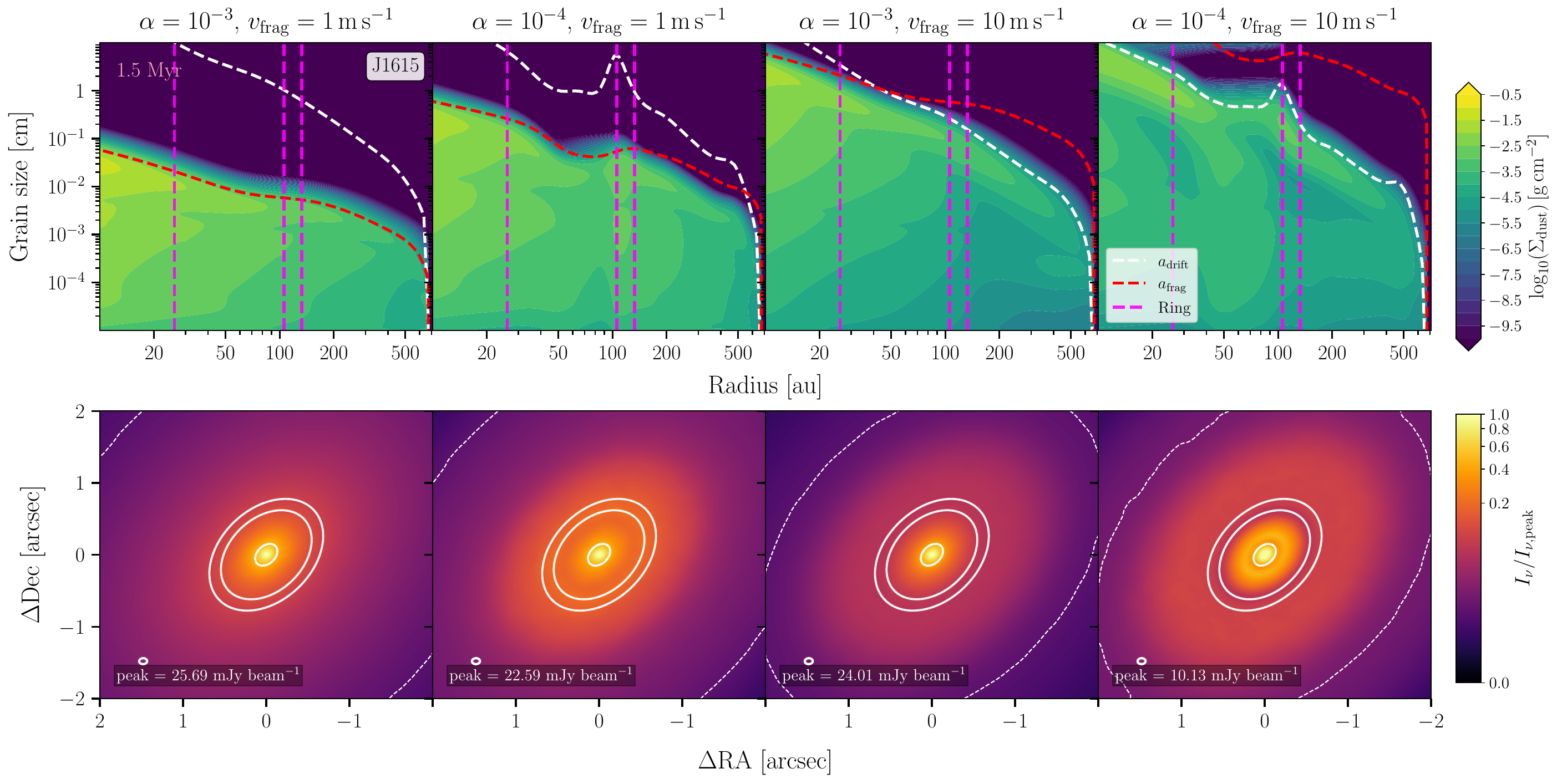}
  \caption{Dust evolution and synthetic Band~7 continuum emission for the late-bump J\,1615 models. The outer gas pressure substructure is introduced after 1\,Myr of dust evolution and the system is subsequently evolved for an additional 0.5\,Myr. The panels show the resulting \dustpy\ dust surface-density distributions together with the beam-convolved \radmc\ synthetic continuum images. White dashed contours indicate the observational $3\sigma$ sensitivity threshold.}
  \label{fig:j1615_latebump}
\end{figure*}

In J\,1615 (Figs.~\ref{fig:j1615_latebump} and~\ref{fig:outcomes_mechanisms}), the late-bump introduction is even less effective. All four configurations retain detectable outer emission after the additional 0.5\,Myr, with much larger $R_{90,\rm syn}$ (300--420\,au) compared with $R_{90,\rm obs}$ (170\,au), and peak outer signal-to-noise ratios of 25--69$\sigma$ (Fig.~\ref{fig:emission_strength}).
 
Overall, these results indicate that delayed bump formation alone is generally insufficient to suppress outer ring formation: the solids are rapidly concentrated once the pressure maximum appears. Suppression by this mechanism would additionally require the outer disc to be depleted of solids at the time the bump forms; this could happen if most of the material has already drifted inward, or if the outer disc is replenished by gas-rich infall carrying very little dust. If outer substructures form after the inner ones, as expected for inside-out planet formation \citep{Lau_2024, Danti_2025}, the outer bump could appear even later than the 1\,Myr assumed in our test. A later bump would encounter a smaller outer solid reservoir, since radial drift continues to deplete the outer disc, whereas an earlier bump would encounter a larger one. The outcome is therefore set by the solid mass outside the bump when it forms; for a sufficiently depleted reservoir, no detectable ring forms, as in our AA\,Tau test.\footnote{We also tested a late-bump scenario for AA\,Tau, where the outer bump contributes a substantial fraction of the reconstructed gas mass and the pre-bump phase therefore begins from a much smaller solid reservoir. No detectable outer ring forms in any configuration, but this reflects the reduced initial dust content rather than suppression of trapping by the bump itself.}
 
\subsubsection{Enhanced outer-disc turbulence}       
\label{sec:results_deadzone}

Lastly, we explored whether enhanced turbulence exterior to the main millimetre disc could suppress outer dust trapping and reduce the resulting continuum emission, using the step-like turbulence profile with $\alphavisc_{\rm out}=10^{-2}$ beyond the transition radius $R_{\rm dz}$ defined in Section~\ref{sec:methods_tests}.
 
\begin{figure*}
  \centering
  \includegraphics[width=0.96\textwidth]{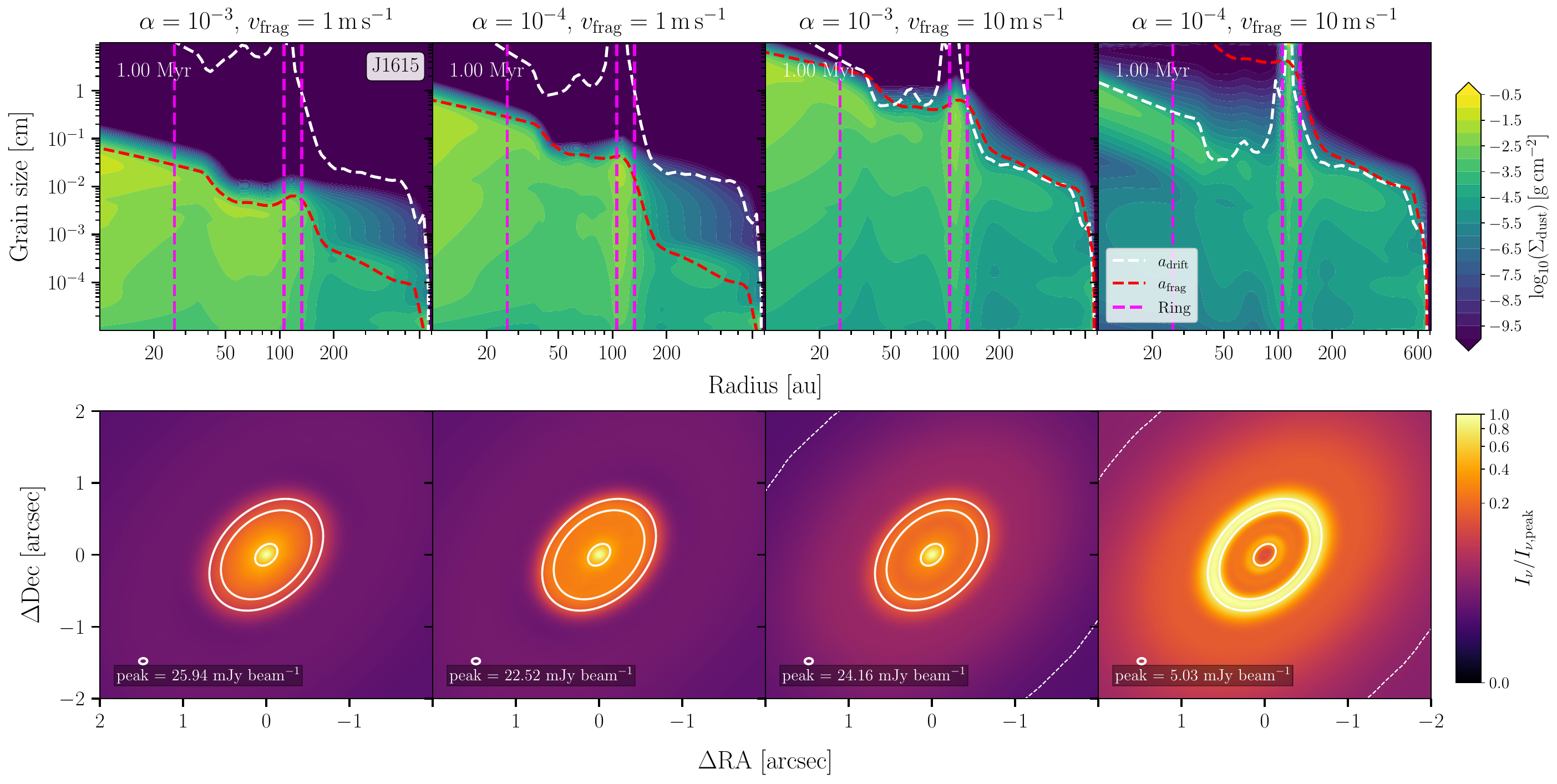}
  \caption{\dustpy\ dust evolution results and synthetic Band~7 continuum emission for the enhanced-turbulence J\,1615 models using the \twelveco\ data. A step-like turbulence profile with $\alphavisc_{\rm out}=10^{-2}$ beyond $R_{\rm dz}$ is applied. White dashed contours indicate the $3\sigma$ observational threshold.}
  \label{fig:j1615_deadzone}
\end{figure*}

\begin{figure*}
  \centering
  \includegraphics[width=0.96\textwidth]{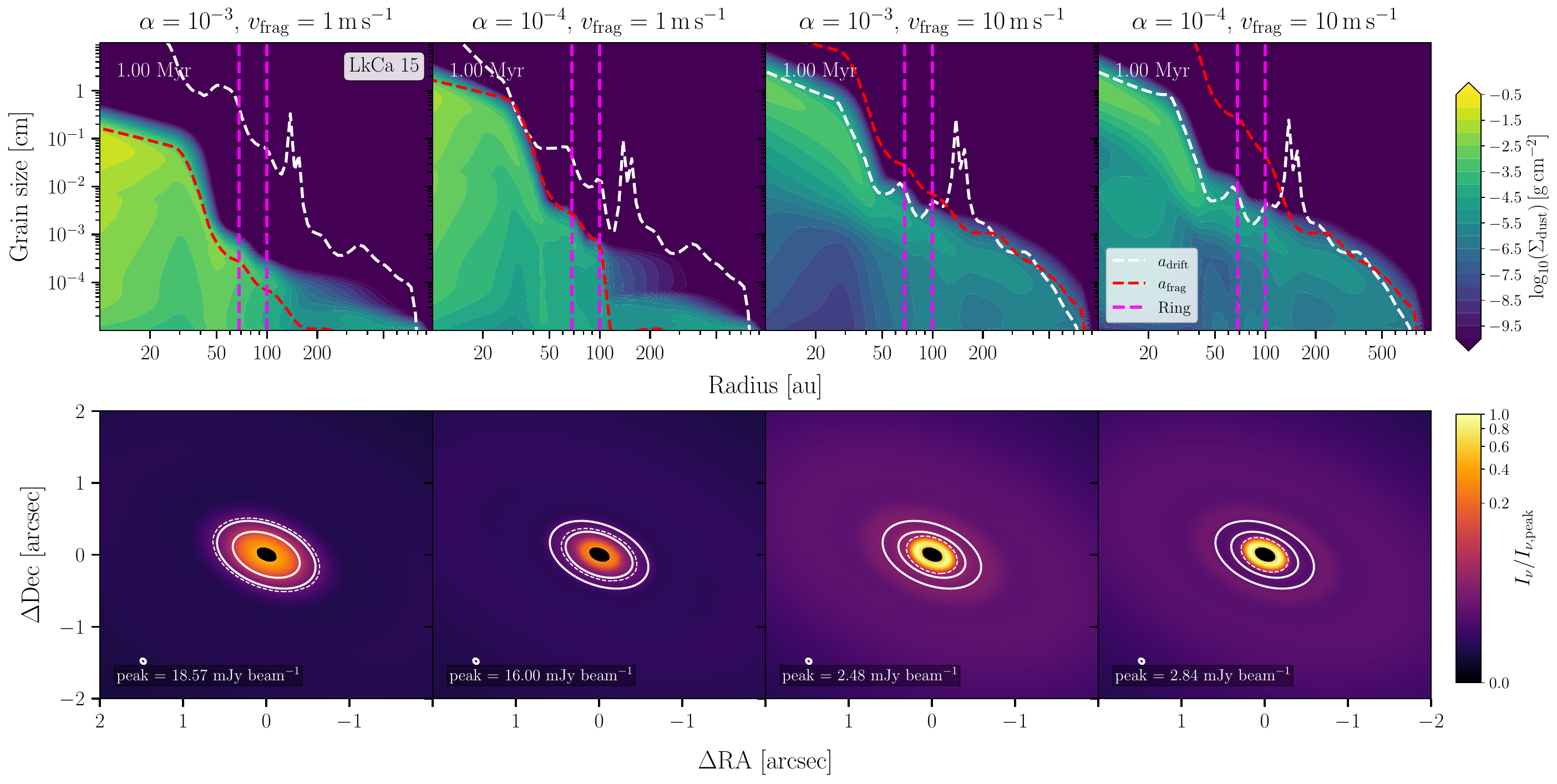}
  \caption{\dustpy\ dust evolution results and synthetic Band~7 continuum emission for the enhanced-turbulence LkCa\,15 models using the \twelveco\ data. None of the four enhanced-turbulence configurations produces detectable outer Band~7 emission beyond the observed disc edge.}
  \label{fig:lkca15_deadzone}
\end{figure*}

Figures~\ref{fig:j1615_deadzone}, \ref{fig:lkca15_deadzone} and~\ref{fig:outcomes_mechanisms} show that this prescription substantially weakens the outer continuum rings relative to the reference models. In LkCa\,15, all four enhanced-turbulence configurations fail to produce detectable outer Band~7 emission beyond the observed continuum disc edge. 
 
For J\,1615 the suppression is only partial: all four enhanced-turbulence configurations retain detectable outer emission, but its prominence is substantially reduced. The peak signal-to-noise ratio of the outer emission falls from 25--69$\sigma$ in the late-bump models to 8--25$\sigma$ with enhanced turbulence, and $R_{90,\rm syn}$ decreases from 347--391\,au in the reference models to 274--311\,au (Fig.~\ref{fig:emission_strength}; Table~\ref{tab:special_models}). The emission is nonetheless classified as halo for every configuration and still extends well beyond $R_{90,\rm obs} = 170$\,au, so enhanced turbulence weakens but does not remove the predicted outer emission in this disc. 
 
We applied the same prescription to AA\,Tau, where the outer emission is confined much closer to the observed disc radius ($R_{90,\rm syn}= 159$--170\,au against $R_{90,\rm obs} = 139$\,au); the extended outer rings present in the reference models are largely removed, although a detectable halo persists in all four configurations (6--11$\sigma$).
 
Among the mechanisms explored in this work, enhanced outer-disc turbulence is the most effective at weakening the predicted outer continuum emission, and in LkCa\,15 it removes it entirely. It is not, however, a general solution: for J\,1615 and AA\,Tau a detectable halo survives in every configuration, and the improvement over the late-bump is mainly quantitative rather than qualitative.

\section{Discussion}
\label{sec:discussion}

\subsection{Standard dust evolution models frequently produce outer continuum rings}
\label{sec:discussion_baseline}

In general the \thirteenco-based reconstructions produce broader or higher-contrast pressure maxima in the outer disc than the corresponding \twelveco\ ones, leading to more efficient dust trapping. Since \thirteenco\ is less optically thick, it should in principle trace deeper, colder disc layers closer to the midplane than \twelveco\ \citep[e.g.][]{Pinte_2018, Teague_2018}, so systematic differences in the inferred gas structure could plausibly also contribute to the stronger outer trapping behaviour recovered in the \thirteenco\ models. However, it is unclear whether the difference in \thirteenco\ and \twelveco\ is physical. It could also reflect the different radial coverage and SNR of the \thirteenco\ and \twelveco\ rotation curves. Differences in optical depth, emitting layers, and the radial extent of the reconstructed gas profiles may thus all influence the resulting dust evolution outcomes.

An important consistency check is provided by the inner disc: many of the pressure bumps \emph{within} the observed continuum disc co-locate well with dust rings in the Band~7 observations \citep{Curone_2025}, indicating that these pressure maxima are efficient, long-lived dust traps: had they been transient they too should fail to retain solids and lack associated rings. The contrast with the outer disc is therefore informative. Rather than these strong outer pressure bumps failing to trap dust, the absence of outer rings point to a physical difference in the inner and outer disc environments, for example: in turbulence level, ionisation, or the lifetime of the outer substructure relative to the local drift timescale of solids. We explored two such possibilities in this work: enhanced outer-disc turbulence (Section~\ref{sec:results_deadzone}), and tested a scenario where the outer bump formed late (Section~\ref{sec:results_latebump}).

We note that the inner disc profiles are (partially) extrapolated analytically rather than constrained by kinematic data, so the inner disc models should not be interpreted as predictions for the observed inner substructures. In fact, our models do not produce many of the observed rings for this reason.

\subsection{Can outer dust remain hidden?}
\label{sec:discussion_hidden_dust}

Although most models explored in this work produce detectable outer continuum rings, a limited subset of configurations suggests that dust may remain partially hidden at millimetre wavelengths despite the presence of an outer gas pressure substructure. LkCa\,15 provides the clearest example (Section~\ref{sec:results_lkca15_aatau}). Its two configurations without detectable outer emission suppress the ring through different mechanisms. In the $(\alphavisc = 10^{-3},\vfrag = 1\,\mathrm{m\,s^{-1}})$ case, high turbulence combined with a low fragmentation threshold keeps grains in the fragmentation-limited regime, preventing efficient growth to mm sizes and allowing small grains to drift inward past the bump rather than accumulating there. We therefore focus on the $\alphavisc = 10^{-4},\vfrag = 10\,\mathrm{m\,s^{-1}}$ case as the more physically interesting hidden-dust scenario, noting that the combination of high turbulence and low fragmentation velocity is also not favoured by recent dust evolution constraints \citep{Luo_2026, Jiang_2024, Tong_2025}.
 
In the $\alphavisc = 10^{-4},\vfrag = 10\,\mathrm{m\,s^{-1}}$ case, the trapped mass is concentrated into a radially narrow ring of efficiently grown grains with low sub-mm opacity. The resulting Band~7 emission is therefore dominated by the co-located, much less massive small-grain population, producing a broad, faint continuum ring rather than a bright ring. Similar behaviour has previously been discussed in the context of evolved grain populations and wavelength-dependent dust-trapping efficiencies \citep{Pinilla_2012, Birnstiel_2024}. Since the outer feature is resolved and is not recovered at any longer wavelength (Section~\ref{sec:results_lkca15_aatau}), its faintness is set by opacity rather than angular resolution.

We note that the substructures considered here differ from the unresolved-trap scenario invoked for Elias\,2-24 by \citet{Das_2026}: our candidate regions are by construction spatially resolved, spanning more than one beam (Section~\ref{sec:methods_candidate_regions}). The LkCa\,15 result shows that even a resolved gas substructure can produce a dust ring that evades detection at Band\,7, because efficient grain growth leaves the emission too faint.
 
The tentative Band~3 detection of an outer ring at $\sim$156\,au in LkCa\,15 by \citet{Sierra_2025} is consistent with this hidden-dust picture: the ring may be present but too grain-growth-dominated to produce a $3\sigma$ detection at Band~7. A similar situation may apply to HD\,34282: \citet{Ma_2026} present multi-wavelength observations in which the dust disc extends tentatively beyond the Band~7 continuum edge, at radii where our \thirteenco\ models independently predict an outer dust ring across all four parameter configurations (Fig.~\ref{fig:outcomes_reference}). We note, however, that the outermost ring reported by \citet{Ma_2026} exhibits a prominent azimuthal asymmetry indicative of a dusty vortex. That ring may therefore arise from azimuthal vortex trapping rather than from the axisymmetric pressure bump assumed in our 1D models, so the comparison should be regarded as qualitative. 
 
Overall, hidden dust remains possible in specific systems, gas tracer and parameter combinations, and the emerging tentative detections in LkCa\,15 and HD\,34282 suggest this scenario may be physically possible in at least some exoALMA discs. However, it does not emerge as a robust generic explanation for the observed absence of outer continuum rings across the full sample.

\subsection{What could suppress the outer rings?}
\label{sec:discussion_turbulence}
Among the mechanisms explored in this work, enhanced outer-disc turbulence provides the most promising route to suppressing observable outer continuum rings. Although our adopted turbulence profile is phenomenological, it is a plausible scenario where changes of disc ionisation lead to MRI activation in the outer parts of the disc \citep{Flock_2015, Delage_2022}. If turbulence increases beyond the main millimetre disc, dust trapping at outer pressure maxima may become substantially less efficient, reducing the likelihood that these structures produce detectable continuum rings. In this picture, the absence of outer dust emission would not necessarily imply the absence of pressure substructures, but rather a change in the local dust-evolution environment. Spatially varying turbulence therefore offers a potential explanation for the apparent disconnect between the outer gas pressure structures inferred from exoALMA kinematics and the observed continuum morphology. This can potentially be tested by measuring disc ionisation using a combination of different molecular lines, such as HCO$^+$ and N$_2$H$^+$, as has been done for the disc around TW\,Hya, where a low-turbulence ``dead zone'' was inferred extending from the inner edge of the disc out to 50--65\,au \citep{Cleeves2015}.
 
The efficiency of this mechanism depends on where the transition radius $R_{\rm dz}$ is placed. Our step-like prescription places $R_{\rm dz}$ just interior to $R_{90,\rm obs}$, which leaves a detectable halo in J\,1615 and AA\,Tau (Fig.~\ref{fig:emission_strength}). Placing the transition further inward, for example at $R_{\rm c}/2$, would expose a larger fraction of the drifting solid reservoir to strong turbulent diffusion and fragmentation, and could plausibly suppress the outer emission more effectively, at the cost of also modifying the inner disc where continuum rings are observed. The required magnitude and location of the turbulence enhancement should therefore be regarded as poorly constrained by the present models; a systematic exploration of $R_{\rm dz}$ is beyond the scope of this work but is a natural next step.
 
Disc ionisation models generally predict a dead zone with low viscosity ($\alphavisc<10^{-2}$) whose outer boundary lies at radii of tens of au \citep{Flock_2015, Delage_2022}, generally interior to the outer bump radii in our sample ($\sim$150--400\,au). The outer disc regions explored here may therefore already reside in an MRI-active regime with high viscosity.
 
The halo emission reported by \citet{Das_2026} in DM\,Tau is consistent with a recently replenished outer disc, supporting the possibility that some outer pressure structures formed at late times. \citet{Das_2026} favour late infall over long-lived dust traps for IM\,Lup and DM\,Tau on the basis of the small grain sizes inferred there, reserving the unresolved-trap interpretation for Elias\,2-24, where cm-sized grains are present in the halo. They further note that dust drift and growth timescales in these regions are shorter than the disc ages, so smooth outer dust reservoirs should not persist under standard radial drift theory. However, our late-bump experiments show that residual solids are rapidly trapped once a pressure maximum appears (Section~\ref{sec:results_latebump}), so delayed bump formation alone is unlikely to explain the absence of outer continuum rings unless the outer disc is already depleted of solids when the bump forms.
 
A complementary possibility is that the outer pressure substructures are not static but dynamic, appearing and disappearing over time \citep{Stadler_2022}. Rather than permanently trapping solids, transient bumps drive episodic variations in the local drift rate; over successive cycles they could enhance the net inward drift, progressively depleting the outer disc of the material needed to build a detectable ring. This scenario would require the innermost pressure maxima to be comparatively long-lived, consistent with their observed continuum rings, while the outer substructures are shorter-lived and leaky. Such a spatial division in bump lifetime could naturally connect to the spatially varying turbulence discussed above, since the change in ionisation or MRI activity that raises the outer turbulence could also influence whether an outer pressure feature persists.

If the outer bumps are carved by embedded planets, the planets themselves could reduce the trapping efficiency. A massive planet stirs the dust in and around its gap, raising the dust scale height and the effective diffusion near the pressure maximum, which makes the trap leakier and the resulting ring fainter \citep{Binkert_2023}. This effect is not captured by our 1D models, in which the trap leakiness is set only by $\alphavisc$, and testing it would require 3D planet--disc simulations.

Several of the mechanisms discussed above, namely enhanced turbulence, transient bumps and planet-induced stirring, would reduce the efficiency with which the outer bumps retain solids, in addition to any effect on the grain size, which raises the question of how permeable a trap must be for its ring to disappear. To quantify the permeability of the outer bumps, we computed the trap blocking efficiency $\mathcal{B}$ of \citet{Houge_2026} for the late-bump models of LkCa\,15 and J\,1615 (Appendix~\ref{app:blocking}). We find no single permeability threshold below which the outer ring disappears: in LkCa\,15 the configuration with a detectable ring has an intermediate $\mathcal{B}\approx0.2$, whereas the most blocking trap ($\mathcal{B}\approx0.75$) remains undetectable because the retained mass resides in large grains, and in J\,1615 a halo is detected although the bump is nearly fully permeable, reflecting the large outer dust reservoir. The visibility of the outer emission is therefore set by the grain sizes of the trapped dust and the mass of the outer reservoir, rather than by the trap permeability alone.

Alternatively, part of the solids accumulated in the outer bumps may have already formed into planetesimals, Dust concentrated in pressure maxima can reach midplane dust-to-gas ratios high enough to trigger the streaming instability, and \citet{Stammler_2019} proposed ongoing planetesimal formation of this nature may explain the moderate optical depths of the DSHARP rings. Three-dimensional simulations showed that planetesimal formation can occur efficiently in dust-trapping rings even in the presence of turbulence \citep{Xu_2022}, and \citet{Zhao_2025} found that infall-induced pressure bumps can likewise form planetesimal belts. Mass that is locked in planetesimals is invisible at (sub-)millimetre wavelengths, so efficient planetesimal formation in the outer bumps would lower the predicted ring brightness. In our models the midplane dust-to-gas ratio exceeds unity beyond $R_{90,\rm obs}$, in regions where the vertically integrated dust-to-gas ratio is also enhanced above its initial value, in 12 out of the 72 reference models, 10 of which have $\alphavisc=10^{-4}$ and $\vfrag=10\,\mathrm{m\,s^{-1}}$, the configuration with the most efficient trapping and the largest and most settled grains. In several of these, the vertically integrated dust-to-gas ratio at the bump approaches or exceeds unity, a regime in which the back-reaction of the dust on the gas which we neglect would become important and planetesimal formation would be expected well before such concentrations are reached. Since we do not include planetesimal formation, the predicted outer emission is an upper estimate in this respect. However only 9 out of 55 models with detectable outer emission reach the conditions, so planetesimal formation could reduce the predicted emission in the most efficient traps but cannot on its own explain the absence of outer rings across the entire sample.

\subsection{Limitations and future work}
\label{sec:discussion_limitations}
A key source of uncertainty in this work arises from the reconstruction of the gas pressure structure itself. The exoALMA pressure-derivative profiles are inferred from deviations in the observed rotation curves and therefore inherit uncertainties associated with finite spatial resolution, beam smearing, noise, and assumptions about the underlying disc geometry and velocity structure. Because our modelling relies on derivatives of the inferred pressure profile, even relatively small uncertainties in the reconstructed velocity field and mid-plane gas pressure derivatives can translate into significant variations in the resulting gas surface density structure \citep{Andrews_2024, Veronesi_2024}.

Converting measurements of $d\ln P/d\ln r$ into gas surface density profiles, $\Sigma_{\rm g}(r)$, is also intrinsically non-unique. In our reconstruction framework, the shape of $\Sigma_{\rm g}(r)$ is determined by the inferred pressure-gradient profile, while the total gas mass is fixed using thermochemical mass estimates \citep{Trapman_2025} and the temperature structure adopted from the exoALMA modelling \citep{Galloway-Sprietsma_2025}. Consequently, uncertainties in the pressure gradients, temperature structure, and treatment of regions outside the kinematically constrained domain can affect both the contrast and radial extent of the reconstructed pressure maxima. This is particularly important because dust trapping efficiency depends strongly on the morphology of these pressure profiles. 
 
All models start with a uniform dust-to-gas ratio of 0.01 out to the edge of the grid. If the dust disc were initially more compact than the gas disc, the smaller outer dust reservoir would generally reduce the predicted outer emission.
 
The dust evolution uses the power-law midplane temperature of \citet{Galloway-Sprietsma_2025}, whereas the synthetic images use the \radmc\ dust temperature (Section~\ref{sec:methods_rt}). For LkCa\,15, adopting the \dustpy\ temperature does not change the classification of the outer emission. Since the Band~7 intensity of the cold outer disc scales approximately linearly with the dust temperature, the models whose outer emission lies close to the $3\sigma$ threshold, such as the LkCa\,15 late-bump models (3--5$\sigma$; Fig.~\ref{fig:emission_strength}), are the most sensitive to this choice, while the strongly detected cases are not. In addition, because the inner disc profiles are partly extrapolated, the shadowing of the outer disc by the modelled inner disc is only approximate.
 
A further limitation is that observational uncertainties in the pressure-gradient profiles are not explicitly propagated through the gas reconstruction and \dustpy\ modelling pipeline. The current models are generated from a single reconstructed $\Sigma_{\rm g}(r)$ profile rather than an ensemble spanning the allowed uncertainty range. A more complete treatment would require propagating rotation-curve uncertainties through the gas reconstruction, dust evolution, and radiative transfer calculations. While computationally prohibitive with full \dustpy\ simulations, future work could employ semi-analytic dust evolution models such as the two-population framework of \citet{Birnstiel_2012} or the three-population \texttt{TriPoD} model \citep{Pfeil_2024} to enable ensemble modelling across the allowed pressure-profile uncertainty range.
 
Another limitation in the present framework is its 1D axisymmetric geometry. Several discs in the sample exhibit azimuthal asymmetries that our models cannot capture: HD\,34282 in particular hosts a bright azimuthal arc consistent with a dust-trapping vortex \citep{Ma_2026}, and the exoALMA kinematic data reveal arc- and spiral-like velocity residuals in this system \citep{Woelfer_2025, Fukagawa_2026}. For any disc where the outer gas substructure is itself azimuthally asymmetric, the 1D assumption will overestimate the radially averaged trapping efficiency, and predicted ring brightnesses should be interpreted accordingly.

\section{Conclusions}
\label{sec:conclusions}

In this work, we used the midplane pressure-derivative profiles published by exoALMA from \twelveco\ and \thirteenco\ observations \citep{Stadler_2025}, together with chemically estimated gas masses \citep{Trapman_2025}, to reconstruct gas surface density profiles for nine protoplanetary discs, and modelled their dust evolution and synthetic continuum emission with \dustpy\ and \radmc. We explored how different combinations of $\alphavisc$ and $\vfrag$ affect the formation and observability of outer dust rings associated with gas pressure substructures beyond the observed millimetre continuum discs, tested two mechanisms that could suppress them, and quantified the permeability of the outer bumps.

Our main conclusions are:

\begin{enumerate}[leftmargin=*]
\item The reconstructed outer pressure substructures should generally be observable. In 21 of the 36 \twelveco-based models and 34 of the 36 \thirteenco-based models, efficient outer dust trapping produces emission above $3\sigma$ beyond the observed continuum disc edge at the resolution and sensitivity of the exoALMA Large Program, in tension with the observations.Trapping also occurs in regions where the velocity-deviation profiles are not formally super-Keplerian, but where variations in the pressure gradient produce traffic-jam pile-ups.

\item The \thirteenco-based reconstructions produce detectable outer emission almost universally, whereas the \twelveco-based ones show greater diversity. Whether this difference is physical or reflects the different radial coverage and signal-to-noise of the two tracers remains unclear.

\item Only a limited set of models suppresses the outer continuum emission, and many of these also produce continuum discs more compact than observed. In LkCa\,15, two configurations show no detectable outer emission: $\alphavisc=10^{-3}$, $\vfrag=1\,\mathrm{m\,s^{-1}}$, where the dust does not grow to (sub-)millimetre sizes, and $\alphavisc=10^{-4}$, $\vfrag=10\,\mathrm{m\,s^{-1}}$, where large grains form and are concentrated in a narrow ring that remains too faint to detect. The latter case shows that some degree of ``hidden dust'' is possible in specific cases, although the outer ring does not become detectable at longer wavelengths in our models. Tentative outer continuum emission has recently been reported in LkCa\,15 at Band~3 \citep{Sierra_2025} and in HD\,34282 \citep{Ma_2026}.

\item We tested two mechanisms for suppressing outer ring formation: a late-time introduction of the outer pressure bump in LkCa\,15 and J\,1615, and enhanced outer-disc turbulence ($\alphavisc_{\rm out}\sim10^{-2}$) in LkCa\,15, J\,1615 and AA\,Tau. Enhanced turbulence is the most effective: it removes the outer emission in LkCa\,15 in all configurations, but a detectable halo persists in J\,1615 and AA\,Tau. Its efficiency depends on the assumed location of the turbulence transition, which we do not explore systematically. Delayed introduction of the outer bump does not robustly suppress ring formation in either system, as residual solids are rapidly swept up once the bump appears.

\item Using the trap blocking efficiency $\mathcal{B}$ of \citet{Houge_2026}, we find no single permeability threshold below which the outer ring disappears. The visibility of the outer emission is therefore set by the grain size distribution of the trapped dust and by the mass of the outer dust reservoir, rather than by the trap permeability alone.
\end{enumerate}

Spatially, as well as, temporarily (though not measurable) varying turbulence remains the most promising explanation for the missing outer rings. It could be tested observationally by measuring the ionisation structure of the outer discs, for example with HCO$^+$ and N$_2$H$^+$, and by deeper continuum observations sensitive to faint, extended outer emission.

\section*{Acknowledgements}
LL would like to thank the Science and Technology Facilities Council (STFC) for funding support through a PhD studentship.

PP acknowledges funding from the UK Research and Innovation (UKRI) under the UK government’s Horizon Europe funding guarantee from ERC (under grant agreement No 101076489). 

We acknowledge Myriam Benisty and Fabiola Antonietta Gerosa for insightful discussions about this project.
We also thank the anonymous referee for their helpful suggestions. This paper makes use of the following ALMA data: 2021.1.01123.L. ALMA is a partnership of ESO (representing its member states), NSF (USA) and NINS (Japan), together with NRC (Canada), MOST and ASIAA (Taiwan), and KASI (Republic of Korea), in cooperation with the Republic of Chile. The Joint ALMA Observatory is operated by ESO, AUI/NRAO and NAOJ.
We acknowledge the use of artificial intelligence (Claude Opus 5.5.) for the purpose of improving the text's readability. We also acknowledge the use of Claude to help draft the python code to plot Figure ~\ref{fig:outcomes_reference} and ~\ref{fig:outcomes_mechanisms}.

\section*{Data Availability}

The data underlying this article are available from the corresponding author upon request.

\bibliographystyle{mnras}
\bibliography{references}

\appendix

\section{Gas surface density reconstruction}
\label{app:reconstruction}        

The pressure-gradient from \citet{Stadler_2025} are derived from deviations of the observed azimuthal velocity from Keplerian rotation. The SNR is typically lower in the inner disc due to beam smearing being more significant at small angular separations and the fractional difference from Keplerian rotation caused by a given pressure perturbation is smaller in the fast-rotating inner disc. Conversely, the outer disc yields more robust kinematic constraints \citep{Stadler_2025}. This is because the radial annulus is resolved by more beams at larger radii, reducing the per-annulus velocity uncertainty and because the pressure scale height increases with radius and therefore more readily resolved in the outer disc.

We reconstruct gas surface density profiles $\Sigma_{\rm g}(r)$
consistent with the observed pressure-derivative constraints and the
chemically estimated gas mass in four steps.

\subsubsection*{Step 1 -- Blending small and large-scale profiles}

The exoALMA data products from \citet{Stadler_2025} provide two
independent measurements of $d\ln P_{\rm g}/d\ln r$ per system: a
high-resolution small-scale profile and a lower-resolution large-scale
profile. In the overlap region we combine them as
\begin{equation}
    \left.\frac{d\ln P_{\rm g}}{d\ln r}\right|_{\rm blend}
    = (1-w)\left.\frac{d\ln P_{\rm g}}{d\ln r}\right|_{\rm small}
    + w\left.\frac{d\ln P_{\rm g}}{d\ln r}\right|_{\rm large},
\end{equation}
where the logistic weight is
\begin{equation}
    w(r) = \frac{1}{1+\exp\!\left[-(r-R_{\rm sw})/\Delta r\right]}.
\end{equation}
Here $R_{\rm sw}$ is the midpoint of the overlap region and $\Delta r$
is 20 \% of its width.
Beyond the outer limit of the observed data, we extend the profile
analytically using the \citet{Lynden-Bell_1974} pressure-derivative
tail,
\begin{equation}
    \left.\frac{d\ln P_{\rm g}}{d\ln r}\right|_{\rm tail}
    = -\gamma - (2-\gamma)\!\left(\frac{r}{R_c}\right)^{\!2-\gamma}
    + \frac{q}{2} - \frac{3}{2},
\end{equation}
with $\gamma = 1$, temperature power-law index $q$ from
\citet{Galloway-Sprietsma_2025}, and characteristic radius $R_c$
from the CO rotation curve modelling of \citet{Longarini_2025}. The inner profile is similarly extended
inward using a \citet{Lynden-Bell_1974} profile that  matches at the innermost observed radius.

\subsubsection*{Step 2 -- Integration}

We obtain the pressure shape by trapezoidal integration,
\begin{equation} \label{eq:integration}
    \ln P_{\rm g}(r) = \int_{r_0}^{r}
    \frac{d\ln P_{\rm g}}{d\ln r'}\,d\ln r',
\end{equation}
Equation~\ref{eq:integration} is integrated with $P_{\rm g}(r_0) \equiv 1$ , where $r_0$ is the innermost radius of the observed (small-scale) pressure-derivative profile. This choice is arbitrary and has no effect on the final profile, since $\Sigma_{\rm g}(r)$ is subsequently rescaled in Step 4 to match the disc mass estimated from chemical tracers \citep{Trapman_2025}.

\subsubsection*{Step 3 -- Conversion to surface density.}

Assuming a Gaussian vertical density profile, the mid-plane gas 
density is
\begin{equation}
    \rho_g = \frac{\Sigma_{\rm g}}{\sqrt{2\pi}\,H},
\end{equation}
so that the midplane pressure is
\begin{equation}
    P_{\rm g} = \rho_g\,c_s^2 
              = \frac{\Sigma_{\rm g}\,c_s^2}{\sqrt{2\pi}\,H},
    \qquad H = \frac{c_s}{\Omega},
\end{equation}
where the isothermal sound speed is
\begin{equation}
    c_s = \sqrt{\frac{k_B\,T_{\rm mid}}{\mu\,m_p}},
\end{equation}
with mean molecular weight $\mu = 2.37$.
The midplane temperature $T_{\rm mid}(r) = T_0\,(r/100\,{\rm au})^q$ is a power-law fit to the disc-specific temperature profiles derived by \citet{Galloway-Sprietsma_2025} from thermochemical modelling of the exoALMA CO isotopologue emission, with $T_0$ and $q$ determined for each system. 

Inverting the pressure equation gives
\begin{equation}
    \Sigma_{\rm g}(r) = \frac{\sqrt{2\pi}\,H(r)\,P_{\rm g}(r)}{c_s^2(r)}.
\end{equation}
\subsubsection*{Step 4 -- Mass normalisation}

The profile is rescaled so that
\begin{equation}
    M_{\rm gas} = 2\pi\int_{r_{\rm min}}^{r_{\rm max}}
    \Sigma_{\rm g}(r)\,r\,dr
\end{equation}
matches the chemically estimated disc mass from \citet{Trapman_2025}.
Uncertainties in $M_{\rm gas}$ and $T_{\rm mid}(r)$ are not propagated
through the pipeline; their impact is discussed in
Section~\ref{sec:discussion_limitations}.

The gas masses from \citet{Trapman_2025} are estimated from 
\ntwohp\ and \ceighteeno\ line fluxes combined with 
thermochemical modelling. We adopt these as our fiducial normalisation as they provide a spatially integrated constraint on the total gas content that is independent of the kinematic modelling used to derive the pressure-gradient profiles. A comparison with the dynamically
measured masses from \citet{Longarini_2025} shows agreement within a factor of 
$\sim$3, with the chemical masses typically a factor of $2.3^{+0.7}_{-1.0}$ lower than the kinematic ones 
\citep{Trapman_2025}. This systematic offset propagates 
directly into the amplitude of $\Sigma_{\rm g}(r)$ and 
therefore into the predicted outer dust ring brightness; since our predicted ring fluxes scale directly with the gas normalisation, models using dynamic masses would predict proportionally brighter rings.

We produce two independent $\Sigma_{\rm g}(r)$ profiles per 
system using the \thirteenco\ and \twelveco\ pressure-derivative 
measurements from \citet{Stadler_2025} as separate model inputs. 
Since \thirteenco\ is less optically thick than \twelveco, it 
traces deeper into the disc, and should in principle more accurately reflect the 
thermal and density structure of the midplane where dust 
trapping occurs \citep{Galloway-Sprietsma_2025}. In practice, we find that the \thirteenco\ based reconstructions systematically produce stronger outer pressure contrasts and more efficient dust trapping than the \twelveco\, ones (see Sect.~\ref{sec:results}). We discuss this further in Section~\ref{sec:discussion_baseline}.

\section{Candidate region selection}
\label{app:candidates}            
We stress that this classification is sensitive to the assumed stellar mass. Whether a region is labelled super- or sub-Keplerian depends on the absolute Keplerian  velocity and hence on $M_{\star}$, whereas the sign of the $\delta v_\phi$ gradient does not. The exoALMA stellar masses are constrained to $\approx 2$--$5\%$ \citep{Longarini_2025}, which corresponds to a $\approx 1$--$2.5 \%$ uncertainty in $v_{\rm Kep}$ since $v_{\rm Kep} \propto M_\star^{1/2}$, so a modest revision of $M_\star$ could shift marginal features across the threshold. We emphasise, however, that this criterion is used only to identify and report candidate outer pressure substructures; it does not enter the dust evolution modelling, which is driven directly by the full mid-plane pressure-gradient profile and therefore captures the drift behaviour, including traffic-jam pile-ups, regardless of how a region is labelled. The stellar-mass sensitivity thus affects only which regions we flag as candidate bumps, not the predicted dust distribution.

Adopting the broad super-Keplerian extent as the region boundary,
rather than only the highly significant pixels, means the
reported $r_{\rm in}$--$r_{\rm out}$ range reflects the full
zone of super-Keplerian rotation and therefore the full spatial
extent over which a pressure bump could plausibly trap drifting
solids. The resulting zones should be interpreted as candidate
gas-pressure substructure regions, not confirmed dust traps (see Appendix Figs.~\ref{fig:velocity_12co} and~\ref{fig:velocity_13co}).

\section{Trap blocking efficiency and pebble fluxes}
\label{app:blocking}

To quantify the permeability of the outer pressure bumps, we adopt the trap blocking efficiency of \citet{Houge_2026},
\begin{equation}
    \mathcal{B}(t) = 1 - \frac{M_{\rm cross}^{\rm bump}(t)}{M_{\rm cross}^{\rm control}(t)},
    \label{eq:blocking}
\end{equation}
where $M_{\rm cross}$ is the dust mass that has crossed a reference radius $r_{\rm gap}$ since time $t_0$ in a simulation with (bump) and without (control) the pressure perturbation. $\mathcal{B}=0$ corresponds to a fully permeable perturbation, $\mathcal{B}=1$ to complete blocking, and $\mathcal{B}<0$ to inward transport faster than in the unperturbed disc.

The late-bump experiments (Section~\ref{sec:methods_tests}) provided us with the required pairs of simulations. After the shared 1\,Myr phase on the no-bump gas profile, each model is continued for 0.5\,Myr both with the outer bump introduced and, as a control, without it. Both runs therefore start from an identical dust distribution and differ only in the outer pressure perturbation. We set $t_0 = 1$\,Myr, when the bump is introduced, and evaluate $\mathcal{B}$ at $t = 1.5$\,Myr. We compute $M_{\rm cross}$ as the decrease of the dust mass outside $r_{\rm gap}$,
\begin{equation}
    M_{\rm cross}(t) = M_{\rm d}(r>r_{\rm gap}, t_0) - M_{\rm d}(r>r_{\rm gap}, t),
\end{equation}
which includes both advective and diffusive transport, and, being computed from mass differences, is insensitive to the snapshot sampling, and the pebble flux as $F_{\rm peb}(r,t) = -\,\mathrm{d}M_{\rm d}(r'>r, t)/\mathrm{d}t$.

The perturbation introduced in the late-bump test consists of a pressure depression interior to the bump, followed by the bump itself. We thus place $r_{\rm gap}$ at the minimum of $P_{\rm bump}/P_{\rm control}$ interior to the bump, at 85\,au for LkCa\,15 and 60\,au for J\,1615, analogous to measuring the flux through the gap of a planet-induced trap. In order to test the sensitivity to this choice, in Table~\ref{tab:blocking} we list $\mathcal{B}$ for a range of $r_{\rm gap}$. For the LkCa\,15 case,  $\mathcal{B}$ is robust to the choice of $r_{\rm gap}$ for the two configurations with $\vfrag = 10\,\mathrm{m\,s^{-1}}$, but varies between $\approx$0.05 and 0.4 for $\alphavisc=10^{-4}$, $\vfrag=1\,\mathrm{m\,s^{-1}}$; in J\,1615, $|\mathcal{B}| \lesssim 0.15$ for all configurations at and inside the depression. Immediately after the bump is introduced, $\mathcal{B}(t)$ shows short transients (including $\mathcal{B}<0$) as the steepened pressure gradient in the depression briefly accelerates the inward drift; the values in Table~\ref{tab:blocking} refer to $t=1.5$\,Myr.

Table~\ref{tab:blocking} compares $\mathcal{B}$ with our classification of the late-bump models. In LkCa\,15, $\mathcal{B}$ ranges from $\approx0$ to $\approx0.75$ but does not map monotonically onto the predicted continuum morphology: the configuration with a detectable outer ring has $\mathcal{B}\approx0.2$, whereas the two configurations without detectable outer emission have $\mathcal{B}\approx0.06$ and $\approx0.75$. The most blocking trap ($\alphavisc=10^{-4}$, $\vfrag=10\,\mathrm{m\,s^{-1}}$) remains undetectable because the retained mass resides in large grains with low Band~7 opacity (Section~\ref{sec:discussion_hidden_dust}). In J\,1615 the bump is nearly fully permeable ($|\mathcal{B}|\lesssim0.15$) in all configurations, yet every model produces a detectable halo, which therefore is indicative of the large outer dust reservoir ($\gtrsim300\,M_\oplus$ beyond 60\,au) rather than efficient trapping. These values are not directly comparable to those of \citet{Houge_2026}, who quote $\mathcal{B}$ after 5\,Myr for planet-induced traps at 5--40\,au. Moreover, we only compute $\mathcal{B}$ for the late-bump models, since for the reference models we had no bump-free counterpart.

\begin{figure*}
  \centering
  \includegraphics[width=0.96\textwidth]{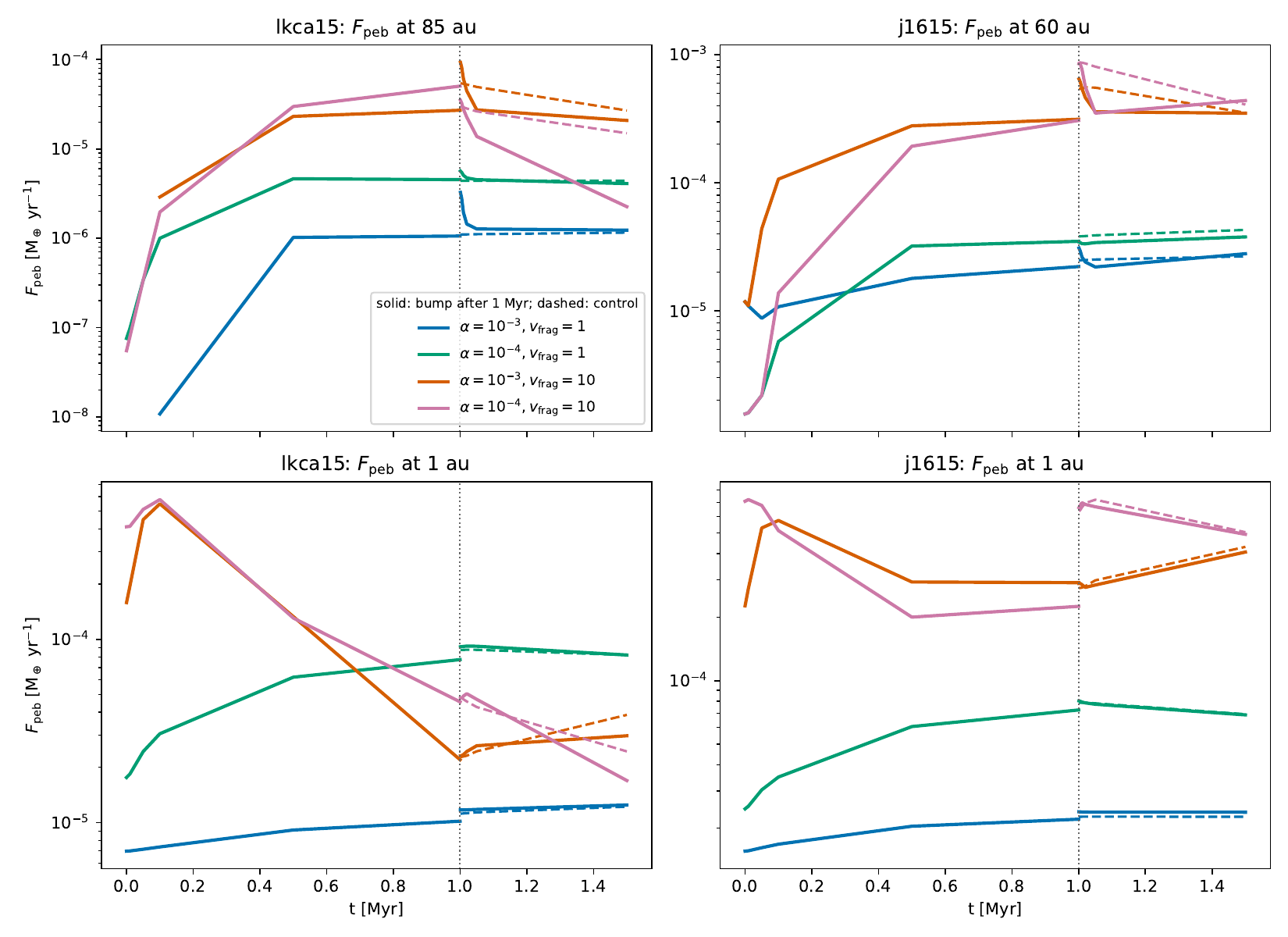}
  \caption{Pebble flux through $r_{\rm gap}$ (top; 85\,au for LkCa\,15, 60\,au for J\,1615) and through 1\,au (bottom) for the \twelveco-based late-bump models. Solid lines show the shared no-bump phase (0--1\,Myr) followed by the run with the outer bump introduced at 1\,Myr (dotted vertical line); dashed lines show the control run continued without the bump. Negative (outward) fluxes are not shown on the logarithmic axis. The apparent discontinuity at 1\,Myr is a sampling effect: the first phase snapshots are separated by up to 0.5\,Myr, so the fluxes shown there are averages over long intervals, whereas the flux rises steeply towards 1\, Myr. Interval-averaged fluxes are continuous across the restart.}
  \label{fig:blocking_flux}
\end{figure*}

Fig.~\ref{fig:blocking_flux} shows the pebble flux at $r_{\rm gap}$ and at 1\,au over the full 1.5\,Myr. At 1\,au, the bump and control runs differ by less than a factor of $\sim$1.5 within the 0.5\,Myr after the bump forms, reflecting the long drift time from the outer disc to the inner disc. Therefore, on this timescale the outer bumps do not measurably change the delivery of pebbles to the inner disc probed by JWST.

\begin{table*}
\centering
\caption{Trap blocking efficiency $\mathcal{B}$ at $t = 1.5$\,Myr (Eq.~\ref{eq:blocking}) for the \twelveco-based late-bump models, for several reference radii $r_{\rm gap}$. Values at the adopted $r_{\rm gap}$ (pressure depression interior to the bump) are in bold. $M_{\rm d}$ is the dust mass outside the adopted $r_{\rm gap}$ when the bump is introduced (1\,Myr). The last column gives the late-bump classification of the outer Band~7 emission (Table~\ref{tab:special_models}).}
\label{tab:blocking}
\begin{tabular}{lccccccl}
\toprule
\multicolumn{8}{l}{LkCa\,15} \\
$(\alphavisc,\ \vfrag/\mathrm{m\,s^{-1}})$ & $M_{\rm d}$ ($M_\oplus$) & 70\,au & \textbf{85\,au} & 100\,au & 120\,au & & Outcome \\
\midrule
$(10^{-3},\ 1)$  & 47  & 0.01 & $\mathbf{-0.10}$ & $-0.05$ & 0.07 & & Halo \\
$(10^{-4},\ 1)$  & 43  & 0.05 & \textbf{0.06}  & 0.25 & 0.40 & & No outer mm \\
$(10^{-3},\ 10)$ & 26  & 0.20 & \textbf{0.22}  & 0.23 & 0.24 & & Ring \\
$(10^{-4},\ 10)$ & 17  & 0.71 & \textbf{0.75}  & 0.81 & 0.87 & & No outer mm \\
\midrule
\multicolumn{8}{l}{J\,1615} \\
$(\alphavisc,\ \vfrag/\mathrm{m\,s^{-1}})$ & $M_{\rm d}$ ($M_\oplus$) & 45\,au & \textbf{60\,au} & 80\,au & 100\,au & & Outcome \\
\midrule
$(10^{-3},\ 1)$  & 551 & 0.15 & $\mathbf{-0.04}$ & $-0.20$ & $-0.20$ & & Halo \\
$(10^{-4},\ 1)$  & 539 & 0.02 & \textbf{0.12}  & 0.38 & 0.63 & & Halo \\
$(10^{-3},\ 10)$ & 306 & 0.04 & \textbf{0.04}  & 0.04 & 0.03 & & Halo \\
$(10^{-4},\ 10)$ & 373 & 0.02 & \textbf{0.01}  & 0.02 & 0.01 & & Halo \\
\bottomrule
\end{tabular}
\end{table*}

\section{Model outcome tables}
\label{app:tables}              
Tables~\ref{tab:12co_summary}--\ref{tab:special_models} list, for every model, the classification of the outer Band~7 emission (Ring, Halo or No outer mm; Section~\ref{sec:methods_rt}), the synthetic disc radius $R_{90,\rm syn}$ compared with the observed $R_{90,\rm obs}$, and the candidate outer pressure-substructure regions inferred from the kinematics (Section~\ref{sec:methods_candidate_regions}). Tables~\ref{tab:12co_summary} and~\ref{tab:13co_summary} give the reference models for the \twelveco- and \thirteenco-based reconstructions, summarised in Fig.~\ref{fig:outcomes_reference}, and Table~\ref{tab:special_models} gives the suppression tests, summarised in Fig.~\ref{fig:outcomes_mechanisms}.
\begin{table*}[t]
\centering
\caption{Summary of the synthetic Band~7 continuum outcomes for the \twelveco-based dust-evolution models. The \emph{outer mm emission} column classifies the morphology of any emission detected above $3\sigma$ beyond the observed continuum disc edge ($R_{90,\rm obs}$, from \citet{Curone_2025}). The final column lists the candidate outer gas pressure-substructure regions inferred from the \twelveco \, kinematics.}
\begin{tabular}{llcccc}
\toprule
Disc & Config & Outer mm emission & $R_{90,\rm obs}$ (au) & $R_{90,\rm syn}$ (au) & Candidate $R_{\rm bump,gas}$ region (au) \\
\midrule
\multirow{4}{*}{AA\,Tau} & \makebox[3.5cm][l]{$\alpha=10^{-3},\,v_{\rm frag}=1\,{\rm m\,s^{-1}}$} & Ring & 139 & 431 & 268--460 \\
 & \makebox[3.5cm][l]{$\alpha=10^{-4},\,v_{\rm frag}=1\,{\rm m\,s^{-1}}$} & Ring &  & 545 &  \\
 & \makebox[3.5cm][l]{$\alpha=10^{-3},\,v_{\rm frag}=10\,{\rm m\,s^{-1}}$} & Ring &  & 552 &  \\
 & \makebox[3.5cm][l]{$\alpha=10^{-4},\,v_{\rm frag}=10\,{\rm m\,s^{-1}}$} & Ring &  & 546 &  \\
\midrule
\multirow{4}{*}{DM\,Tau} & \makebox[3.5cm][l]{$\alpha=10^{-3},\,v_{\rm frag}=1\,{\rm m\,s^{-1}}$} & No outer mm & 202 & 44 & 200--340 \\
 & \makebox[3.5cm][l]{$\alpha=10^{-4},\,v_{\rm frag}=1\,{\rm m\,s^{-1}}$} & No outer mm &  & 32 &  \\
 & \makebox[3.5cm][l]{$\alpha=10^{-3},\,v_{\rm frag}=10\,{\rm m\,s^{-1}}$} & No outer mm &  & 34 &  \\
 & \makebox[3.5cm][l]{$\alpha=10^{-4},\,v_{\rm frag}=10\,{\rm m\,s^{-1}}$} & No outer mm &  & 26 &  \\
\midrule
\multirow{4}{*}{HD\,34282} & \makebox[3.5cm][l]{$\alpha=10^{-3},\,v_{\rm frag}=1\,{\rm m\,s^{-1}}$} & No outer mm & 239 & 83 & 243--382 \\
 & \makebox[3.5cm][l]{$\alpha=10^{-4},\,v_{\rm frag}=1\,{\rm m\,s^{-1}}$} & No outer mm &  & 58 &  \\
 & \makebox[3.5cm][l]{$\alpha=10^{-3},\,v_{\rm frag}=10\,{\rm m\,s^{-1}}$} & No outer mm &  & 48 &  \\
 & \makebox[3.5cm][l]{$\alpha=10^{-4},\,v_{\rm frag}=10\,{\rm m\,s^{-1}}$} & No outer mm &  & 50 &  \\
\midrule
\multirow{4}{*}{J1615} & \makebox[3.5cm][l]{$\alpha=10^{-3},\,v_{\rm frag}=1\,{\rm m\,s^{-1}}$} & Halo & 170 & 384 & 170--252 \\
 & \makebox[3.5cm][l]{$\alpha=10^{-4},\,v_{\rm frag}=1\,{\rm m\,s^{-1}}$} & Halo &  & 347 &  \\
 & \makebox[3.5cm][l]{$\alpha=10^{-3},\,v_{\rm frag}=10\,{\rm m\,s^{-1}}$} & Halo &  & 369 &  \\
 & \makebox[3.5cm][l]{$\alpha=10^{-4},\,v_{\rm frag}=10\,{\rm m\,s^{-1}}$} & Ring &  & 391 &  \\
\midrule
\multirow{4}{*}{J1842} & \makebox[3.5cm][l]{$\alpha=10^{-3},\,v_{\rm frag}=1\,{\rm m\,s^{-1}}$} & Halo & 85 & 206 & 85--110 \\
 & \makebox[3.5cm][l]{$\alpha=10^{-4},\,v_{\rm frag}=1\,{\rm m\,s^{-1}}$} & Ring &  & 226 &  \\
 & \makebox[3.5cm][l]{$\alpha=10^{-3},\,v_{\rm frag}=10\,{\rm m\,s^{-1}}$} & Halo &  & 100 &  \\
 & \makebox[3.5cm][l]{$\alpha=10^{-4},\,v_{\rm frag}=10\,{\rm m\,s^{-1}}$} & No outer mm &  & --- &  \\
\midrule
\multirow{4}{*}{J1852} & \makebox[3.5cm][l]{$\alpha=10^{-3},\,v_{\rm frag}=1\,{\rm m\,s^{-1}}$} & Ring & 70 & 275 & --- \\
 & \makebox[3.5cm][l]{$\alpha=10^{-4},\,v_{\rm frag}=1\,{\rm m\,s^{-1}}$} & Ring &  & 203 &  \\
 & \makebox[3.5cm][l]{$\alpha=10^{-3},\,v_{\rm frag}=10\,{\rm m\,s^{-1}}$} & Halo &  & 169 &  \\
 & \makebox[3.5cm][l]{$\alpha=10^{-4},\,v_{\rm frag}=10\,{\rm m\,s^{-1}}$} & Ring &  & 215 &  \\
\midrule
\multirow{4}{*}{LkCa\,15} & \makebox[3.5cm][l]{$\alpha=10^{-3},\,v_{\rm frag}=1\,{\rm m\,s^{-1}}$} & No outer mm & 156 & 66 & 159--254; 336--395 \\
 & \makebox[3.5cm][l]{$\alpha=10^{-4},\,v_{\rm frag}=1\,{\rm m\,s^{-1}}$} & Ring &  & 164 &  \\
 & \makebox[3.5cm][l]{$\alpha=10^{-3},\,v_{\rm frag}=10\,{\rm m\,s^{-1}}$} & Ring &  & 160 &  \\
 & \makebox[3.5cm][l]{$\alpha=10^{-4},\,v_{\rm frag}=10\,{\rm m\,s^{-1}}$} & No outer mm &  & 36 &  \\
\midrule
\multirow{4}{*}{SY\,Cha} & \makebox[3.5cm][l]{$\alpha=10^{-3},\,v_{\rm frag}=1\,{\rm m\,s^{-1}}$} & No outer mm & 198 & 63 & 198--334 \\
 & \makebox[3.5cm][l]{$\alpha=10^{-4},\,v_{\rm frag}=1\,{\rm m\,s^{-1}}$} & No outer mm &  & 39 &  \\
 & \makebox[3.5cm][l]{$\alpha=10^{-3},\,v_{\rm frag}=10\,{\rm m\,s^{-1}}$} & No outer mm &  & 41 &  \\
 & \makebox[3.5cm][l]{$\alpha=10^{-4},\,v_{\rm frag}=10\,{\rm m\,s^{-1}}$} & No outer mm &  & 41 &  \\
\midrule
\multirow{4}{*}{V4046\,Sgr} & \makebox[3.5cm][l]{$\alpha=10^{-3},\,v_{\rm frag}=1\,{\rm m\,s^{-1}}$} & Halo & 61 & 279 & 72--212 \\
 & \makebox[3.5cm][l]{$\alpha=10^{-4},\,v_{\rm frag}=1\,{\rm m\,s^{-1}}$} & Ring &  & 295 &  \\
 & \makebox[3.5cm][l]{$\alpha=10^{-3},\,v_{\rm frag}=10\,{\rm m\,s^{-1}}$} & Ring &  & 169 &  \\
 & \makebox[3.5cm][l]{$\alpha=10^{-4},\,v_{\rm frag}=10\,{\rm m\,s^{-1}}$} & Ring &  & 257 &  \\
\bottomrule
\end{tabular}
\label{tab:12co_summary}
\end{table*}

\begin{table*}[t]
\centering
\caption{Summary of the synthetic Band~7 continuum outcomes for the $^{13}$CO-based dust-evolution models. The \emph{outer mm emission} column classifies the morphology of any emission detected above $3\sigma$ beyond the observed continuum disc edge ($R_{90,\rm obs}$, from \citet{Curone_2025}). The final column lists the candidate outer gas pressure-substructure regions inferred from the $^{13}$CO kinematics.}
\begin{tabular}{llcccc}
\toprule
Disc & Config & Outer mm emission & $R_{90,\rm obs}$ (au) & $R_{90,\rm syn}$ (au) & Candidate $R_{\rm bump,gas}$ region (au) \\
\midrule
\multirow{4}{*}{AA\,Tau} & \makebox[3.5cm][l]{$\alpha=10^{-3},\,v_{\rm frag}=1\,{\rm m\,s^{-1}}$} & Ring & 139 & 304 & --- \\
 & \makebox[3.5cm][l]{$\alpha=10^{-4},\,v_{\rm frag}=1\,{\rm m\,s^{-1}}$} & Ring &  & 286 &  \\
 & \makebox[3.5cm][l]{$\alpha=10^{-3},\,v_{\rm frag}=10\,{\rm m\,s^{-1}}$} & Ring &  & 285 &  \\
 & \makebox[3.5cm][l]{$\alpha=10^{-4},\,v_{\rm frag}=10\,{\rm m\,s^{-1}}$} & Ring &  & 281 &  \\
\midrule
\multirow{4}{*}{DM\,Tau} & \makebox[3.5cm][l]{$\alpha=10^{-3},\,v_{\rm frag}=1\,{\rm m\,s^{-1}}$} & No outer mm & 202 & 186 & 200--448 \\
 & \makebox[3.5cm][l]{$\alpha=10^{-4},\,v_{\rm frag}=1\,{\rm m\,s^{-1}}$} & No outer mm &  & 80 &  \\
 & \makebox[3.5cm][l]{$\alpha=10^{-3},\,v_{\rm frag}=10\,{\rm m\,s^{-1}}$} & Ring &  & 296 &  \\
 & \makebox[3.5cm][l]{$\alpha=10^{-4},\,v_{\rm frag}=10\,{\rm m\,s^{-1}}$} & Ring &  & 432 &  \\
\midrule
\multirow{4}{*}{HD\,34282} & \makebox[3.5cm][l]{$\alpha=10^{-3},\,v_{\rm frag}=1\,{\rm m\,s^{-1}}$} & Ring & 239 & 562 & 243--521 \\
 & \makebox[3.5cm][l]{$\alpha=10^{-4},\,v_{\rm frag}=1\,{\rm m\,s^{-1}}$} & Ring &  & 553 &  \\
 & \makebox[3.5cm][l]{$\alpha=10^{-3},\,v_{\rm frag}=10\,{\rm m\,s^{-1}}$} & Ring &  & 552 &  \\
 & \makebox[3.5cm][l]{$\alpha=10^{-4},\,v_{\rm frag}=10\,{\rm m\,s^{-1}}$} & Ring &  & 543 &  \\
\midrule
\multirow{4}{*}{J1615} & \makebox[3.5cm][l]{$\alpha=10^{-3},\,v_{\rm frag}=1\,{\rm m\,s^{-1}}$} & Halo & 170 & 413 & --- \\
 & \makebox[3.5cm][l]{$\alpha=10^{-4},\,v_{\rm frag}=1\,{\rm m\,s^{-1}}$} & Ring &  & 365 &  \\
 & \makebox[3.5cm][l]{$\alpha=10^{-3},\,v_{\rm frag}=10\,{\rm m\,s^{-1}}$} & Ring &  & 340 &  \\
 & \makebox[3.5cm][l]{$\alpha=10^{-4},\,v_{\rm frag}=10\,{\rm m\,s^{-1}}$} & Ring &  & 359 &  \\
\midrule
\multirow{4}{*}{J1842} & \makebox[3.5cm][l]{$\alpha=10^{-3},\,v_{\rm frag}=1\,{\rm m\,s^{-1}}$} & Ring & 85 & 181 & --- \\
 & \makebox[3.5cm][l]{$\alpha=10^{-4},\,v_{\rm frag}=1\,{\rm m\,s^{-1}}$} & Ring &  & 178 &  \\
 & \makebox[3.5cm][l]{$\alpha=10^{-3},\,v_{\rm frag}=10\,{\rm m\,s^{-1}}$} & Ring &  & 176 &  \\
 & \makebox[3.5cm][l]{$\alpha=10^{-4},\,v_{\rm frag}=10\,{\rm m\,s^{-1}}$} & Ring &  & 178 &  \\
\midrule
\multirow{4}{*}{J1852} & \makebox[3.5cm][l]{$\alpha=10^{-3},\,v_{\rm frag}=1\,{\rm m\,s^{-1}}$} & Ring & 70 & 180 & --- \\
 & \makebox[3.5cm][l]{$\alpha=10^{-4},\,v_{\rm frag}=1\,{\rm m\,s^{-1}}$} & Halo &  & 137 &  \\
 & \makebox[3.5cm][l]{$\alpha=10^{-3},\,v_{\rm frag}=10\,{\rm m\,s^{-1}}$} & Halo &  & 104 &  \\
 & \makebox[3.5cm][l]{$\alpha=10^{-4},\,v_{\rm frag}=10\,{\rm m\,s^{-1}}$} & Ring &  & 145 &  \\
\midrule
\multirow{4}{*}{LkCa\,15} & \makebox[3.5cm][l]{$\alpha=10^{-3},\,v_{\rm frag}=1\,{\rm m\,s^{-1}}$} & Ring & 156 & 453 & --- \\
 & \makebox[3.5cm][l]{$\alpha=10^{-4},\,v_{\rm frag}=1\,{\rm m\,s^{-1}}$} & Ring &  & 547 &  \\
 & \makebox[3.5cm][l]{$\alpha=10^{-3},\,v_{\rm frag}=10\,{\rm m\,s^{-1}}$} & Ring &  & 461 &  \\
 & \makebox[3.5cm][l]{$\alpha=10^{-4},\,v_{\rm frag}=10\,{\rm m\,s^{-1}}$} & Ring &  & 506 &  \\
\midrule
\multirow{4}{*}{SY\,Cha} & \makebox[3.5cm][l]{$\alpha=10^{-3},\,v_{\rm frag}=1\,{\rm m\,s^{-1}}$} & Halo & 198 & 259 & --- \\
 & \makebox[3.5cm][l]{$\alpha=10^{-4},\,v_{\rm frag}=1\,{\rm m\,s^{-1}}$} & Ring &  & 355 &  \\
 & \makebox[3.5cm][l]{$\alpha=10^{-3},\,v_{\rm frag}=10\,{\rm m\,s^{-1}}$} & Ring &  & 333 &  \\
 & \makebox[3.5cm][l]{$\alpha=10^{-4},\,v_{\rm frag}=10\,{\rm m\,s^{-1}}$} & Ring &  & 334 &  \\
\midrule
\multirow{4}{*}{V4046\,Sgr} & \makebox[3.5cm][l]{$\alpha=10^{-3},\,v_{\rm frag}=1\,{\rm m\,s^{-1}}$} & Halo & 61 & 247 & --- \\
 & \makebox[3.5cm][l]{$\alpha=10^{-4},\,v_{\rm frag}=1\,{\rm m\,s^{-1}}$} & Ring &  & 253 &  \\
 & \makebox[3.5cm][l]{$\alpha=10^{-3},\,v_{\rm frag}=10\,{\rm m\,s^{-1}}$} & Ring &  & 148 &  \\
 & \makebox[3.5cm][l]{$\alpha=10^{-4},\,v_{\rm frag}=10\,{\rm m\,s^{-1}}$} & Ring &  & 265 &  \\
\bottomrule
\end{tabular}
\label{tab:13co_summary}
\end{table*}

\begin{table*}[t]
\centering
\caption{Summary of the special-case dust-evolution models (enhanced outer-disc turbulence and late-bump introduction) for LkCa\,15, and J1615, and including the enhanced $\alpha$ of AA\,Tau. Outer mm emission classifies the morphology of emission detected above 3$\sigma$ beyond $R_{90,\rm obs}$ as defined in Section~\ref{sec:methods_rt}. \emph{No*} (No outer mm) indicates no emission extending meaningfully beyond $R_{90,\rm obs}$.  }
\begin{tabular}{llcccc}
\toprule
Disc & Config & Outer mm emission & $R_{90,\rm obs}$ (au) & $R_{90,\rm syn}$ (au) & Candidate $R_{\rm bump,gas}$ region (au) \\
\midrule
\multirow{4}{*}{AA\,Tau} & \makebox[4cm][l]{Enhanced-$\alpha$, $\alpha=10^{-3},\,v_{\rm frag}=1\,{\rm m\,s^{-1}}$} & Halo & 139 & 159 & 268--460 \\
 & \makebox[4cm][l]{Enhanced-$\alpha$, $\alpha=10^{-4},\,v_{\rm frag}=1\,{\rm m\,s^{-1}}$} & Halo &  & 166 &  \\
 & \makebox[4cm][l]{Enhanced-$\alpha$, $\alpha=10^{-3},\,v_{\rm frag}=10\,{\rm m\,s^{-1}}$} & Halo &  & 167 &  \\
 & \makebox[4cm][l]{Enhanced-$\alpha$, $\alpha=10^{-4},\,v_{\rm frag}=10\,{\rm m\,s^{-1}}$} & Halo &  & 170 &  \\
\midrule

\multirow{4}{*}{LkCa\,15} & \makebox[4cm][l]{Enhanced-$\alpha$, $\alpha=10^{-3},\,v_{\rm frag}=1\,{\rm m\,s^{-1}}$} & No* & 156 & 69 & 159--254; 336--395 \\
 & \makebox[4cm][l]{Enhanced-$\alpha$, $\alpha=10^{-4},\,v_{\rm frag}=1\,{\rm m\,s^{-1}}$} & No* &  & 47 &  \\
 & \makebox[4cm][l]{Enhanced-$\alpha$, $\alpha=10^{-3},\,v_{\rm frag}=10\,{\rm m\,s^{-1}}$} & No* &  & 44 &  \\
 & \makebox[4cm][l]{Enhanced-$\alpha$, $\alpha=10^{-4},\,v_{\rm frag}=10\,{\rm m\,s^{-1}}$} & No* &  & 41 &  \\
\midrule
\multirow{4}{*}{LkCa\,15} & \makebox[4cm][l]{Late bump, $\alpha=10^{-3},\,v_{\rm frag}=1\,{\rm m\,s^{-1}}$} & Halo & 156 & 160 & 159--254; 336--395 \\
 & \makebox[4cm][l]{Late bump, $\alpha=10^{-4},\,v_{\rm frag}=1\,{\rm m\,s^{-1}}$} & No* &  & 150 &  \\
 & \makebox[4cm][l]{Late bump, $\alpha=10^{-3},\,v_{\rm frag}=10\,{\rm m\,s^{-1}}$} & Ring &  & 188 &  \\
 & \makebox[4cm][l]{Late bump, $\alpha=10^{-4},\,v_{\rm frag}=10\,{\rm m\,s^{-1}}$} & No* &  & 54 &  \\
\midrule
\multirow{4}{*}{J1615} & \makebox[4cm][l]{Enhanced-$\alpha$, $\alpha=10^{-3},\,v_{\rm frag}=1\,{\rm m\,s^{-1}}$} & Halo & 170 & 279 & 170--252 \\
 & \makebox[4cm][l]{Enhanced-$\alpha$, $\alpha=10^{-4},\,v_{\rm frag}=1\,{\rm m\,s^{-1}}$} & Halo &  & 274 &  \\
 & \makebox[4cm][l]{Enhanced-$\alpha$, $\alpha=10^{-3},\,v_{\rm frag}=10\,{\rm m\,s^{-1}}$} & Halo &  & 291 &  \\
 & \makebox[4cm][l]{Enhanced-$\alpha$, $\alpha=10^{-4},\,v_{\rm frag}=10\,{\rm m\,s^{-1}}$} & Halo &  & 311 &  \\
\midrule
\multirow{4}{*}{J1615} & \makebox[4cm][l]{Late bump, $\alpha=10^{-3},\,v_{\rm frag}=1\,{\rm m\,s^{-1}}$} & Halo & 170 & 420 & 170--252 \\
 & \makebox[4cm][l]{Late bump, $\alpha=10^{-4},\,v_{\rm frag}=1\,{\rm m\,s^{-1}}$} & Halo &  & 363 &  \\
 & \makebox[4cm][l]{Late bump, $\alpha=10^{-3},\,v_{\rm frag}=10\,{\rm m\,s^{-1}}$} & Halo &  & 300 &  \\
 & \makebox[4cm][l]{Late bump, $\alpha=10^{-4},\,v_{\rm frag}=10\,{\rm m\,s^{-1}}$} & Halo &  & 314 &  \\
\bottomrule
\end{tabular}
\label{tab:special_models}
\end{table*}

\section{Additional Targets and Supporting Figures}
\label{sec:appendix}
We provide additional results for the exoALMA systems not shown in the main text. For each system we include the dust evolution models and the resulting Band~7 RADMC3D images for the four $(\alphavisc,\vfrag)$ combinations. 

\begin{figure*}
    \centering
    \includegraphics[width=\textwidth]{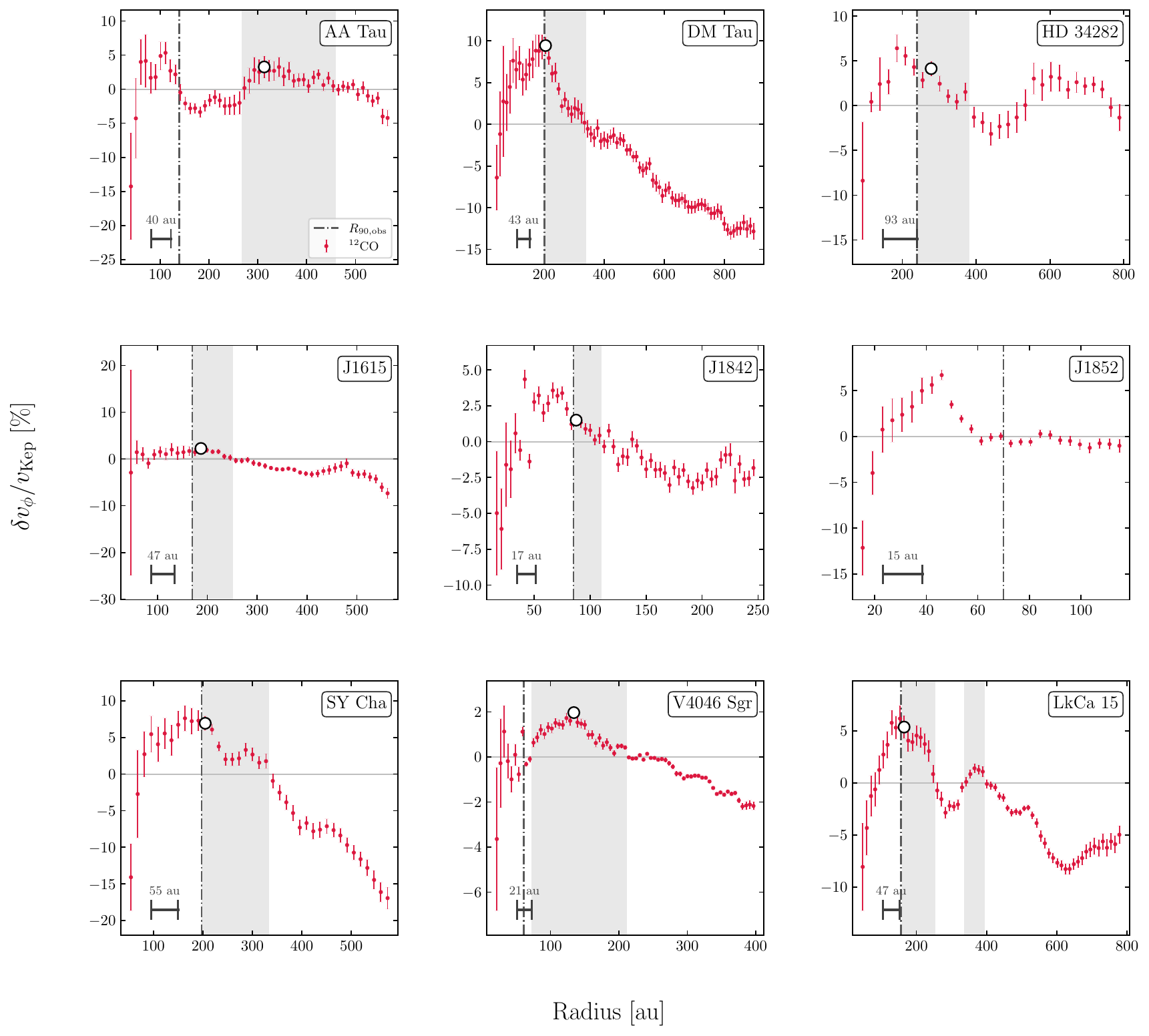}
    \caption{Radial profiles of the fractional azimuthal velocity
    deviation $\delta v_\phi / v_{\rm Kep}$ [\%] for the nine
    exoALMA discs modelled in this work, derived from the
    \twelveco\ rotation curves of \citet{Stadler_2025}.
    Error bars show the $1\sigma$ uncertainties on the measured
    velocity residuals. The dash-dotted vertical line marks
    $R_{90,\rm obs}$, the radius enclosing 90~per~cent of the
    Band~7 continuum flux \citep{Curone_2025}. The red scale bar
    in the lower-left corner of each panel indicates the effective
    beam size $\theta_{\rm beam}$ in au. The open circle marks
    the peak velocity deviation within each candidate region.
    Grey shaded regions indicate the candidate outer
    pressure-substructure zones identified as described in
    Section~\ref{sec:methods_candidate_regions}; candidate regions are
    identified in 8 of the 9 systems from the \twelveco\ data. Note that the super-/sub-Keplerian classification depends on the assumed stellar mass, regions near the threshold should be regarded as marginal.}
    \label{fig:velocity_12co}
\end{figure*}

\begin{figure*}
    \centering
    \includegraphics[width=\textwidth]{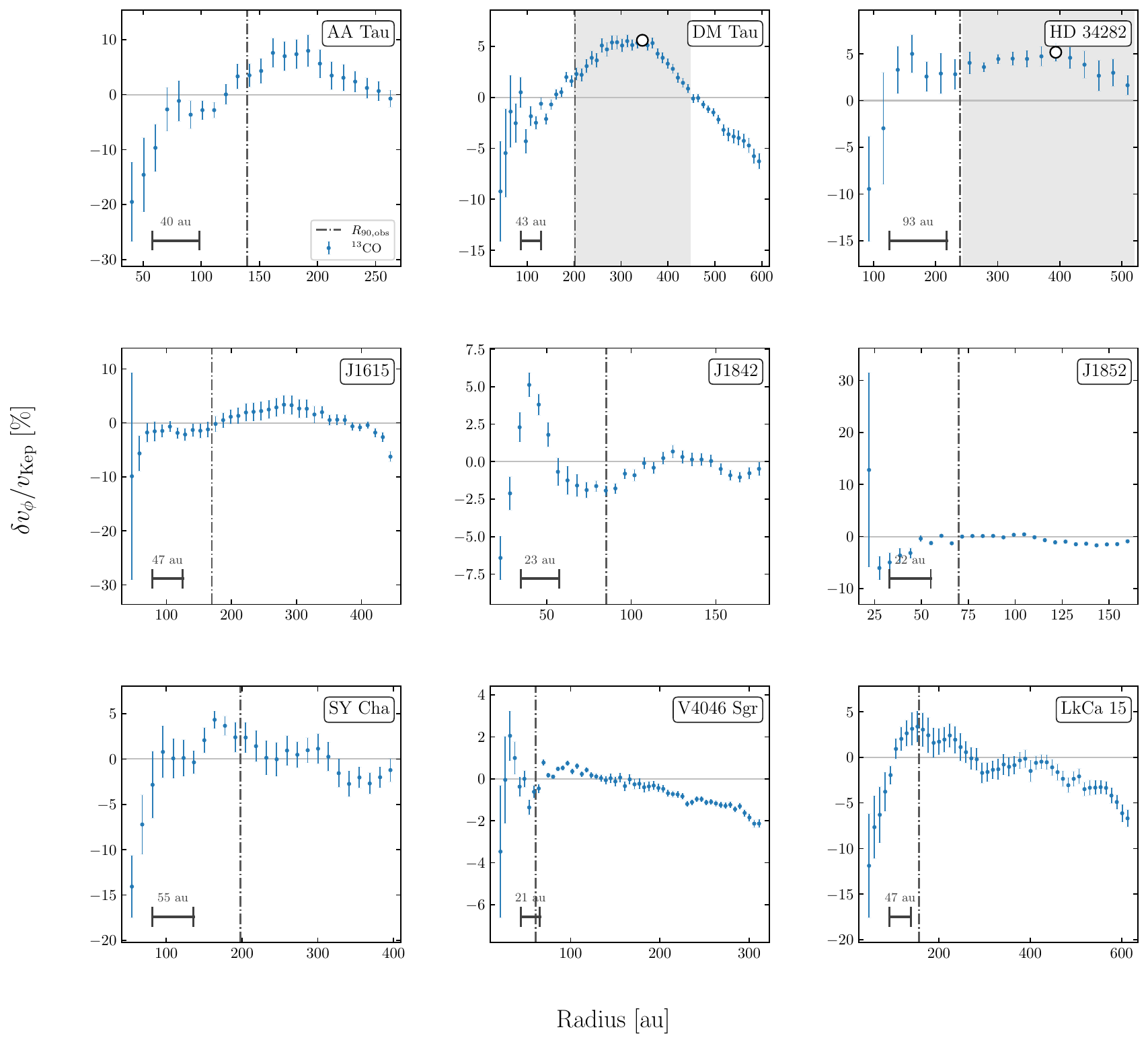}
    \caption{As Fig.~\ref{fig:velocity_12co}, but derived from the
    \thirteenco\ rotation curves. Candidate outer pressure-substructure
    regions satisfying the significance threshold ($\Delta v_\phi >
    1$~per~cent, $S > 3$) are identified in only 2 of the 9 systems
    (DM\,Tau and HD\,34282). In some of the remaining systems, velocity
    deviations are present but fall below the adopted significance
    threshold, and are therefore not selected as candidate regions
    by our criterion.Note that the super-/sub-Keplerian classification depends on the assumed stellar mass, regions near the threshold should be regarded as marginal.}
    \label{fig:velocity_13co}
\end{figure*}

\begin{figure*}
  \centering
  \includegraphics[width=0.96\textwidth]{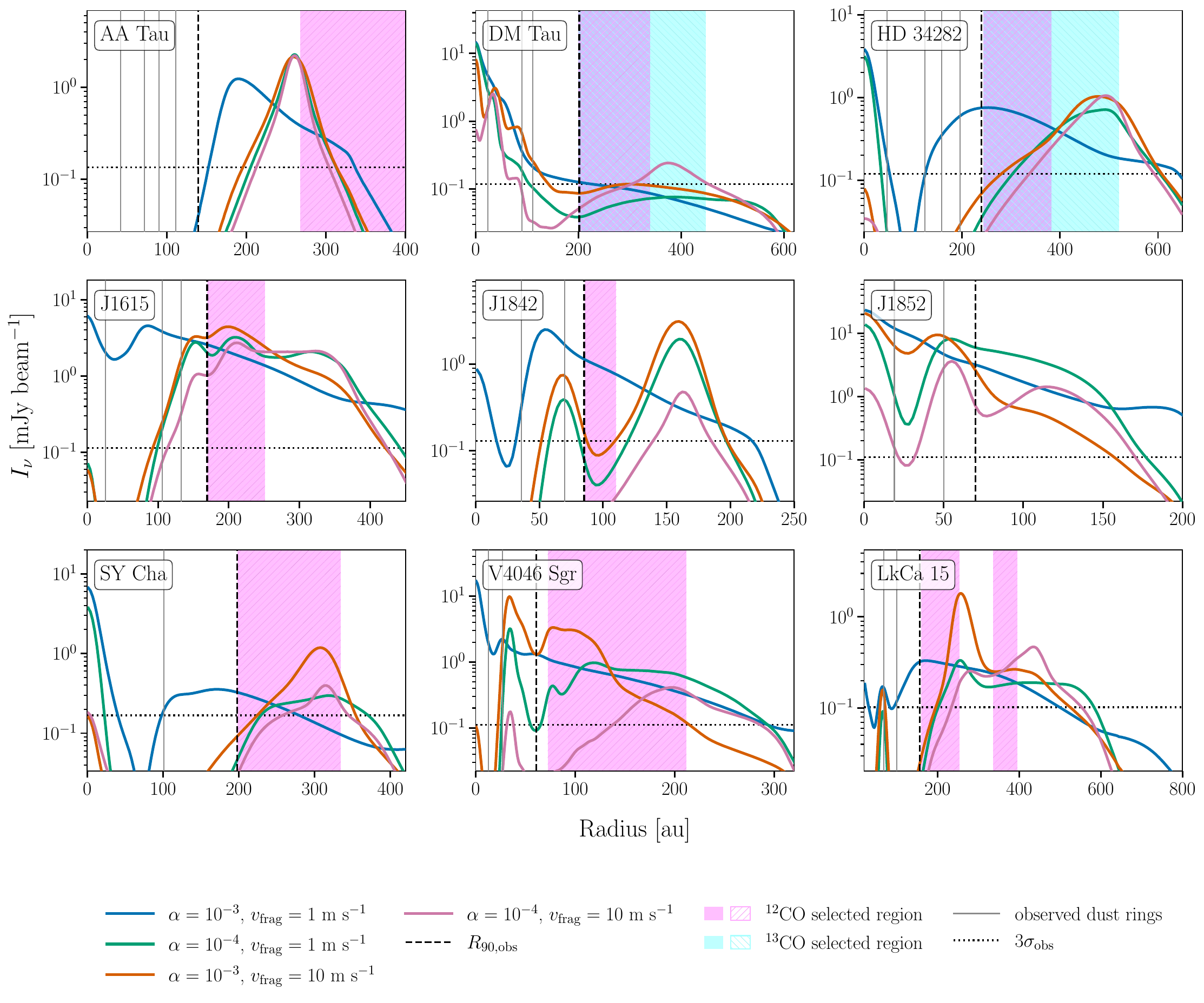}
  \caption{As Fig.~\ref{fig:survey_profiles}, but for the \thirteenco-based reconstructions.}

  \label{fig:survey_profiles_13_co}
\end{figure*}

\begin{figure*}
    \centering
    \includegraphics[width=\textwidth]{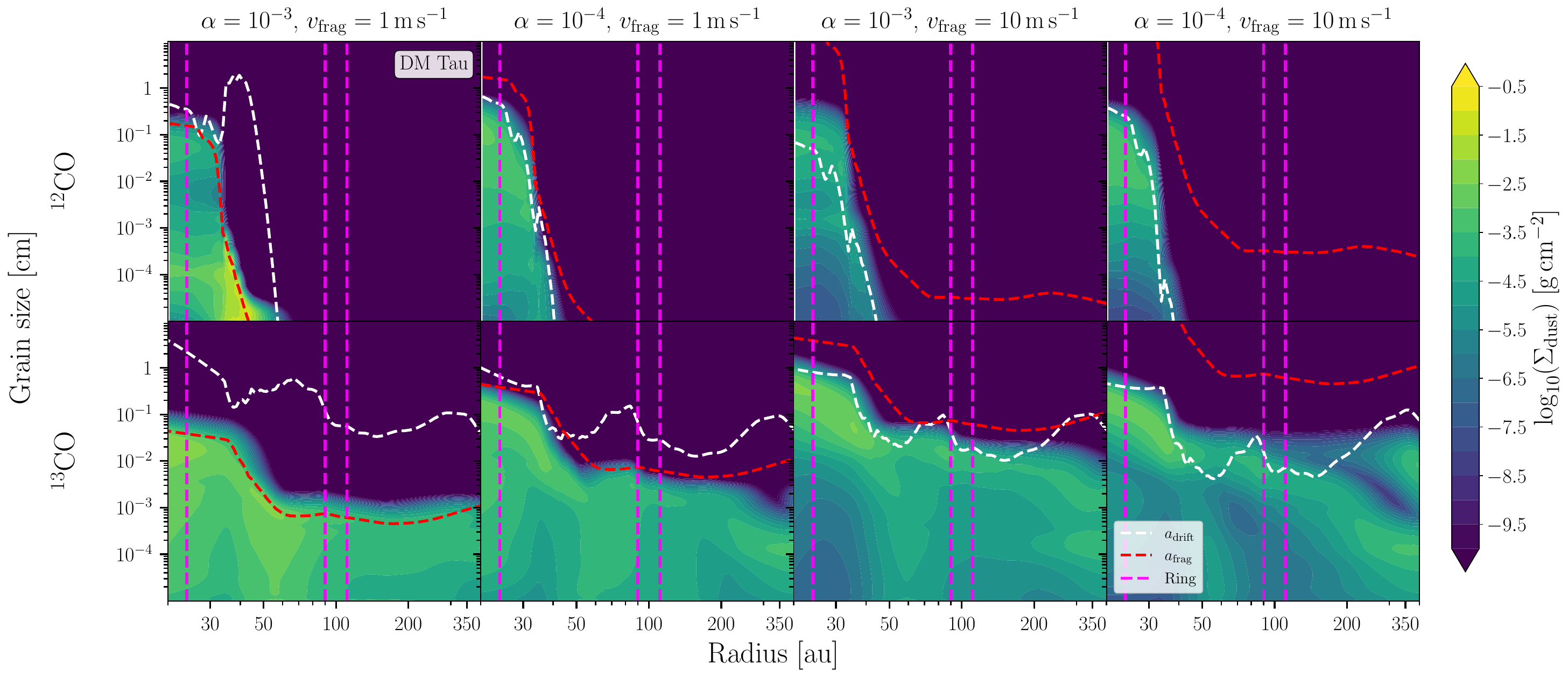}
    \caption{Dust density distribution after 1~Myr for DM\,Tau, using the exoALMA gas surface density reconstructions derived from the \twelveco~(top) and \thirteenco~(bottom) observations. White and red dashed curves indicate the drift and fragmentation barriers, respectively. Vertical magenta dashed lines mark dust rings identified in the Band~7 continuum observations \citep{Curone_2025}. The four columns represent different dust evolution model configurations spanning $\alpha \in \{10^{-4},\,10^{-3}\}$ and $v_{\mathrm{frag}} \in \{1,\,10\}\,\mathrm{m\,s^{-1}}$.
    }
    \label{fig:dustpy_dmtau}
\end{figure*}

\begin{figure*}
    \centering
    \includegraphics[width=\textwidth]{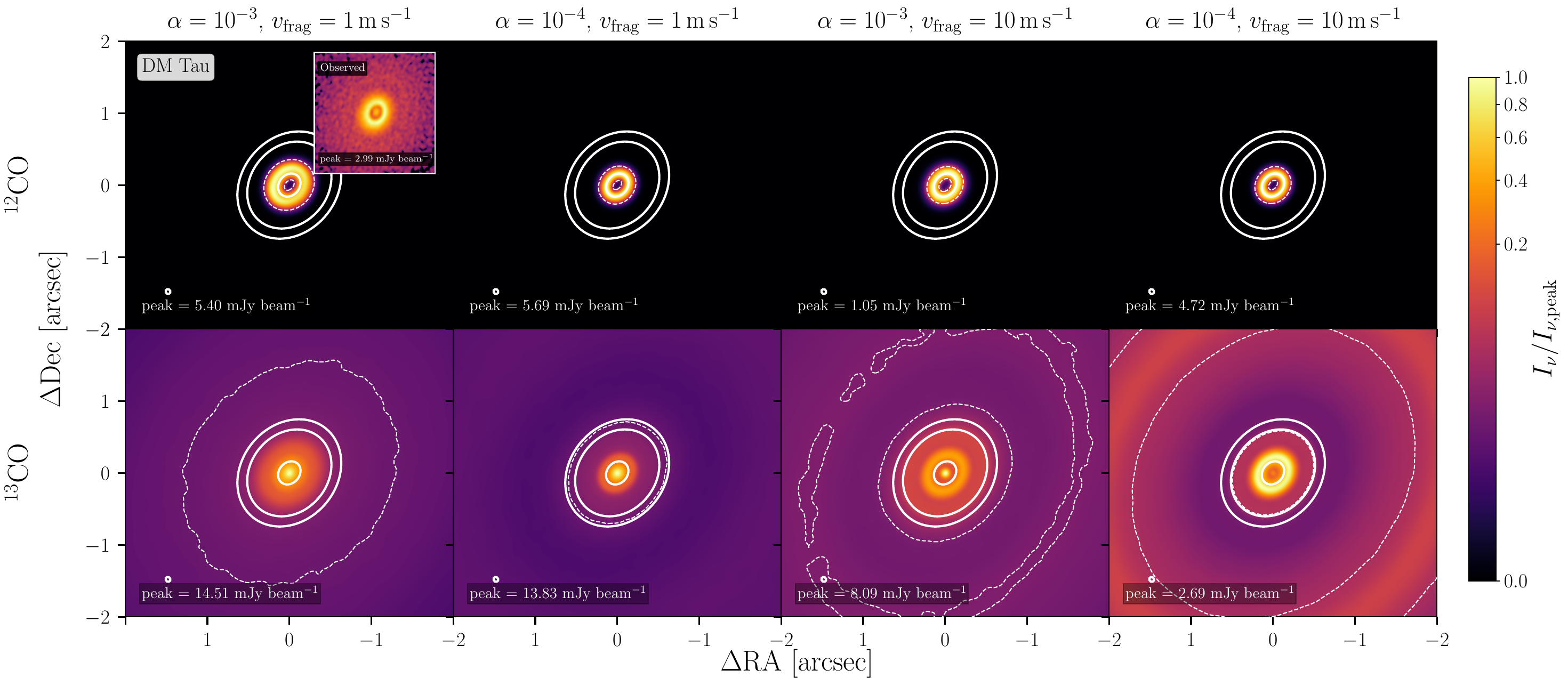}
    \caption{
    \texttt{RADMC} synthetic observations of the dustpy $^{12}$CO and $^{13}$CO models. The fluxes are normalised to the peak. The inset shows the observation in B7 \citep{Curone_2025}. The white ellipses indicate dust rings identified in the Band~7 continuum observations \citep{Curone_2025}. The white dashed lines indicate the 3$\sigma$ threshold based on the sensitivity of the observations. The four columns represent different dust evolution model configurations spanning $\alphavisc \in \{10^{-4},10^{-3}\}$ and $\vfrag \in \{1,10\}\,\mathrm{m\,s^{-1}}$. 
    }
    \label{fig:radmc_dmtau}
\end{figure*}

\begin{figure*}
    \centering
    \includegraphics[width=\textwidth]{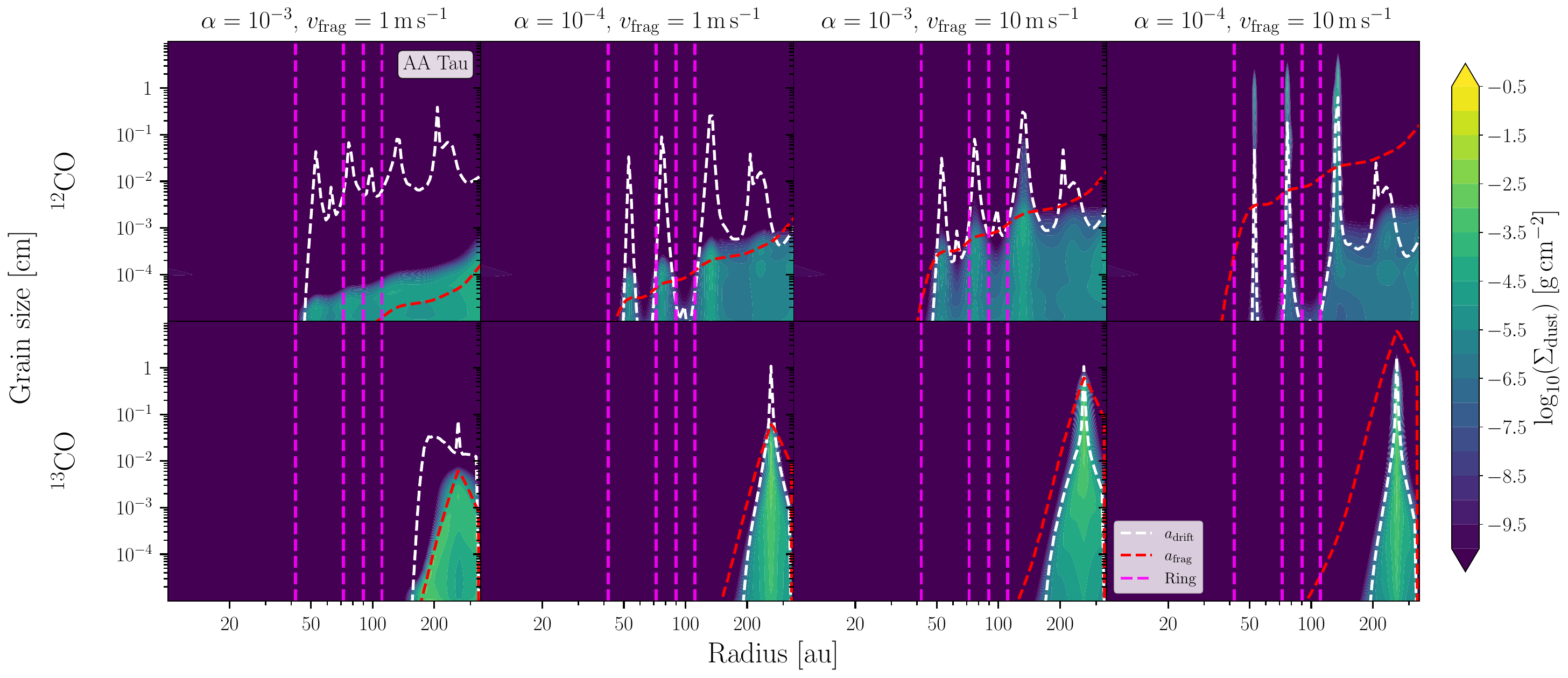}
    \caption{Dust density distribution after 1~Myr for AA\,Tau, using the exoALMA gas surface density reconstructions derived from the \twelveco~(top) and \thirteenco~(bottom) observations. White and red dashed curves indicate the drift and fragmentation barriers, respectively. Vertical magenta dashed lines mark dust rings identified in the Band~7 continuum observations \citep{Curone_2025}. The four columns represent different dust evolution model configurations spanning $\alpha \in \{10^{-4},\,10^{-3}\}$ and $v_{\mathrm{frag}} \in \{1,\,10\}\,\mathrm{m\,s^{-1}}$.
    }
    \label{fig:dustpy_aatau}
\end{figure*}

\begin{figure*}
    \centering
    \includegraphics[width=\textwidth]{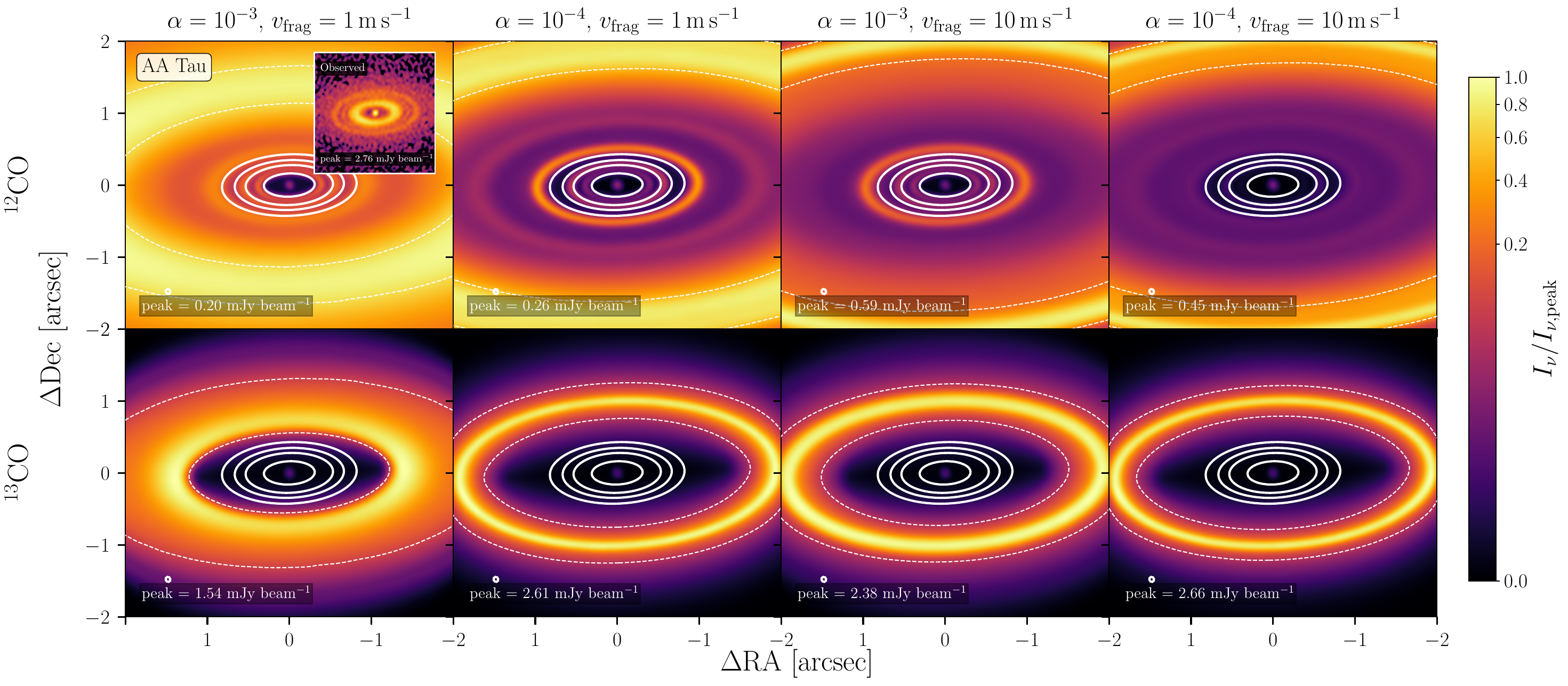}
    \caption{
    \texttt{RADMC} synthetic observations of the dustpy $^{12}$CO and $^{13}$CO models. The fluxes are normalised to the peak. The inset shows the observation in B7 \citep{Curone_2025}. The white ellipses indicate dust rings identified in the Band~7 continuum observations \citep{Curone_2025}. The white dashed lines indicate the 3$\sigma$ threshold based on the sensitivity of the observations. The four columns represent different dust evolution model configurations spanning $\alphavisc \in \{10^{-4},10^{-3}\}$ and $\vfrag \in \{1,10\}\,\mathrm{m\,s^{-1}}$. 
    }
    \label{fig:radmc_aatau}
\end{figure*}

\begin{figure*}
    \centering
    \includegraphics[width=\textwidth]{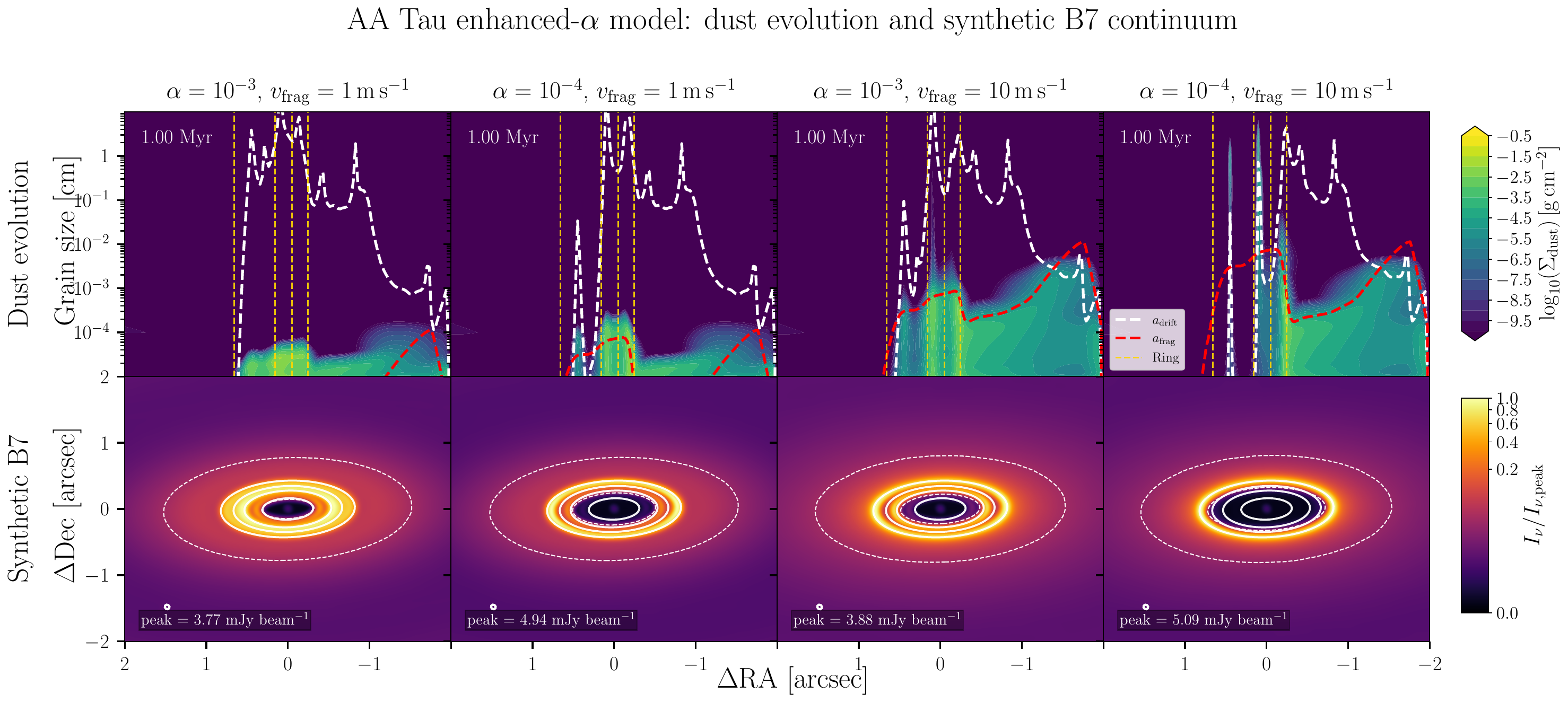}
    \caption{\dustpy\ dust evolution results and synthetic Band~7 
continuum emission for the enhanced-turbulence AA\,Tau 
models using the \twelveco\, data. A step-like turbulence profile with 
$\alphavisc_{\rm out}=10^{-2}$ beyond $R_{\rm dz}$ is 
applied. White dashed contours indicate the $3\sigma$ 
observational threshold.
    }
    \label{fig:radmc_aatau_enhanced}
\end{figure*}

\begin{figure*}
    \centering
    \includegraphics[width=\textwidth]{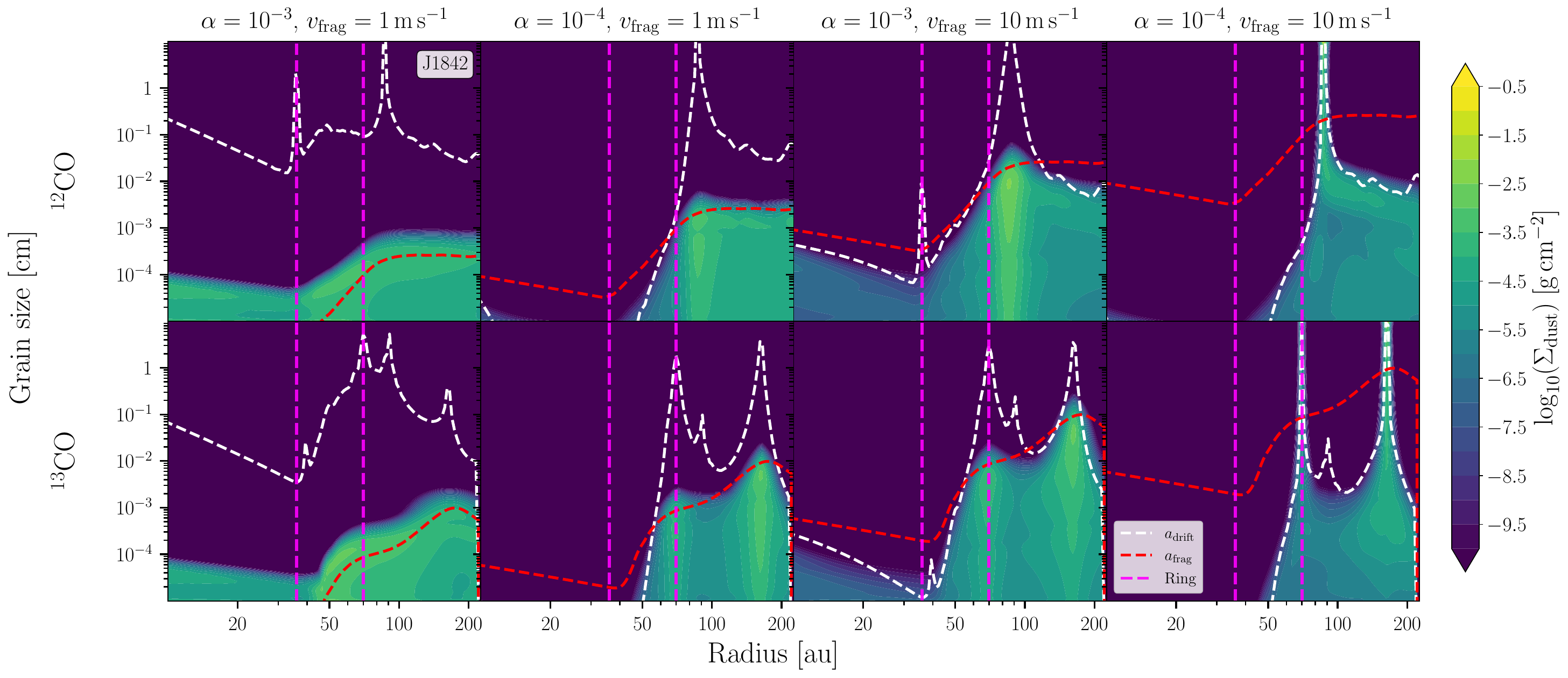}
    \caption{Dust density distribution after 1~Myr for J\,1842, using the exoALMA gas surface density reconstructions derived from the \twelveco~(top) and \thirteenco~(bottom) observations. White and red dashed curves indicate the drift and fragmentation barriers, respectively. Vertical magenta dashed lines mark dust rings identified in the Band~7 continuum observations \citep{Curone_2025}. The four columns represent different dust evolution model configurations spanning $\alpha \in \{10^{-4},\,10^{-3}\}$ and $v_{\mathrm{frag}} \in \{1,\,10\}\,\mathrm{m\,s^{-1}}$.
    }
    \label{fig:dustpy_j1842}
\end{figure*}

\begin{figure*}
    \centering
    \includegraphics[width=\textwidth]{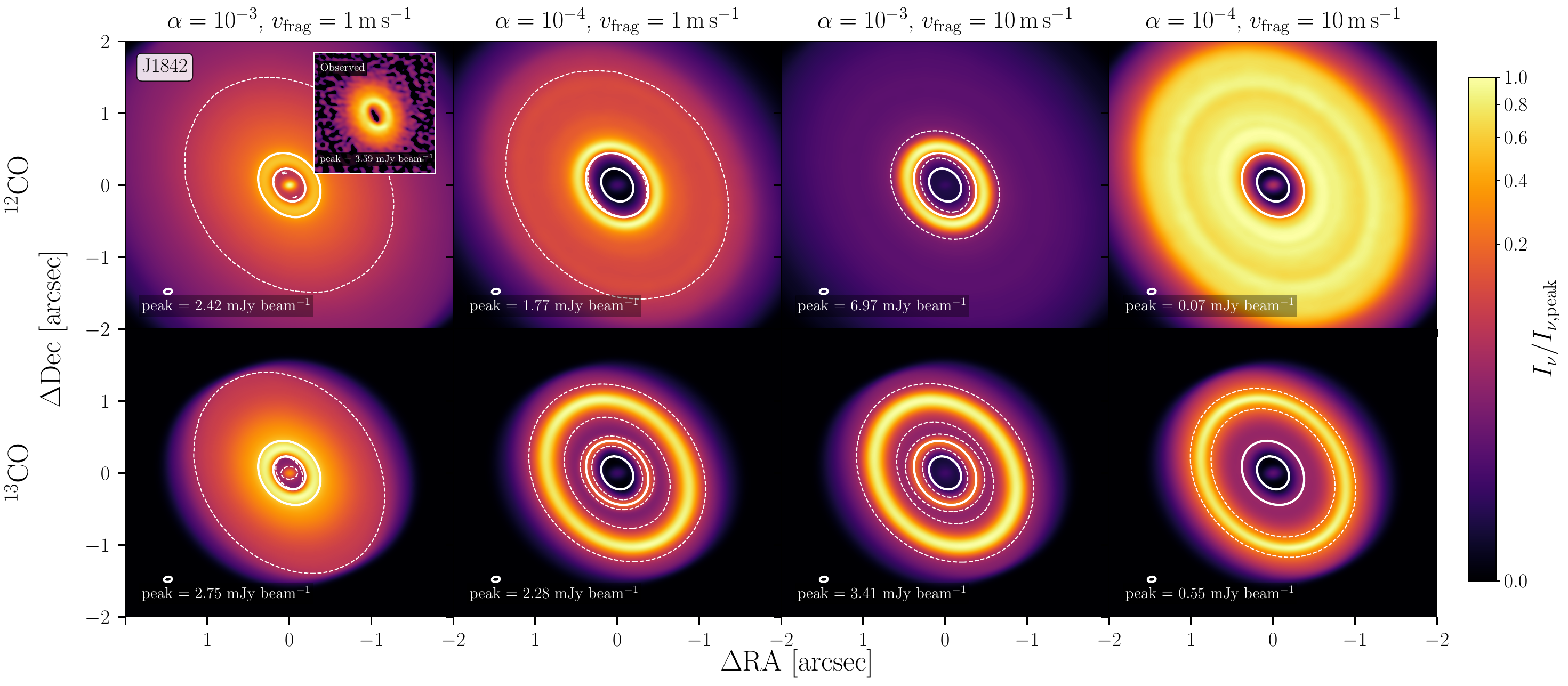}
    \caption{\texttt{RADMC} synthetic observations of the dustpy $^{12}$CO and $^{13}$CO models. The fluxes are normalised to the peak. The inset shows the observation in B7 \citep{Curone_2025}. The white ellipses indicate dust rings identified in the Band~7 continuum observations \citep{Curone_2025}. The white dashed lines indicate the 3$\sigma$ threshold based on the sensitivity of the observations. The four columns represent different dust evolution model configurations spanning $\alphavisc \in \{10^{-4},10^{-3}\}$ and $\vfrag \in \{1,10\}\,\mathrm{m\,s^{-1}}$. }
    \label{fig:radmc_j1842}
\end{figure*}

\begin{figure*}
    \centering
    \includegraphics[width=\textwidth]{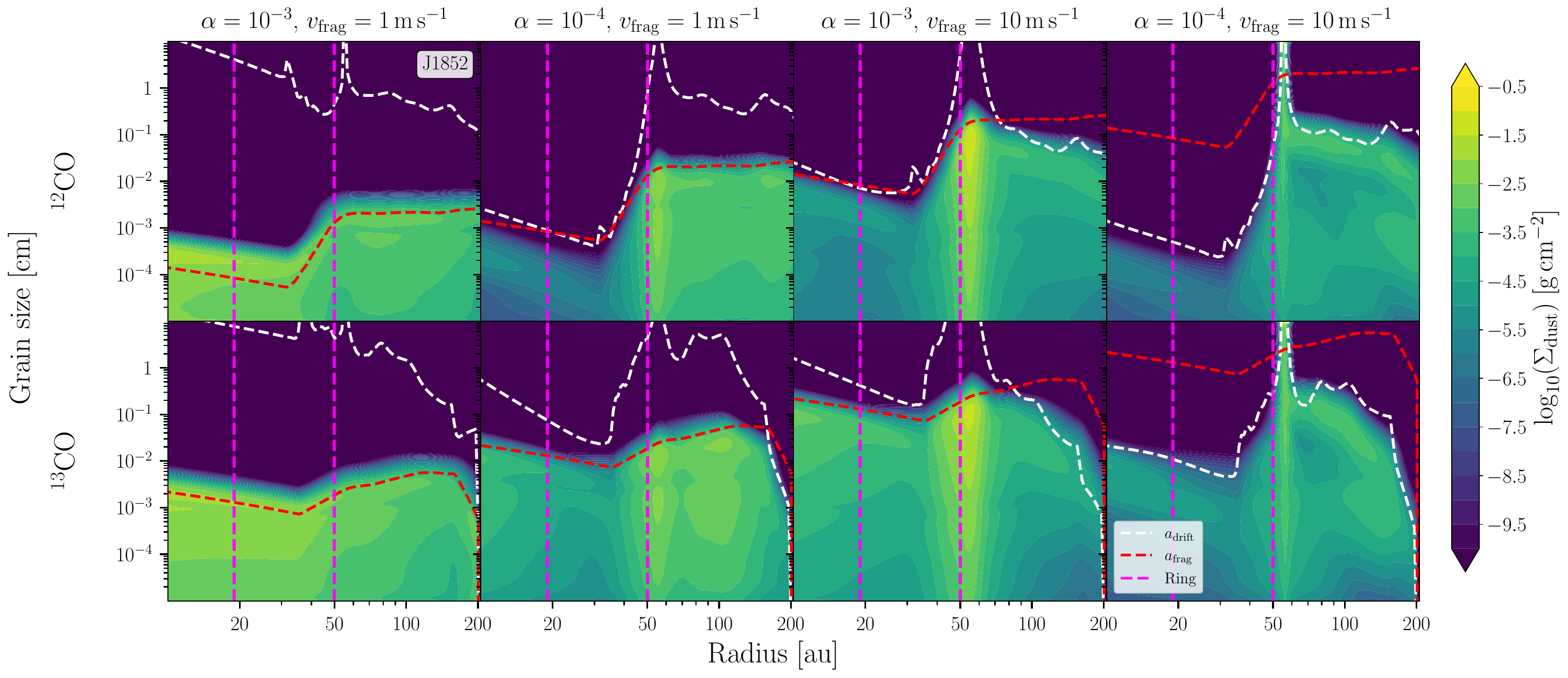}
    \caption{Dust density distribution after 1~Myr for J\,1852, using the exoALMA gas surface density reconstructions derived from the \twelveco~(top) and \thirteenco~(bottom) observations. White and red dashed curves indicate the drift and fragmentation barriers, respectively. Vertical magenta dashed lines mark dust rings identified in the Band~7 continuum observations \citep{Curone_2025}. The four columns represent different dust evolution model configurations spanning $\alpha \in \{10^{-4},\,10^{-3}\}$ and $v_{\mathrm{frag}} \in \{1,\,10\}\,\mathrm{m\,s^{-1}}$.
    }
    \label{fig:dustpy_j1852}
\end{figure*}

\begin{figure*}
    \centering
    \includegraphics[width=\textwidth]{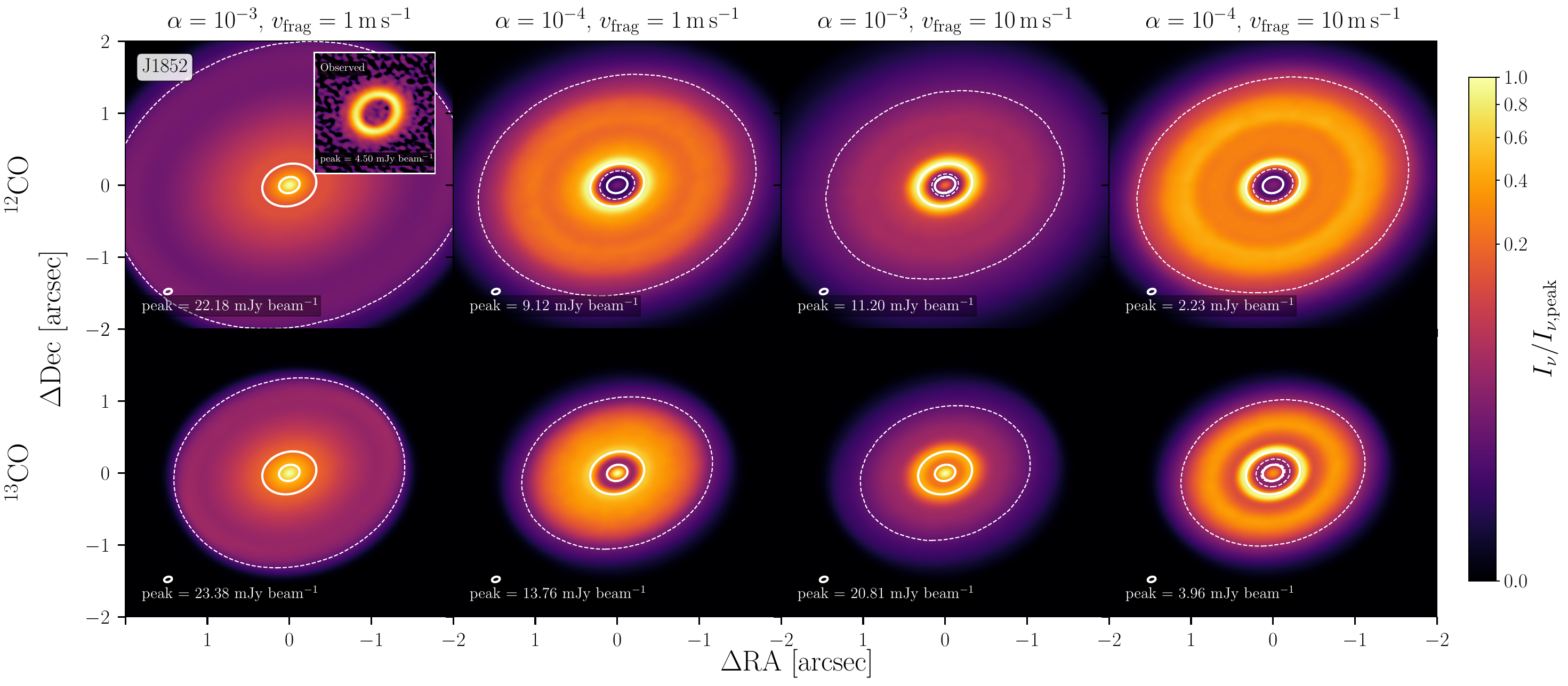}
    \caption{\texttt{RADMC} synthetic observations of the dustpy $^{12}$CO and $^{13}$CO models. The fluxes are normalised to the peak. The inset shows the observation in B7 \citep{Curone_2025}. The white ellipses indicate dust rings identified in the Band~7 continuum observations \citep{Curone_2025}. The white dashed lines indicate the 3$\sigma$ threshold based on the sensitivity of the observations. The four columns represent different dust evolution model configurations spanning $\alphavisc \in \{10^{-4},10^{-3}\}$ and $\vfrag \in \{1,10\}\,\mathrm{m\,s^{-1}}$. }
    \label{fig:radmc_j1852}
\end{figure*}

\begin{figure*}
    \centering
    \includegraphics[width=\textwidth]{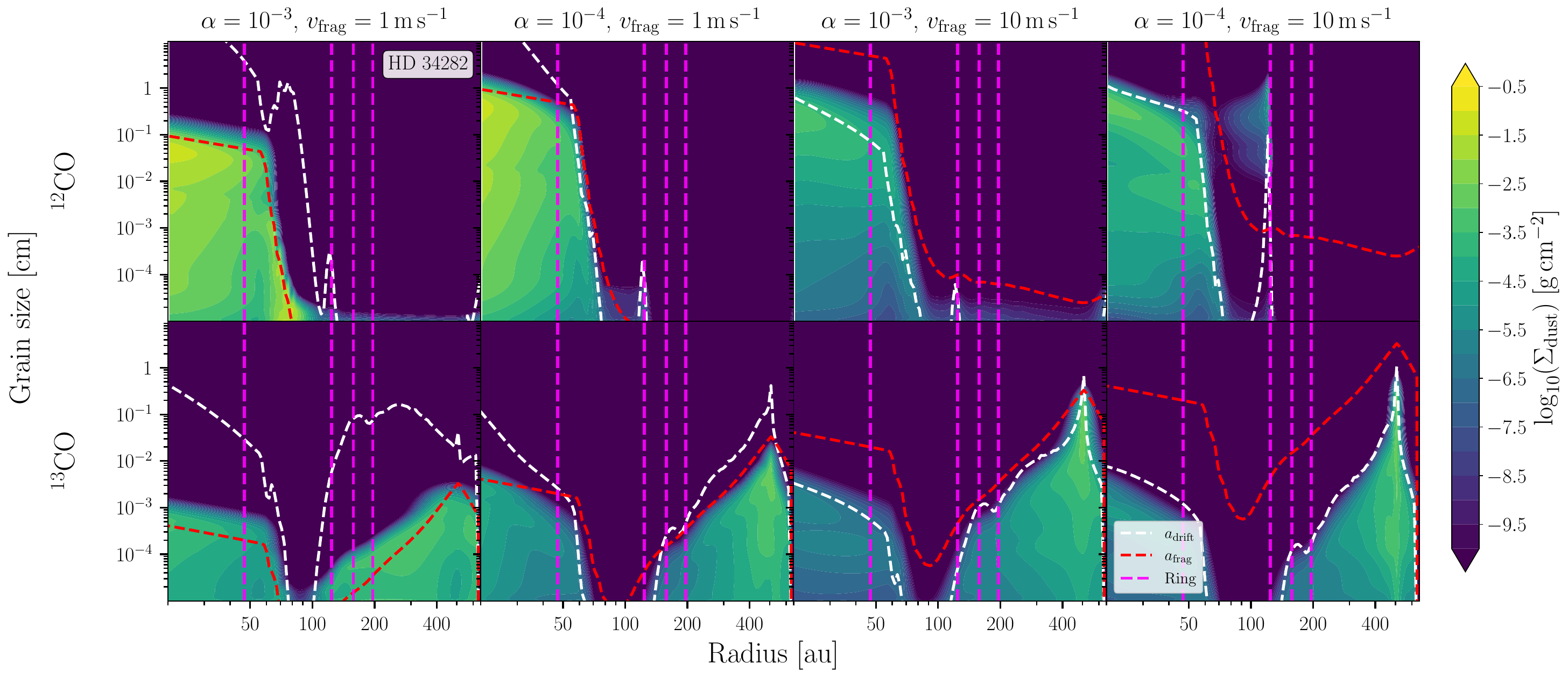}
    \caption{Dust density distribution after 1~Myr for HD\,34282, using the exoALMA gas surface density reconstructions derived from the \twelveco~(top) and \thirteenco~(bottom) observations. White and red dashed curves indicate the drift and fragmentation barriers, respectively. Vertical magenta dashed lines mark dust rings identified in the Band~7 continuum observations \citep{Curone_2025}. The four columns represent different dust evolution model configurations spanning $\alpha \in \{10^{-4},\,10^{-3}\}$ and $v_{\mathrm{frag}} \in \{1,\,10\}\,\mathrm{m\,s^{-1}}$.
    }
    \label{fig:dustpy_hd34282}
\end{figure*}

\begin{figure*}
    \centering
    \includegraphics[width=\textwidth]{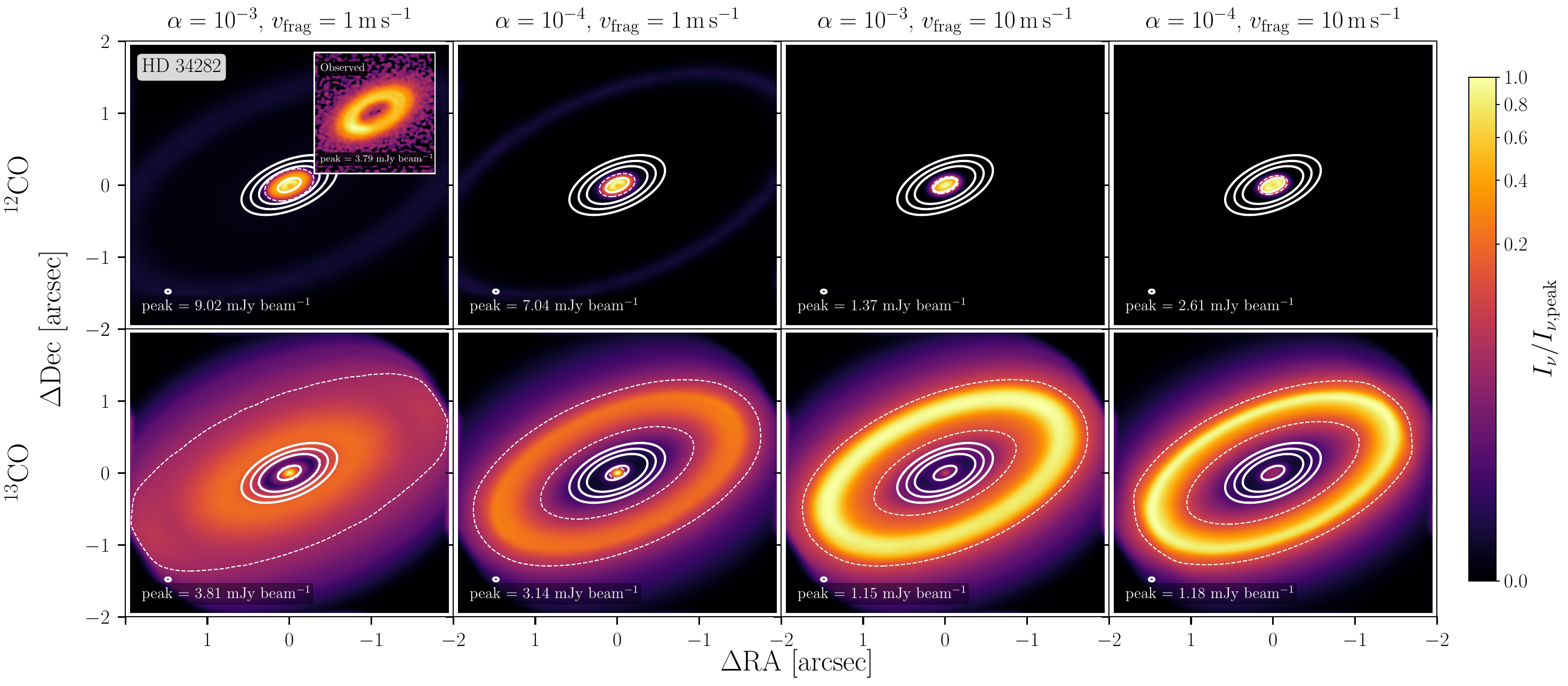}
    \caption{\texttt{RADMC} synthetic observations of the dustpy $^{12}$CO and $^{13}$CO models. The fluxes are normalised to the peak. The inset shows the observation in B7 \citep{Curone_2025}. The white ellipses indicate dust rings identified in the Band~7 continuum observations \citep{Curone_2025}. The white dashed lines indicate the 3$\sigma$ threshold based on the sensitivity of the observations. The four columns represent different dust evolution model configurations spanning $\alphavisc \in \{10^{-4},10^{-3}\}$ and $\vfrag \in \{1,10\}\,\mathrm{m\,s^{-1}}$. }
    \label{fig:radmc_hd34282}
\end{figure*}

\begin{figure*}
    \centering
    \includegraphics[width=\textwidth]{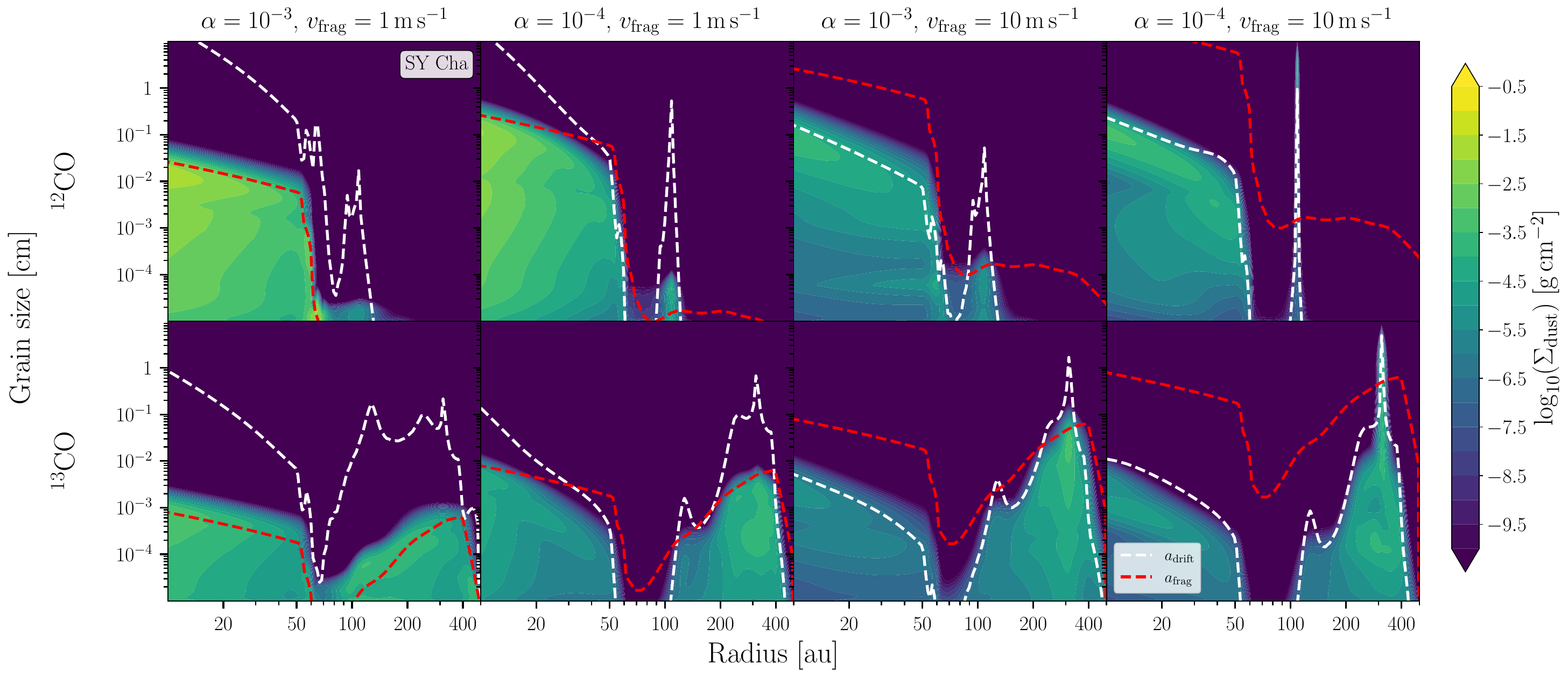}
    \caption{Dust density distribution after 1~Myr for SY\,Cha, using the exoALMA gas surface density reconstructions derived from the \twelveco~(top) and \thirteenco~(bottom) observations. White and red dashed curves indicate the drift and fragmentation barriers, respectively. Vertical magenta dashed lines mark dust rings identified in the Band~7 continuum observations \citep{Curone_2025}. The four columns represent different dust evolution model configurations spanning $\alpha \in \{10^{-4},\,10^{-3}\}$ and $v_{\mathrm{frag}} \in \{1,\,10\}\,\mathrm{m\,s^{-1}}$.
    }
    \label{fig:dustpy_sycha}
\end{figure*}

\begin{figure*}
    \centering
    \includegraphics[width=\textwidth]{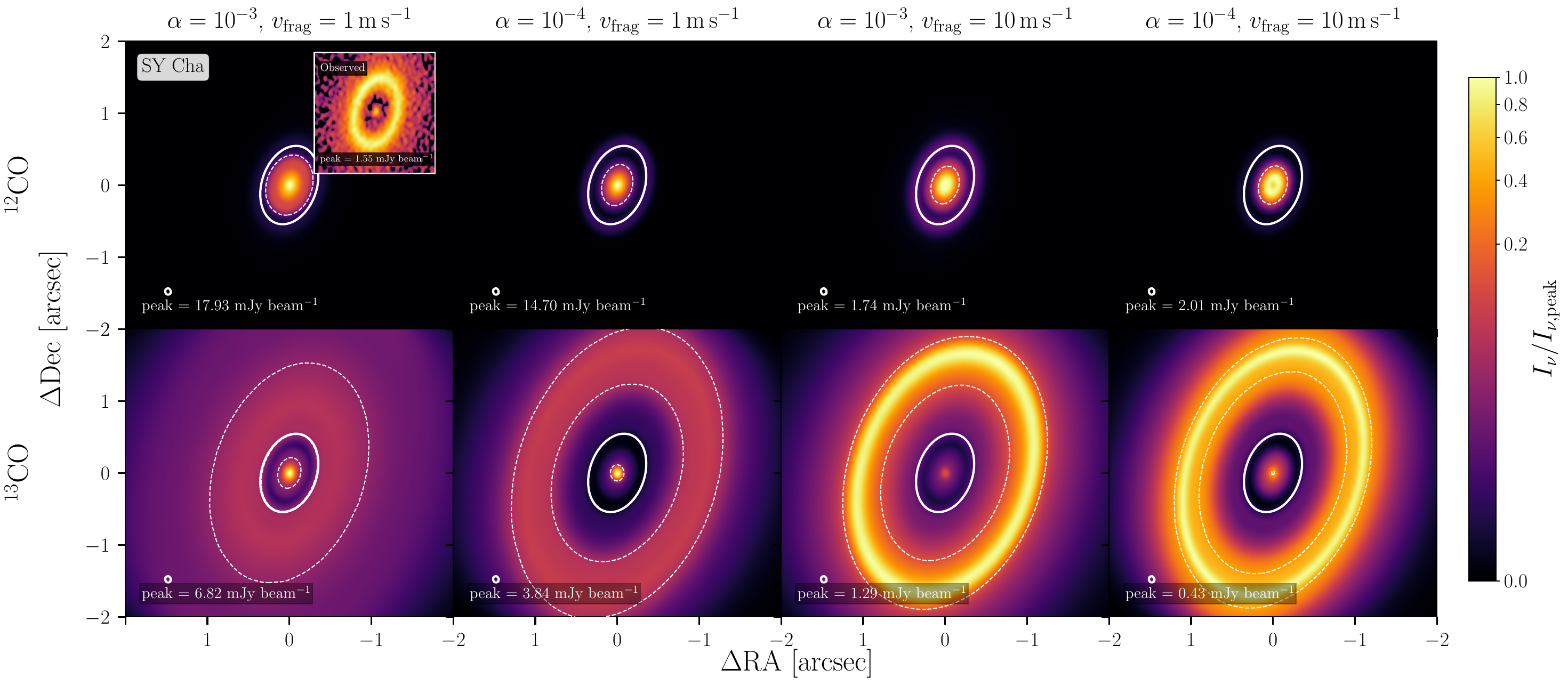}
    \caption{\texttt{RADMC} synthetic observations of the dustpy $^{12}$CO and $^{13}$CO models. The fluxes are normalised to the peak. The inset shows the observation in B7 \citep{Curone_2025}. The white ellipses indicate dust rings identified in the Band~7 continuum observations \citep{Curone_2025}. The white dashed lines indicate the 3$\sigma$ threshold based on the sensitivity of the observations. The four columns represent different dust evolution model configurations spanning $\alphavisc \in \{10^{-4},10^{-3}\}$ and $\vfrag \in \{1,10\}\,\mathrm{m\,s^{-1}}$. }
    \label{fig:radmc_sycha}
\end{figure*}

\begin{figure*}
    \centering
    \includegraphics[width=\textwidth]{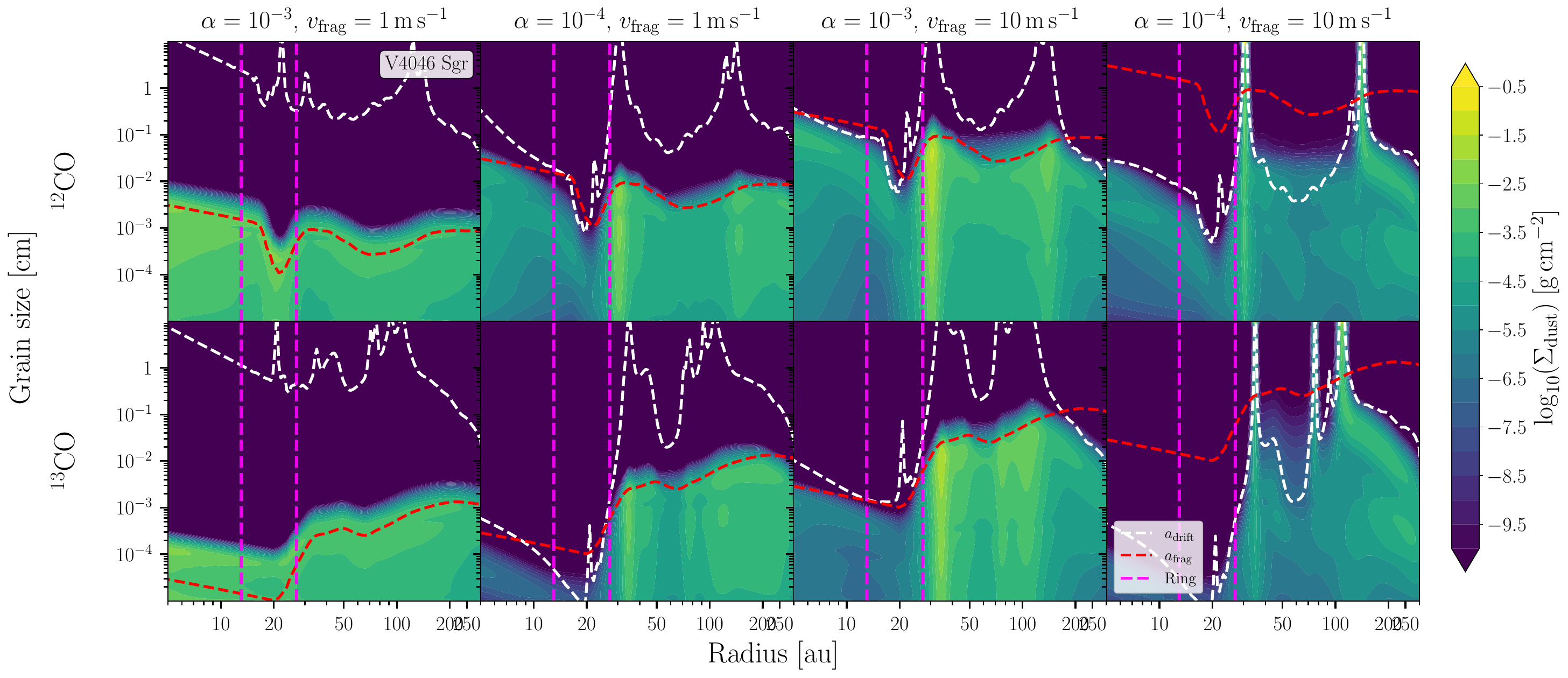}
    \caption{Dust density distribution after 1~Myr for V\,4046, using the exoALMA gas surface density reconstructions derived from the \twelveco~(top) and \thirteenco~(bottom) observations. White and red dashed curves indicate the drift and fragmentation barriers, respectively. Vertical magenta dashed lines mark dust rings identified in the Band~7 continuum observations \citep{Curone_2025}. The four columns represent different dust evolution model configurations spanning $\alpha \in \{10^{-4},\,10^{-3}\}$ and $v_{\mathrm{frag}} \in \{1,\,10\}\,\mathrm{m\,s^{-1}}$.
    }
    \label{fig:dustpy_v4046}
\end{figure*}

\begin{figure*}
    \centering
    \includegraphics[width=\textwidth]{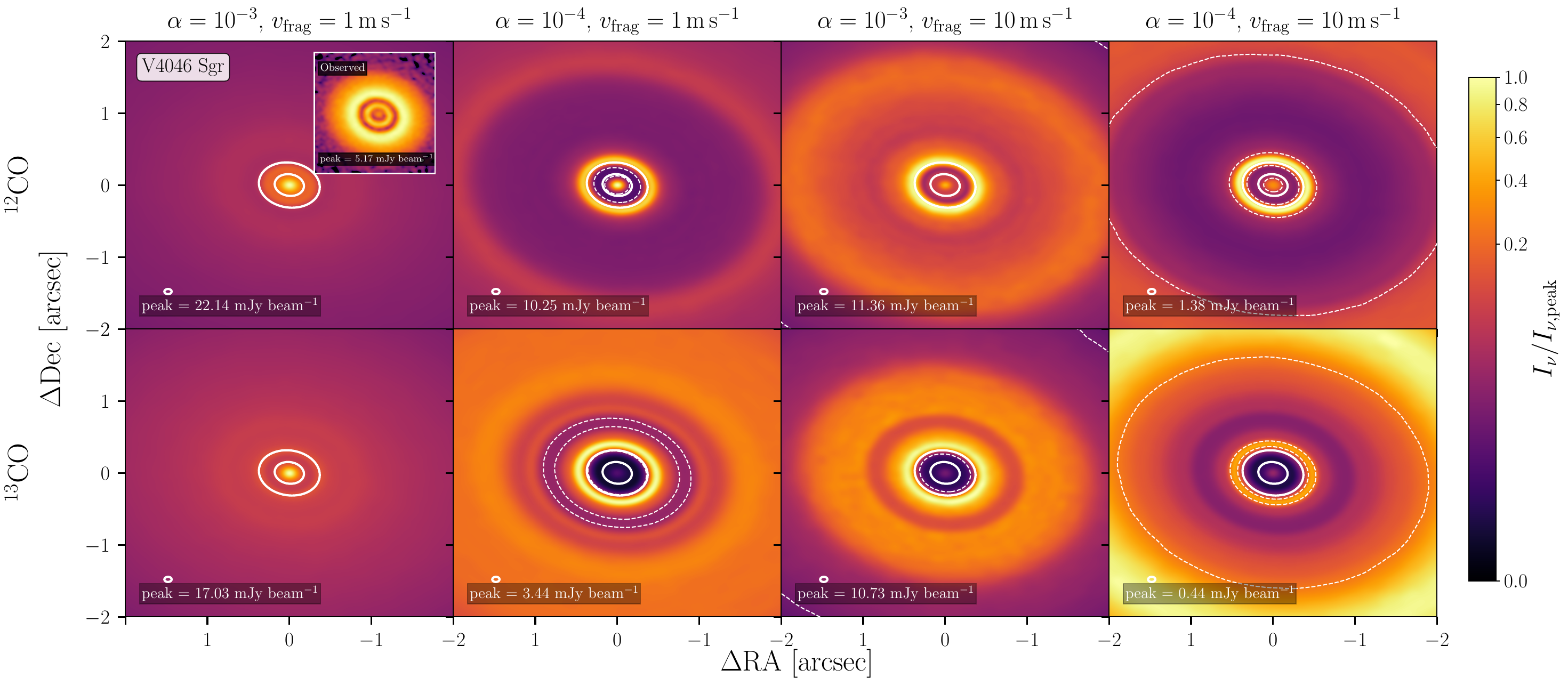}
    \caption{\texttt{RADMC} synthetic observations of the dustpy $^{12}$CO and $^{13}$CO models. The fluxes are normalised to the peak. The inset shows the observation in B7 \citep{Curone_2025}. The white ellipses indicate dust rings identified in the Band~7 continuum observations \citep{Curone_2025}. The white dashed lines indicate the 3$\sigma$ threshold based on the sensitivity of the observations. The four columns represent different dust evolution model configurations spanning $\alphavisc \in \{10^{-4},10^{-3}\}$ and $\vfrag \in \{1,10\}\,\mathrm{m\,s^{-1}}$. }
    \label{fig:radmc_v4046}
\end{figure*}

\begin{figure*}
    \centering
    \includegraphics[
        width=\textwidth,
        height=0.9\textheight,
        keepaspectratio
    ]{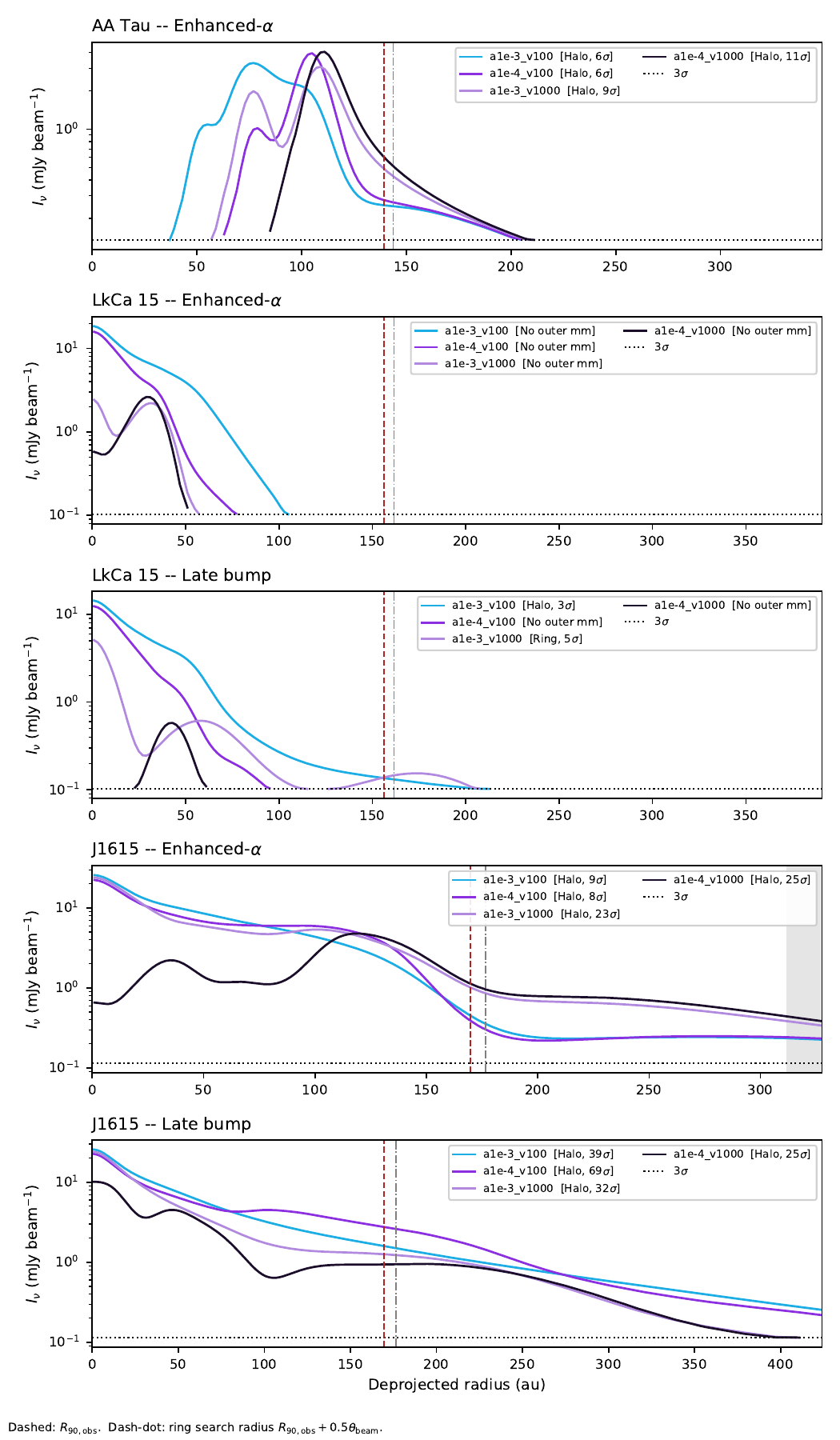}
    \caption{
Azimuthally averaged Band~7 radial intensity profiles for the enhanced-$\alpha$ and late-bump models. The dashed and dash-dotted vertical lines indicate $R_{90,\rm obs}$ and the minimum search radius for outer emission ($R_{90,\rm obs}+0.5\theta_{\rm beam}$), respectively, while the dotted horizontal line shows the $3\sigma$ detection threshold. The legend reports the model classification and the peak signal-to-noise ratio of the outer emission, showing the detectability and relative strength of the predicted outer dust emission.
}
\label{fig:emission_strength}
\end{figure*}

\label{lastpage}
\end{document}